\documentclass[superscriptaddress,aps,prc,showkeys,showpacs,twocolumn,10pt]{revtex4}
\usepackage[OT1,T1]{fontenc}
\usepackage{graphicx}
\usepackage{rotating}
\begin{document}

\title{On the Production of an Isotensor Dibaryon in the $pp \to pp\pi^+\pi^-$ Reaction}  
\date{\today}

\newcommand*{\IKPUU}{Division of Nuclear Physics, Department of Physics and 
 Astronomy, Uppsala University, Box 516, 75120 Uppsala, Sweden}
\newcommand*{\ASWarsN}{Nuclear Physics Division, National Centre for 
 Nuclear Research, ul.\ Hoza~69, 00-681, Warsaw, Poland}
\newcommand*{\IPJ}{Institute of Physics, Jagiellonian University, prof.\ 
 Stanis{\l}awa {\L}ojasiewicza~11, 30-348 Krak\'{o}w, Poland}
\newcommand*{\Edinb}{Department of Physics, University of York, Heslington,
  York, YO10 5DD, UK
}
\newcommand*{\MS}{Institut f\"ur Kernphysik, Westf\"alische 
 Wilhelms--Universit\"at M\"unster, Wilhelm--Klemm--Str.~9, 48149 M\"unster, 
 Germany}
\newcommand*{\ASWarsH}{High Energy Physics Division, National Centre for 
 Nuclear Research, ul.\ Hoza~69, 00-681, Warsaw, Poland}
\newcommand*{\Budker}{Budker Institute of Nuclear Physics of SB RAS, 
 11~Acad.\ Lavrentieva Pr., Novosibirsk, 630090 Russia}
\newcommand*{\Novosib}{Novosibirsk State University, 2~Pirogova Str., 
 Novosibirsk, 630090 Russia}
\newcommand*{\PGI}{Peter Gr\"unberg Institut, PGI--6 Elektronische 
 Eigenschaften, Forschungszentrum J\"ulich, 52425 J\"ulich, Germany}
\newcommand*{\DUS}{Institut f\"ur Laser-- und Plasmaphysik, Heinrich Heine 
 Universit\"at D\"usseldorf, Universit\"atsstr.~1, 40225 D\"usseldorf, Germany}
\newcommand*{\IFJ}{The Henryk Niewodnicza{\'n}ski Institute of Nuclear 
 Physics, Polish Academy of Sciences, ul.\ Radzikowskiego~152, 31-342 
 Krak\'{o}w, Poland}
\newcommand*{\PITue}{Physikalisches Institut, Eberhard Karls Universit\"at 
 T\"ubingen, Auf der Morgenstelle~14, 72076 T\"ubingen, Germany}
\newcommand*{\Kepler}{Kepler Center for Astro and Particle Physics,
  Physikalisches Institut der Universit\"at T\"ubingen, Auf der 
 Morgenstelle~14, 72076 T\"ubingen, Germany}
\newcommand*{\IKPJ}{Institut f\"ur Kernphysik, Forschungszentrum J\"ulich, 
 52425 J\"ulich, Germany}
\newcommand*{\ZELJ}{Zentralinstitut f\"ur Engineering, Elektronik und 
 Analytik, Forschungszentrum J\"ulich, 52425 J\"ulich, Germany}
\newcommand*{\Erl}{Physikalisches Institut, Friedrich--Alexander Universit\"at
 Erlangen--N\"urnberg, Erwin--Rommel-Str.~1, 91058 Erlangen, Germany}
\newcommand*{\ITEP}{Institute for Theoretical and Experimental Physics named 
 by A.I.\ Alikhanov of National Research Centre ``Kurchatov Institute'', 
 25~Bolshaya Cheremushkinskaya Str., Moscow, 117218 Russia}
\newcommand*{\Giess}{II.\ Physikalisches Institut, 
 Justus--Liebig--Universit\"at Gie{\ss}en, Heinrich--Buff--Ring~16, 35392 
 Giessen, Germany}
\newcommand*{\IITI}{Discipline of Physics, Indian Institute of Technology 
 Indore, Khandwa Road, Indore, Madhya Pradesh 453 552, India}
\newcommand*{\HepGat}{High Energy Physics Division, Petersburg Nuclear Physics 
 Institute named by B.P.\ Konstantinov of National Research Centre ``Kurchatov 
 Institute'', 1~mkr.\ Orlova roshcha, Leningradskaya Oblast, Gatchina, 188300
 Russia}
\newcommand*{\HeJINR}{Veksler and Baldin Laboratory of High Energiy Physics, 
 Joint Institute for Nuclear Physics, 6~Joliot--Curie, Dubna, 141980 Russia}
\newcommand*{\Katow}{August Che{\l}kowski Institute of Physics, University of 
  Silesia, ul.\ 75 Pu{\l}ku Piechoty 1, 41-500 Chorz\'{o}w, Poland}
\newcommand*{\NITJ}{Department of Physics, Malaviya National Institute of 
 Technology Jaipur, JLN Marg, Jaipur, Rajasthan 302 017, India}
\newcommand*{\JARA}{JARA--FAME, J\"ulich Aachen Research Alliance, 
 Forschungszentrum J\"ulich, 52425 J\"ulich, and RWTH Aachen, 52056 Aachen, 
 Germany}
\newcommand*{\Bochum}{Institut f\"ur Experimentalphysik I, Ruhr--Universit\"at 
 Bochum, Universit\"atsstr.~150, 44780 Bochum, Germany}
\newcommand*{\IITB}{Department of Physics, Indian Institute of Technology 
 Bombay, Powai, Mumbai, Maharashtra 400 076, India}
\newcommand*{\Tomsk}{Department of Physics, Tomsk State University, 36~Lenin 
 Ave., Tomsk, 634050 Russia}
\newcommand*{\KEK}{High Energy Accelerator Research Organisation KEK, Tsukuba, 
 Ibaraki 305--0801, Japan} 
\newcommand*{\ASLodz}{Astrophysics Division, National Centre for Nuclear
 Research, Box~447, 90-950 {\L}\'{o}d\'{z}, Poland}

\author{P.~Adlarson}    \affiliation{\IKPUU}
\author{W.~Augustyniak} \affiliation{\ASWarsN}
\author{W.~Bardan}      \affiliation{\IPJ}
\author{M.~Bashkanov}   \affiliation{\Edinb}
\author{F.S.~Bergmann}  \affiliation{\MS}
\author{M.~Ber{\l}owski}\affiliation{\ASWarsH}
\author{A.~Bondar}      \affiliation{\Budker}\affiliation{\Novosib}
\author{M.~B\"uscher}   \affiliation{\PGI}\affiliation{\DUS}
\author{H.~Cal\'{e}n}   \affiliation{\IKPUU}
\author{I.~Ciepa{\l}}   \affiliation{\IFJ}
\author{H.~Clement}     \affiliation{\PITue}\affiliation{\Kepler}
\author{E.~Czerwi{\'n}ski}\affiliation{\IPJ}
\author{K.~Demmich}     \affiliation{\MS}
\author{R.~Engels}      \affiliation{\IKPJ}
\author{A.~Erven}       \affiliation{\ZELJ}
\author{W.~Erven}       \affiliation{\ZELJ}
\author{W.~Eyrich}      \affiliation{\Erl}
\author{P.~Fedorets}    \affiliation{\IKPJ}\affiliation{\ITEP}
\author{K.~F\"ohl}      \affiliation{\Giess}
\author{K.~Fransson}    \affiliation{\IKPUU}
\author{F.~Goldenbaum}  \affiliation{\IKPJ}
\author{A.~Goswami}     \affiliation{\IKPJ}\affiliation{\IITI}
\author{K.~Grigoryev}   \affiliation{\IKPJ}\affiliation{\HepGat}
\author{L.~Heijkenskj\"old}\altaffiliation[present address: ]{\Mainz}\affiliation{\IKPUU}
\author{V.~Hejny}       \affiliation{\IKPJ}
\author{N.~H\"usken}    \affiliation{\MS}
\author{L.~Jarczyk}     \affiliation{\IPJ}
\author{T.~Johansson}   \affiliation{\IKPUU}
\author{B.~Kamys}       \affiliation{\IPJ}
\author{G.~Kemmerling}\altaffiliation[present address: ]{\JCNS}\affiliation{\ZELJ}
\author{A.~Khoukaz}     \affiliation{\MS}
\author{O.~Khreptak}    \affiliation{\IPJ}
\author{D.A.~Kirillov}  \affiliation{\HeJINR}
\author{S.~Kistryn}     \affiliation{\IPJ}
\author{H.~Kleines}\altaffiliation[present address: ]{\JCNS}\affiliation{\ZELJ}
\author{B.~K{\l}os}     \affiliation{\Katow}
\author{W.~Krzemie{\'n}}\affiliation{\ASWarsH}
\author{P.~Kulessa}     \affiliation{\IFJ}
\author{A.~Kup\'{s}\'{c}}\affiliation{\IKPUU}\affiliation{\ASWarsH}
\author{K.~Lalwani}     \affiliation{\NITJ}
\author{D.~Lersch}\altaffiliation[present address: ]{\FSU}\affiliation{\IKPJ}
\author{B.~Lorentz}     \affiliation{\IKPJ}
\author{A.~Magiera}     \affiliation{\IPJ}
\author{R.~Maier}       \affiliation{\IKPJ}\affiliation{\JARA}
\author{P.~Marciniewski}\affiliation{\IKPUU}
\author{B.~Maria{\'n}ski}\affiliation{\ASWarsN}
\author{H.--P.~Morsch}  \affiliation{\ASWarsN}
\author{P.~Moskal}      \affiliation{\IPJ}
\author{H.~Ohm}         \affiliation{\IKPJ}
\author{W.~Parol}       \affiliation{\IFJ}
\author{E.~Perez del Rio}\altaffiliation[present address: ]{\INFN}\affiliation{\PITue}\affiliation{\Kepler}
\author{N.M.~Piskunov}  \affiliation{\HeJINR}
\author{D.~Prasuhn}     \affiliation{\IKPJ}
\author{D.~Pszczel}     \affiliation{\IKPUU}\affiliation{\ASWarsH}
\author{K.~Pysz}        \affiliation{\IFJ}
\author{J.~Ritman}\affiliation{\IKPJ}\affiliation{\JARA}\affiliation{\Bochum}
\author{A.~Roy}         \affiliation{\IITI}
\author{Z.~Rudy}        \affiliation{\IPJ}
\author{O.~Rundel}      \affiliation{\IPJ}
\author{S.~Sawant}      \affiliation{\IITB}
\author{S.~Schadmand}   \affiliation{\IKPJ}
\author{I.~Sch\"atti--Ozerianska}\affiliation{\IPJ}
\author{T.~Sefzick}     \affiliation{\IKPJ}
\author{V.~Serdyuk}     \affiliation{\IKPJ}
\author{B.~Shwartz}     \affiliation{\Budker}\affiliation{\Novosib}
\author{T.~Skorodko}\affiliation{\PITue}\affiliation{\Kepler}\affiliation{\Tomsk}
\author{M.~Skurzok}     \affiliation{\IPJ}
\author{J.~Smyrski}     \affiliation{\IPJ}
\author{V.~Sopov}       \affiliation{\ITEP}
\author{R.~Stassen}     \affiliation{\IKPJ}
\author{J.~Stepaniak}   \affiliation{\ASWarsH}
\author{E.~Stephan}     \affiliation{\Katow}
\author{G.~Sterzenbach} \affiliation{\IKPJ}
\author{H.~Stockhorst}  \affiliation{\IKPJ}
\author{H.~Str\"oher}   \affiliation{\IKPJ}\affiliation{\JARA}
\author{A.~Szczurek}    \affiliation{\IFJ}
\author{A.~Trzci{\'n}ski}\affiliation{\ASWarsN}
\author{M.~Wolke}       \affiliation{\IKPUU}
\author{A.~Wro{\'n}ska} \affiliation{\IPJ}
\author{P.~W\"ustner}   \affiliation{\ZELJ}
\author{A.~Yamamoto}    \affiliation{\KEK}
\author{J.~Zabierowski} \affiliation{\ASLodz}
\author{M.J.~Zieli{\'n}ski}\affiliation{\IPJ}
\author{J.~Z{\l}oma{\'n}czuk}\affiliation{\IKPUU}
\author{J.~Z{\l}oma{\'n}czuk}\affiliation{\IKPUU}
\author{P.~{\.Z}upra{\'n}ski}\affiliation{\ASWarsN}
\author{M.~{\.Z}urek}   \affiliation{\IKPJ}

\newcommand*{\Mainz}{Institut f\"ur Kernphysik, Johannes 
 Gutenberg Universit\"at Mainz, Johann--Joachim--Becher Weg~45, 55128 Mainz, 
 Germany}
\newcommand*{\JCNS}{J\"ulich Centre for Neutron Science JCNS, 
 Forschungszentrum J\"ulich, 52425 J\"ulich, Germany}
\newcommand*{\FSU}{Department of Physics, Florida State University,
  77~Chieftan Way, Tallahassee, FL~32306-4350, USA}
\newcommand*{\INFN}{INFN, Laboratori Nazionali di Frascati, Via E. Fermi, 40, 
 00044 Frascati (Roma), Italy}

\collaboration{WASA-at-COSY Collaboration}\noaffiliation

\begin{abstract}

The quasi-free $pp \to pp\pi^+\pi^-$ reaction has been measured by means of
$pd$ collisions at $T_p$ = 1.2 GeV using the WASA detector setup at COSY
enabling exclusive and kinematically complete measurements. Total and
differential cross sections have been extracted for the energy region $T_p =
1.08 - 1.36$ GeV ($\sqrt s$ = 2.35 -2.46 GeV) covering thus the regions of
$N^*(1440)$ and $\Delta(1232)\Delta(1232)$ resonance excitations. Calculations
describing these excitations by $t$-channel meson exchange as well as isospin
relations based on the $pp \to pp\pi^0\pi^0$ data underpredict substantially
the measured total cross section. The calculations are also at variance with
specific experimental differential cross sections. An isotensor $\Delta N$
dibaryon resonance  
 with $I(J^P) = 2(1^+)$ produced associatedly with a pion is able to overcome
 these deficiencies. Such a dibaryon was predicted by Dyson and Xuong
 and more recently calculated by Gal and Garcilazo.

\end{abstract}







\maketitle

\section{Introduction}

Early measurements of two-pion production initiated by nucleon-nucleon ($NN$) 
collisions were conducted with bubble-chambers, where due to low statistics
primarily only results for total cross sections were obtained
\cite{Dakhno,Brunt,Shimizu,Sarantsev,Eisner,Pickup,kek}.

In recent years the two-pion production has been measured from threshold up to
incident energies of $T_p$ = 1.4 GeV with high-accuracy by exclusive and
kinematically complete experiments conducted at CELSIUS
\cite{WB2,JJ,JP,TS,iso,FK,deldel,nnpipi,prl2009}, COSY
\cite{tt,evd,AE,prl2011,poldpi0pi0,isofus,pp0-,np00}, GSI \cite{Hades} and
JINR \cite{Jerus}. Whereas initially proton-proton ($pp$) induced two-pion
production was the primary aim of these measurements
\cite{WB2,JJ,JP,TS,iso,FK,deldel,nnpipi,tt,evd,AE}, the interest moved later
to proton-neutron ($pn$) induced reaction channels -- after the first
clear-cut evidence for a dibaryon resonance with $I(J^P) = 0(3^+)$ had been 
observed in the $pn \to d\pi^0\pi^0$ reaction
\cite{prl2009,prl2011,poldpi0pi0}. Subsequent measurements of the $pn \to
d\pi^+\pi^-$\cite{isofus}, $pn \to pp\pi^0\pi^-$ \cite{pp0-}, $np \to
np\pi^0\pi^0$ \cite{np00} and $pn \to pn \pi^+\pi^-$ \cite{Hades,np+-}
reactions revealed that all two-pion production channels, which contain
isoscalar 
contributions, exhibit a signal of this resonance --- now called $d^*(2380)$
after observation of its pole in $pn$ scattering \cite{np,npfull,RWnew}. 

Aside from the dibaryon resonance phenomenon the standard theoretical
description of the two-pion production process at the energies of interest here
is dominated by $t$-channel meson exchange leading to excitation and decay of
the Roper resonance $N^*(1440)$ and of the $\Delta(1232)\Delta(1232)$ system
\cite{Luis,Zou}. At lower 
incident energies the Roper excitation dominates. At incident energies beyond
1 GeV the $\Delta\Delta$ process takes over. Such calculations give quite a
reasonable description of the data --- with the big exception of the $pp \to
pp\pi^0\pi^0$ cross section above 1 GeV. After readjusting the decay
branching of the Roper resonance used in these calculations to that obtained
in recent analyses of data on pion- and photon-induced two-pion production
\cite{SarantsevRoper,PDG}, a quantitative description of total and
differential cross section data was achieved for both the $pp \to pp\pi^0\pi^0$
and the $pp \to pp\pi^+\pi^-$ reactions at incident energies below 0.9 GeV
\cite{WB2,JJ,JP,TS,AE}, where the Roper excitation dominates. 

For a quantitative description of the $pp \to pp\pi^0\pi^0$ data above 1 GeV,
however, the
calculation of the $\Delta\Delta$ process as used originally in
Ref. \cite{Luis} had to be modified \cite{deldel}, in particular the $\rho$
exchange contribution had to be strongly reduced. Also, the strength of
the Roper excitation had to be reduced in accord with isospin decomposition
\cite{iso}, and in order to describe
the $pp \to nn\pi^+\pi^+$ reaction quantitatively, a contribution from a
higher-lying broad $\Delta$ resonance, {\it e.g.}, the $\Delta(1600)$, had to
be assumed 
\cite{nnpipi}. These calculations, called now "modified Valencia" calculations
give a good description of all data in $pp$-induced and also of $pn$-induced
channels -- if in the latter the $d^*(2380)$ resonance is taken into account
-- with one striking exception: the $pp \to pp\pi^+\pi^-$ total cross section
data beyond 0.9 GeV are strongly underpredicted (see
dashed line in Fig.~\ref{fig4}). This
problem was already noted in the isospin decomposition of $pp$-induced
two-pion production \cite{iso}. However, since all the $pp \to pp\pi^+\pi^-$
data beyond 0.8 GeV originate from early low-statistics bubble-chamber
measurements, it appears appropriate to reinvestigate this region by exclusive
and kinematically complete measurements.

There is yet another point of interest in this reaction at energies above 0.8
GeV. Dyson and Xuong \cite{Dyson} were the first, who properly predicted the
dibaryon resonances $d^*(2380)$ (called $D_{IJ} = D_{03}$ by Dyson and Xuong) and
$D_{12}$, the latter denoting a slightly 
bound $\Delta N$ threshold state with $I(J^P) = 1(2^+)$. For a recent
discussion about that state see {\it e.g.} Ref. \cite{hcl} and references
therein. According to 
Dyson and Xuong, as well as to recent Faddeev calculations performed by Gal and
Garcilazo \cite{GG}, there should exist still another $\Delta N$ threshold state
with $I(J^P) = 2(1^+)$, called $D_{21}$ in Ref. \cite{Dyson}. Very recently
also another theoretical study appeared stating that if existent $D_{21}$
should have a somewhat larger mass than $D_{12}$ based on the
spin-isospin splittings observed for baryons \cite{Nanjing}.

Because of its
large isospin of $I = 2$, this state cannot be excited directly by incident
$NN$ collisions, but only associatedly -- favorably by production of an
additional pion, which carries away one unit of isospin. By isospin selection
the decay of an isotensor $\Delta N$ state will dominantly proceed into the
purely isotensor $pp\pi^+$ channel. Hence the $pp \to pp\pi^+\pi^-$ reaction
is the ideal place to look for the process $pp \to D_{21}^{+++} \pi^- \to
pp\pi^+\pi^-$ -- as already suggested in Ref. \cite{Dyson}. The main results
of this investigation have been communicated recently in a Letter \cite{short}.

\section{Experiment}

Exclusive and kinematically complete measurements of the $pp \to pp\pi^+\pi^-$
reaction have been achieved by utilizing the quasifree
process in $pd$ collisions. The experiment was carried out with the WASA
detector at COSY (Forschungszentrum J\"ulich) having a proton beam of energy 
$T_p$~=~1.2~GeV hit a deuterium pellet target \cite{barg,wasa}. By use of the
quasi-free scattering process $p d \to pp\pi^+\pi^- + n_{spectator}$, we can
exploit the Fermi motion in the target deuteron and cover thus the energy
region $T_p = 1.08 - 1.36$ GeV ($\sqrt s = 2.35 - 2.46 GeV$) of the $pp \to
pp\pi^+\pi^-$ reaction. 

The data analysis used a hardware trigger, which required two  
charged hits in the forward detector and two charged hits in the
central detector.  

In the offline analysis the reaction $p d \to pp \pi^+\pi^- + n$
was selected by requiring two proton tracks in the forward detector in
addition to one $\pi^+$ and one $\pi^-$ track in the central detector.   

The unmeasured neutron four-momentum could be reconstructed that way by a
kinematic fit with one over-constraint. The achieved resolution in $\sqrt s$
was about 20 MeV.

For the identification of the charged particles registered in the segmented
forward detector of WASA we applied the $\Delta E - E$ energy loss method
using all combinations of signals stemming from the five layers of the forward 
range hodoscope. In the central detector the charged particles have been
identified by their curved track in the magnetic field as well as by their
energy loss in the surrounding plastic scintillator barrel and electromagnetic
calorimeter. 

\begin{figure} 
\centering
\includegraphics[width=0.89\columnwidth]{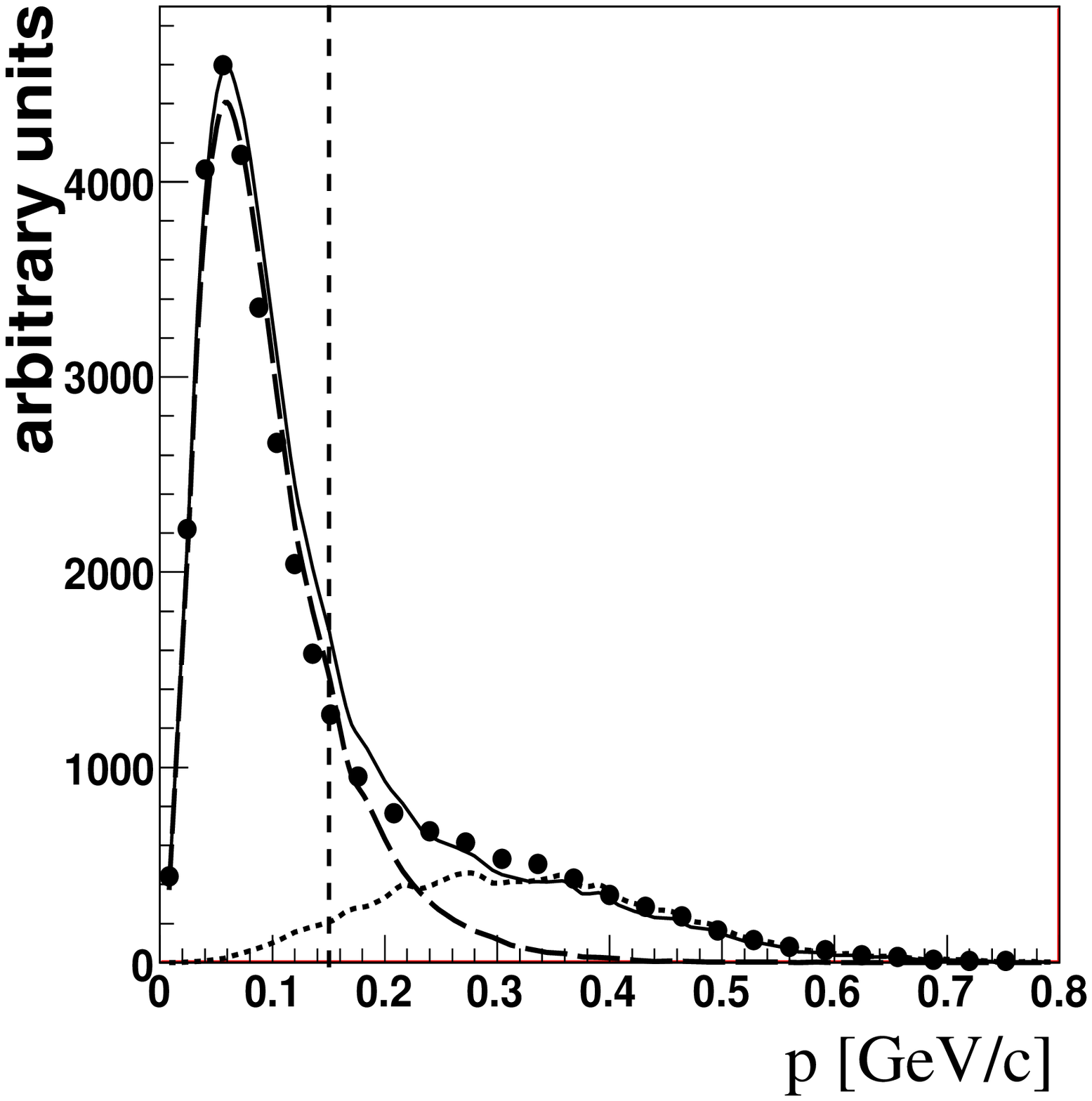}
\includegraphics[width=0.89\columnwidth]{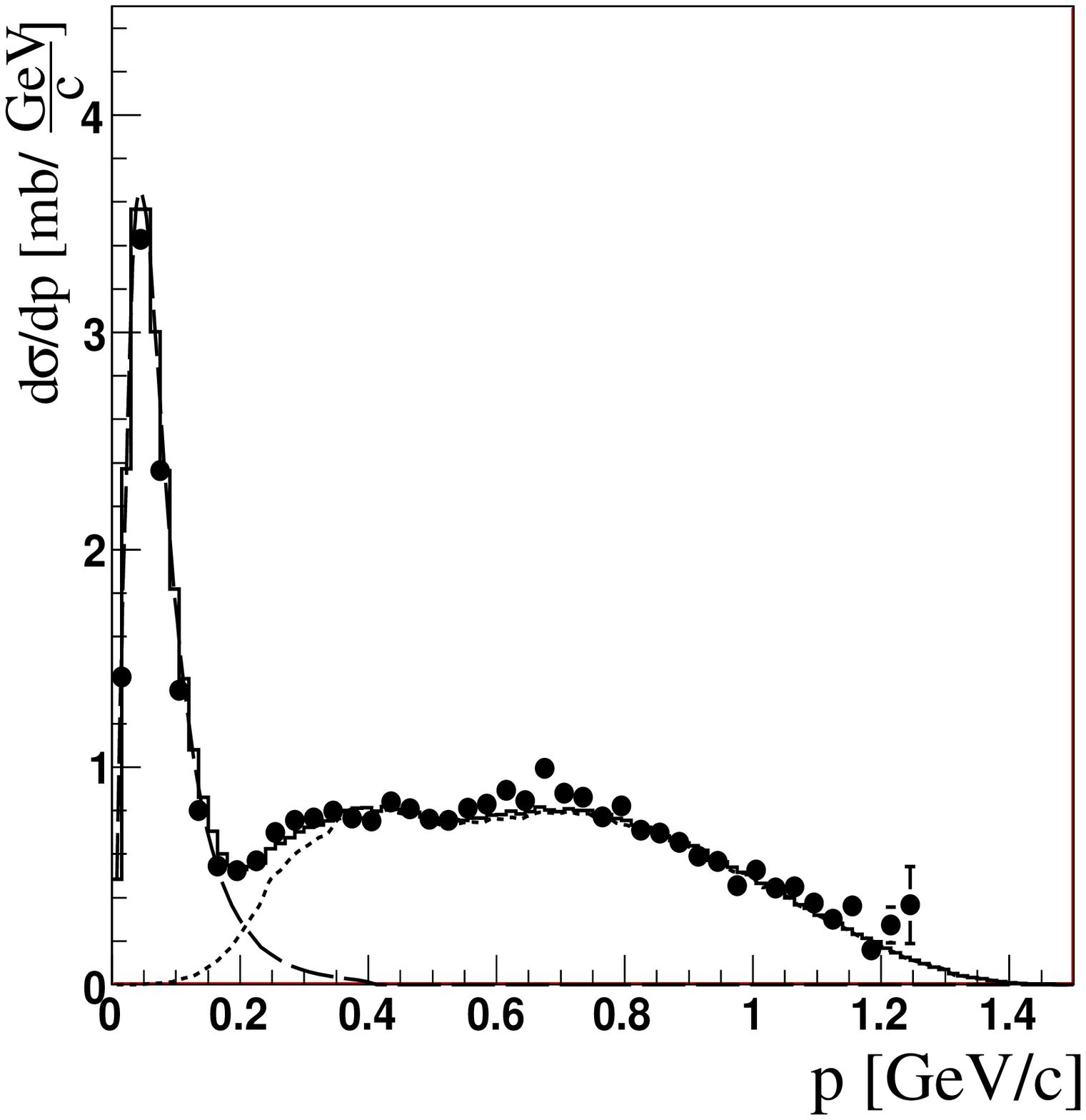}
\caption{\small (Color online)
  Distribution of the neutron momenta in the $pd \to npp\pi^+\pi^-$
  reaction before (top) and after acceptance and efficiency correction
  (bottom). Data are given by solid dots. The dashed line shows the expected
  distribution for the quasifree process  $pd \to pp\pi^+\pi^- +
  n_{spectator}$ based on the CD Bonn potential \cite{mach} deuteron
  wavefunction. The vertical line indicates the region $p <$ 0.15 GeV/c used
  for the evaluation of the quasifree process. The dotted line 
  gives the modeling of the non-quasifree reaction
  process. The solid line is the incoherent sum of both processes.
}
\label{fig1}
\end{figure}

The momentum distribution of the reconstructed neutron is shown in
Fig.~\ref{fig1}. The dashed curve gives the expected momentum distribution for
a spectator neutron according to the deuteron wavefunction based on the CD
Bonn potential \cite{mach}. Compared to previous 
measurements on $d\pi^0\pi^0$ \cite{prl2011}, $d\pi^+\pi^-$ \cite{isofus},
$np\pi^0\pi^0$ \cite{np00} and $pp\pi^0\pi^-$ \cite{pp0-} channels we find
here a somewhat enhanced background from non-spectator contributions. In order
to keep these background contributions smaller than 2$\%$, we would need
to restrict the spectator momentum range to $p <$ 0.10 GeV/c. But such a cut
would severely reduce the covered energy range to 1.15 GeV $< T_p <$ 1.3 GeV
(2.38 GeV $< \sqrt s <$ 2.44 GeV). To reliably evaluate
the data up to $p$ = 0.15 GeV/c for the quasifree reaction -- as done in our
previous analyses  -- we decided to perform a proper background correction by
analyzing additionally the non-spectator reaction 
process by evaluating the data in the non-overlap region $p_n >$
0.25 GeV/c.  

The instrumental acceptance has been 30$\%$ in case of the quasifree process
and about 5$\%$ in case of the non-quasifree reaction due to the requirement
that the two protons have to be in the angular range covered by the forward
detector and that $\pi^+$ and $\pi^-$ have to be in the angular range of the
central detector. The total reconstruction efficiency including all
cuts and conditions has been 1.1$\%$ for the quasifree process and
about 0.2$\%$ for the non-quasifree process. 
In total a sample of about 26000 events has been selected meeting all cuts
and conditions for the quasifree process $pd \to  pp\pi^+\pi^- +
n_{spectator}$.
For $p >$ 0.25 GeV/c in the region of the non-quasifree process this number
is about 20000.

Efficiency and acceptance corrections of the data have been performed by MC
simulations of reaction process and detector setup. For the MC simulations
both pure phase-space and model descriptions have been used, which will be
discussed below. Since WASA does not cover the full reaction
phase space, albeit a large fraction of it, these 
corrections are not fully model independent. The hatched grey histograms in 
Figs. \ref{fig2} - \ref{fig3} and \ref{fig5} - \ref{fig9} give an estimate for
these systematic uncertainties. As a measure of 
these we have taken the difference between model corrected results and those
obtained by assuming pure phase space for the acceptance corrections in case
of the non-spectator background reaction. In case of the quasifree $pp \to
pp\pi^+\pi^- + n_{spectator}$ reaction we
use the difference between the results obtained with the final model and those
using the "modified Valencia" model for the acceptance correction. Compared to
the uncertainties in these corrections, 
systematic errors associated with modeling the reconstruction of particles
are negligible.

The absolute normalization
of the data has been obtained by comparing the 
quasi-free single pion production process $pd \to pp \pi^0 + n_{spectator}$
to previous bubble-chamber results for the $pp
\to pp \pi^0$ reaction \cite{Shimizu,Eisner}. That way, the uncertainty in the
absolute normalization of our data is essentially that of the previous $pp \to
pp \pi^0$ data, {\it i.e.} in the order of 5 - 15$\%$.

\section{The non-spectator background process $pd \to ppn \pi^+\pi^-$}

\begin{figure} 
\centering
\includegraphics[width=0.49\columnwidth]{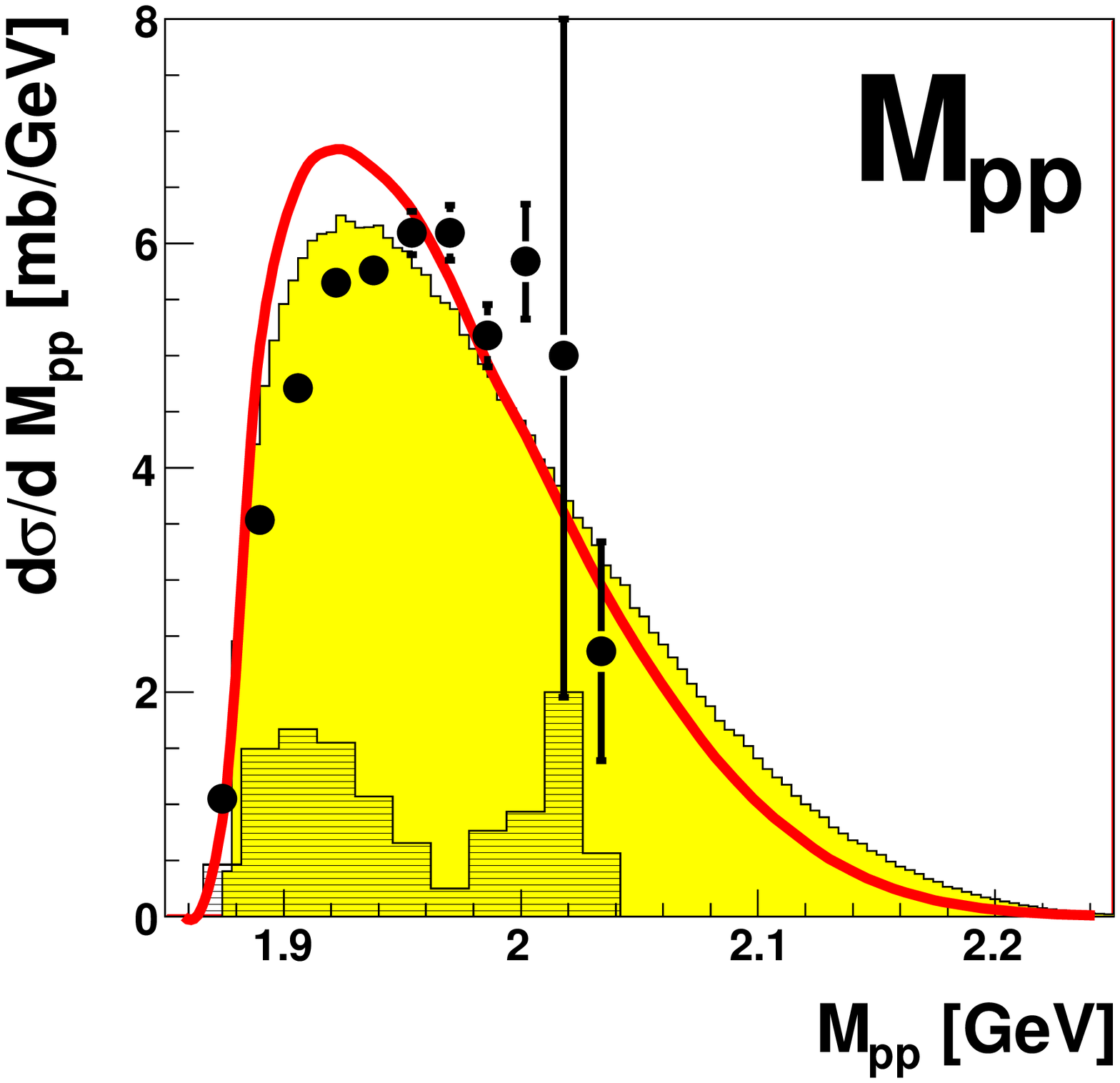}
\includegraphics[width=0.49\columnwidth]{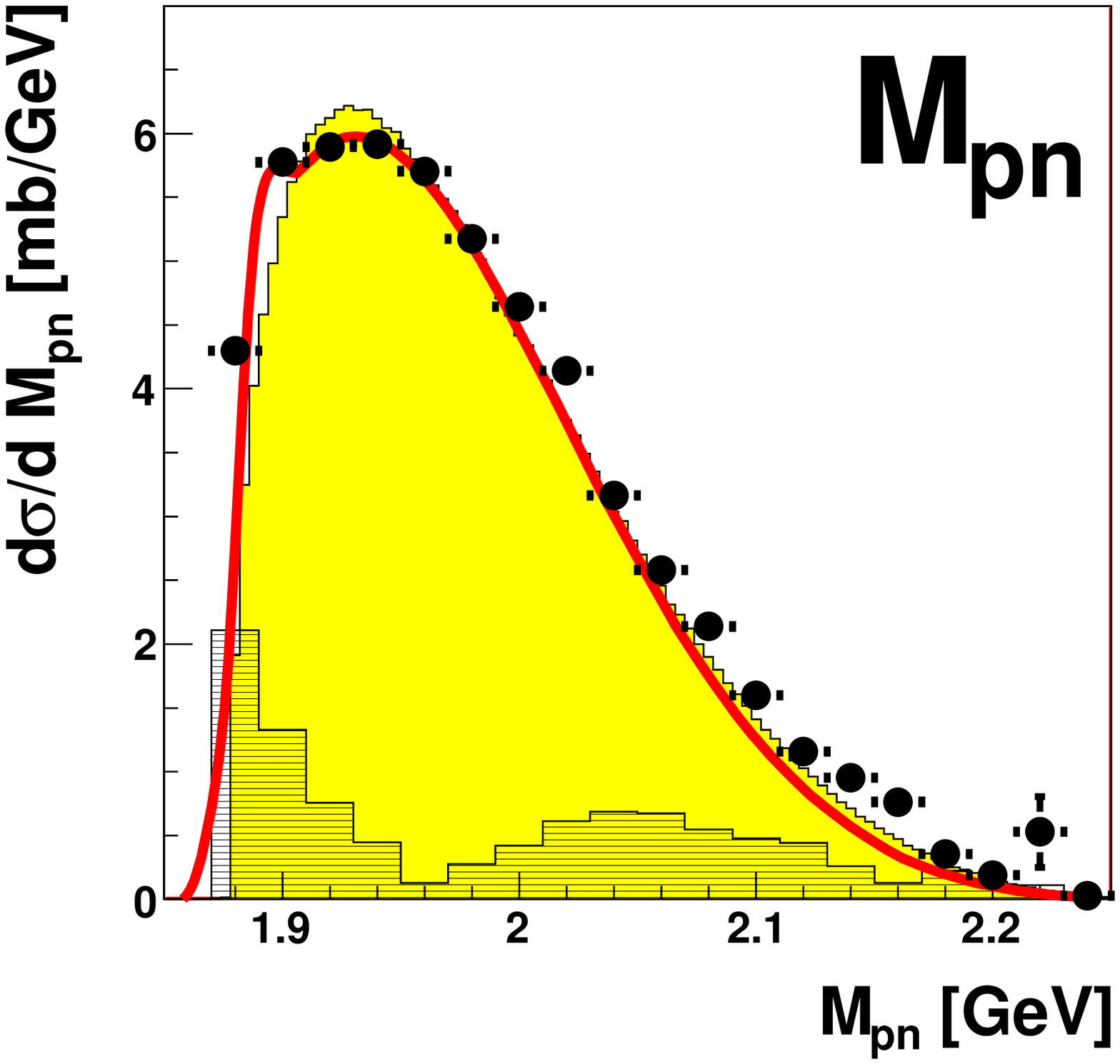}
\includegraphics[width=0.49\columnwidth]{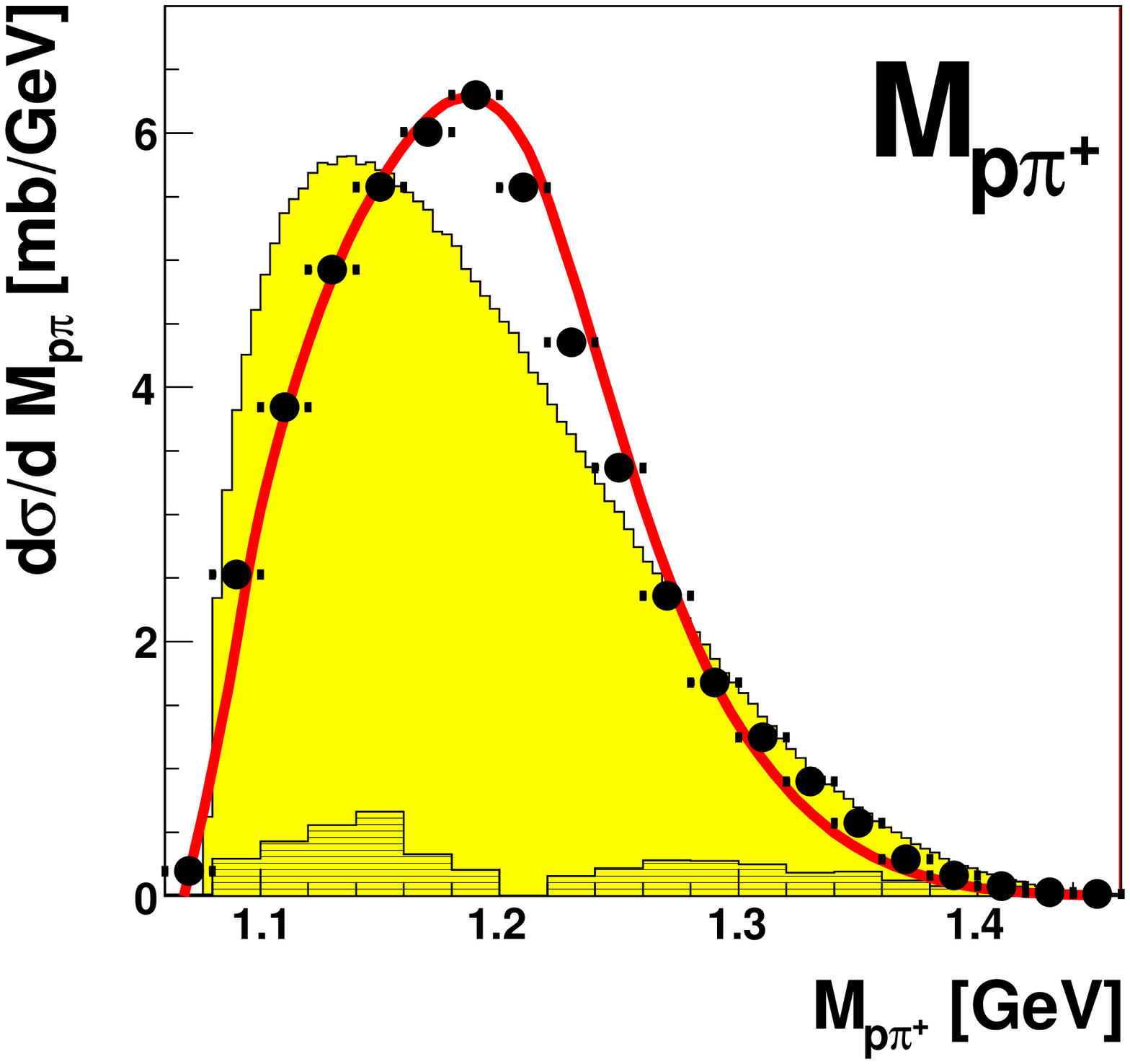}
\includegraphics[width=0.49\columnwidth]{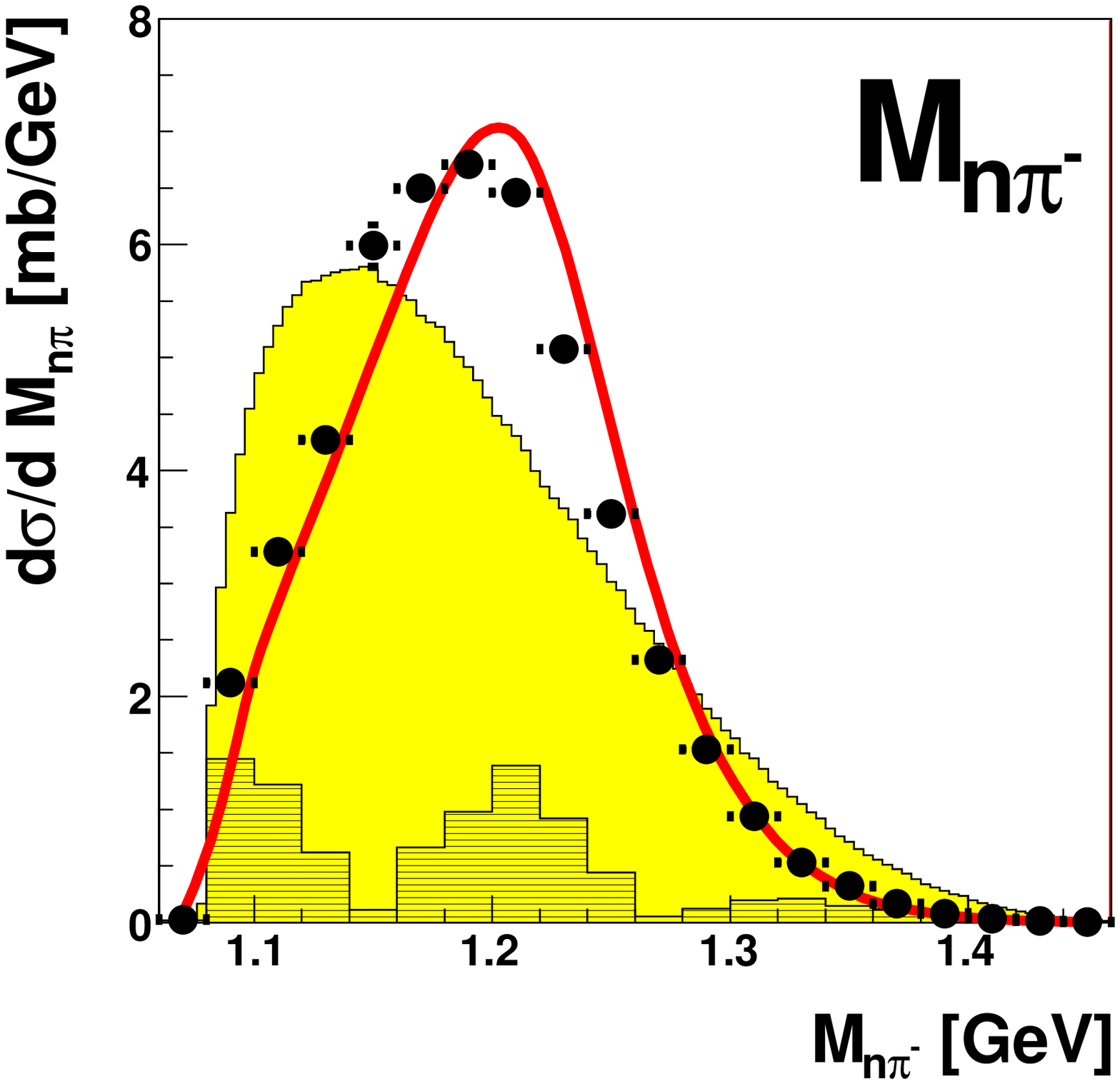}
\includegraphics[width=0.49\columnwidth]{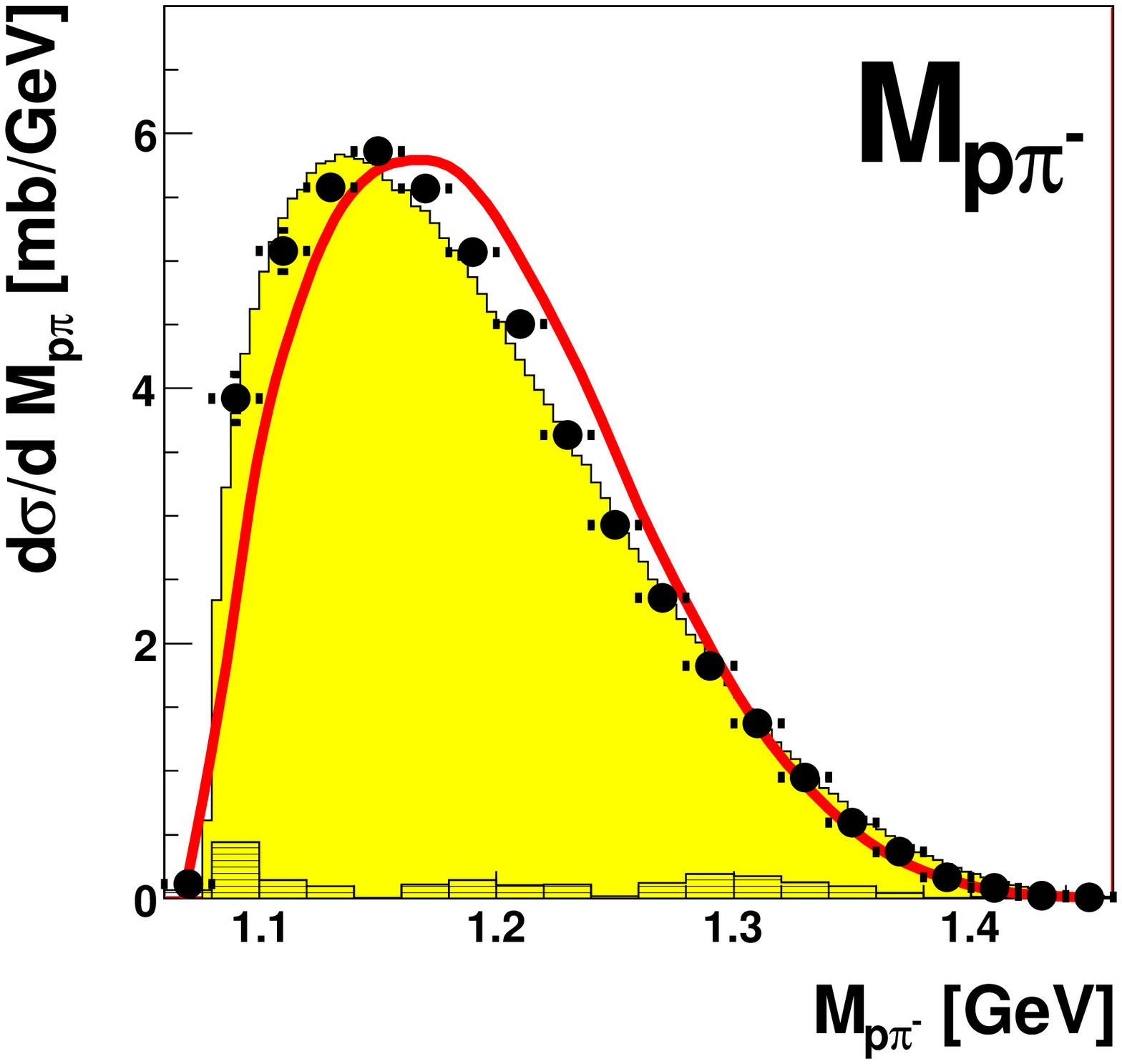}
\includegraphics[width=0.49\columnwidth]{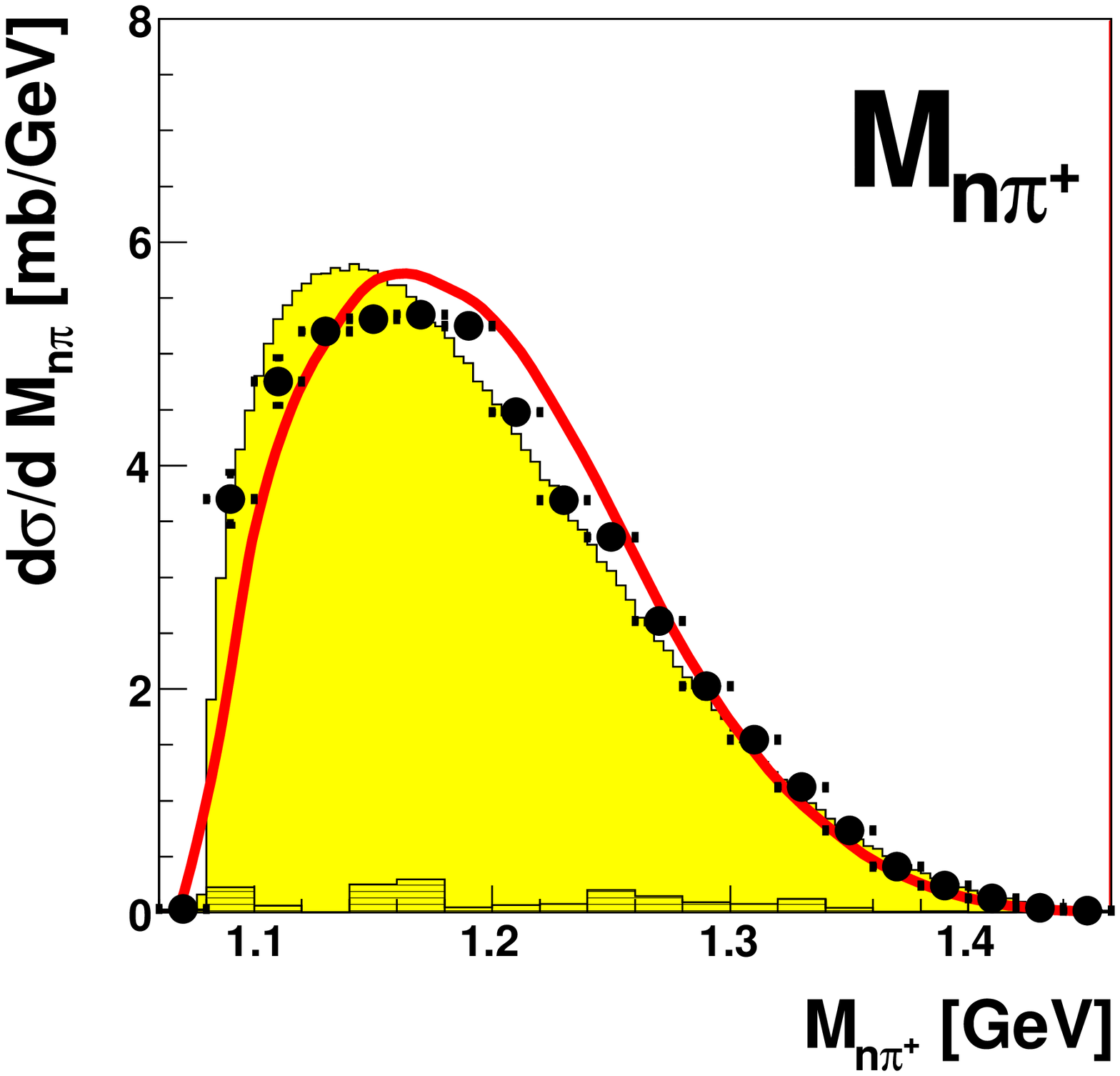}
\includegraphics[width=0.49\columnwidth]{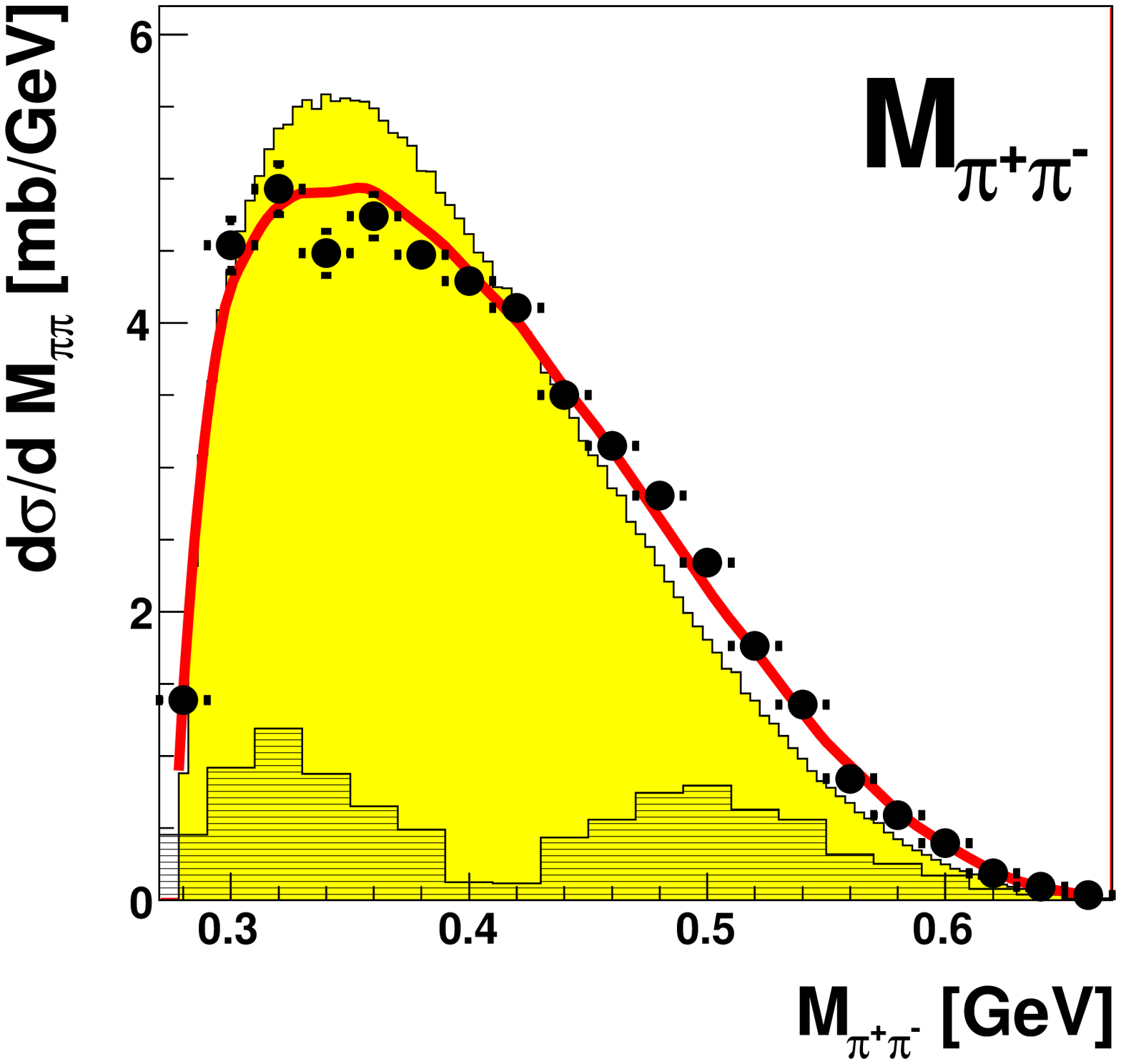}
\includegraphics[width=0.49\columnwidth]{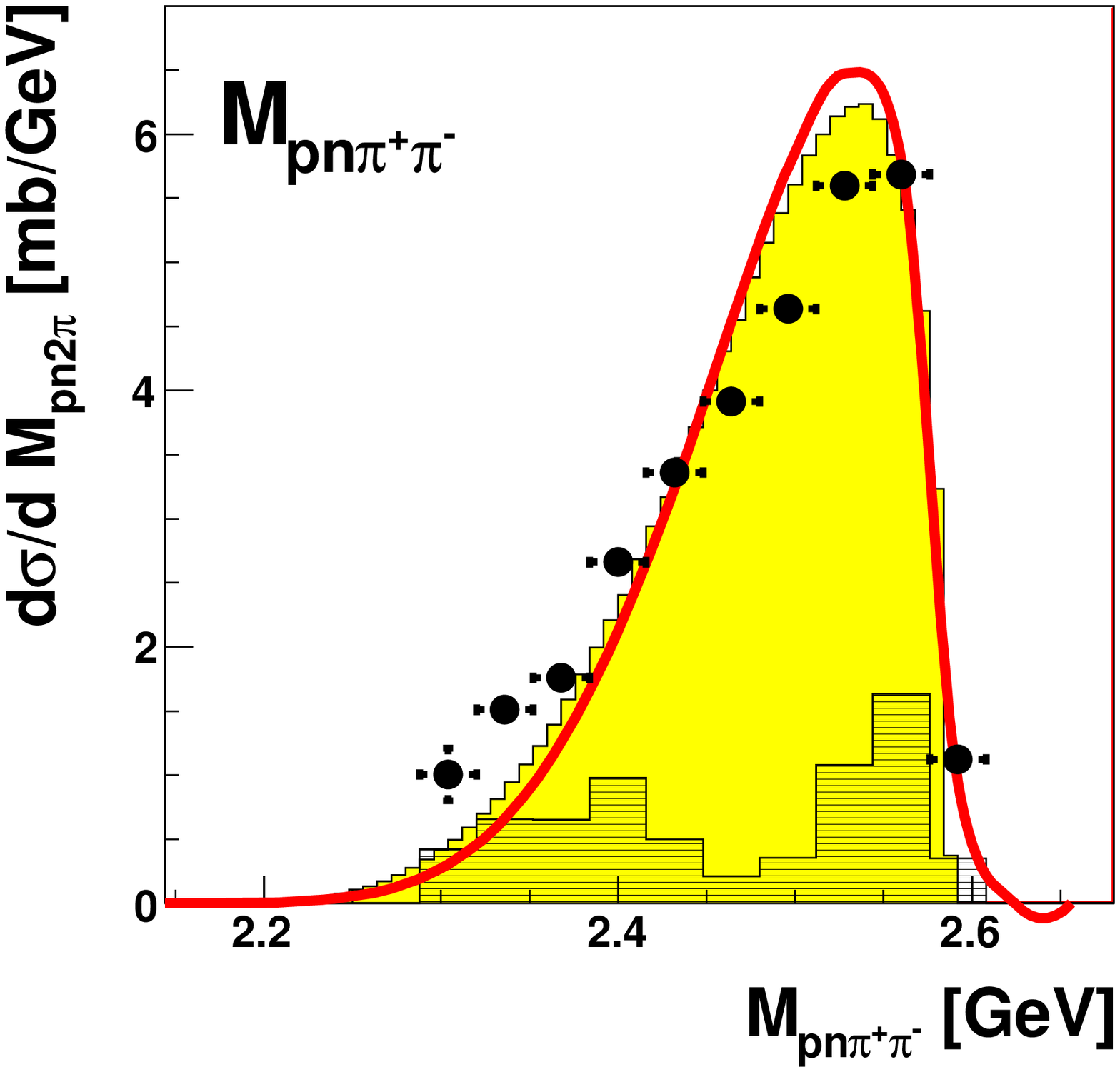}
\caption{\small (Color online)
  Differential distributions of the non-spectator background reaction $pd \to
  ppn\pi^+\pi^-$  
  for the invariant mass distributions $M_{pp}$, $M_{pn}$,
  $M_{p\pi^+}$, $M_{n\pi^-}$, $M_{p\pi^-}$, $M_{n\pi^+}$, $M_{\pi^+\pi^-}$ and
  $M_{pn\pi^+\pi^-}$ for $p_n >$ 0.2 GeV/c. The shaded areas represent pure
  phase-space distributions. The hatched areas indicate systematic
  uncertainties due to the restricted phase-space coverage in the
  measurement. The solid curves give a modeling of the
  process $pd \to ppn\pi^+\pi^-$.
}
\label{fig2}
\end{figure}

\begin{figure} 
\centering
\includegraphics[width=0.49\columnwidth]{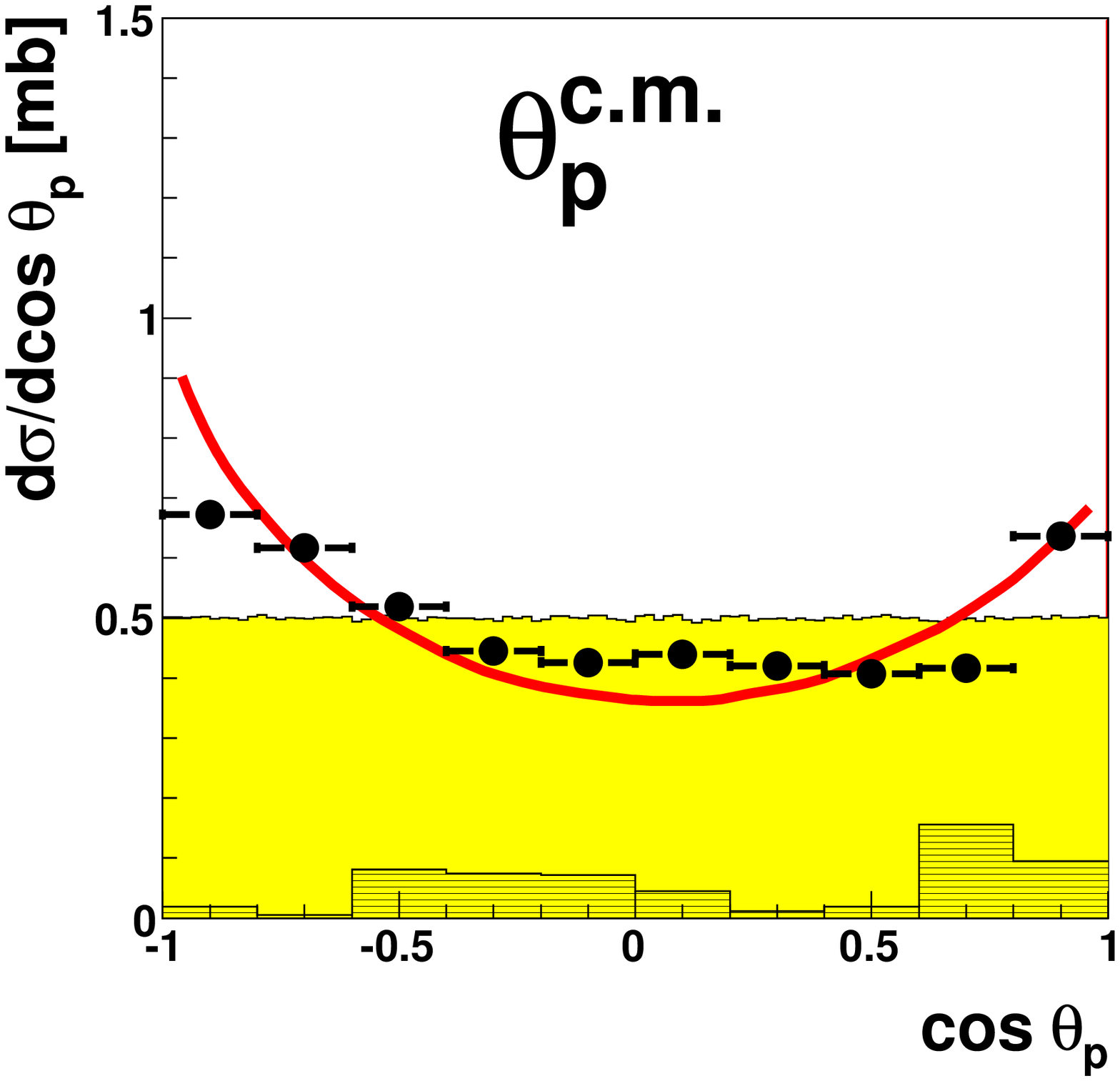}
\includegraphics[width=0.49\columnwidth]{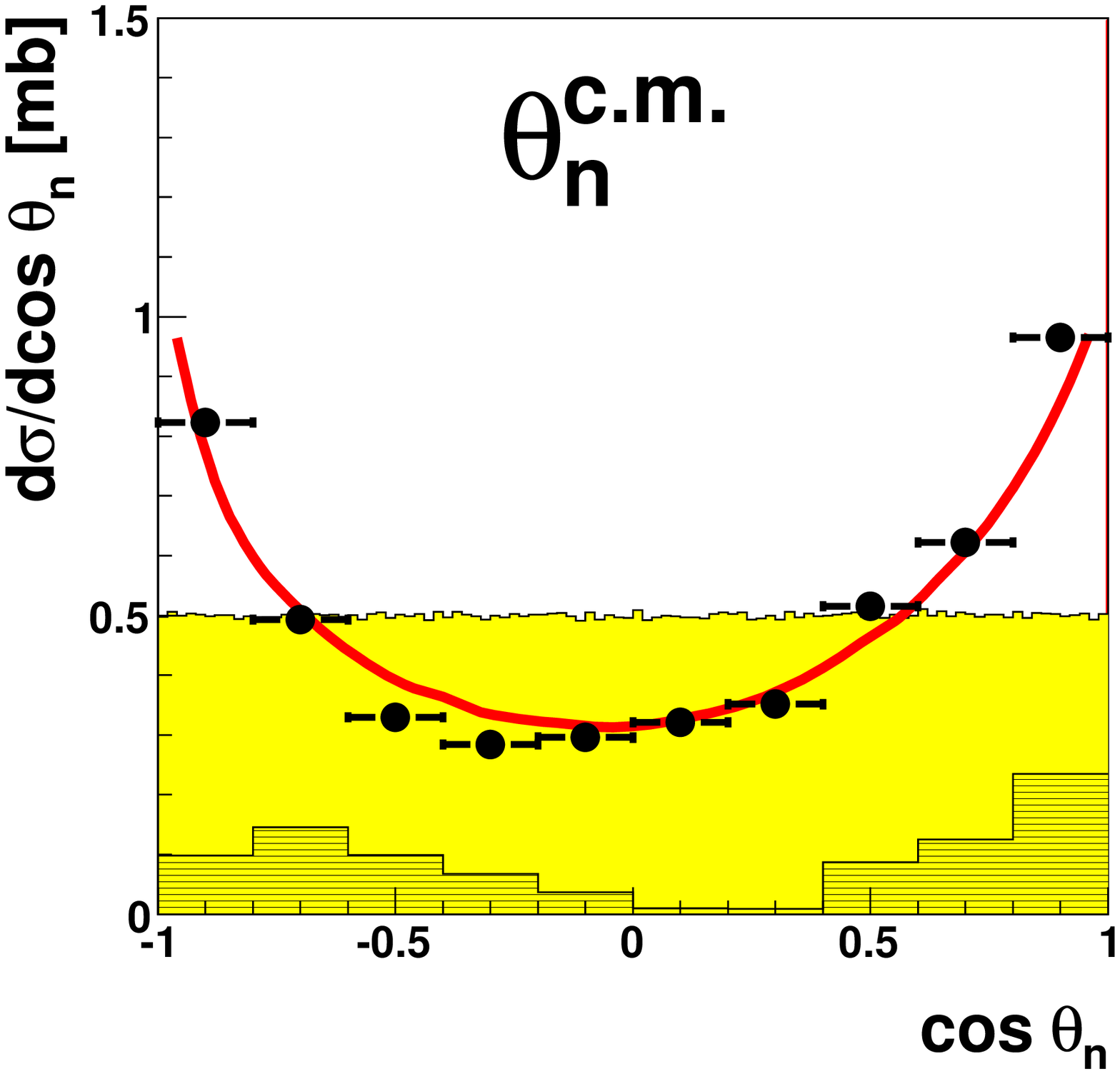}
\includegraphics[width=0.49\columnwidth]{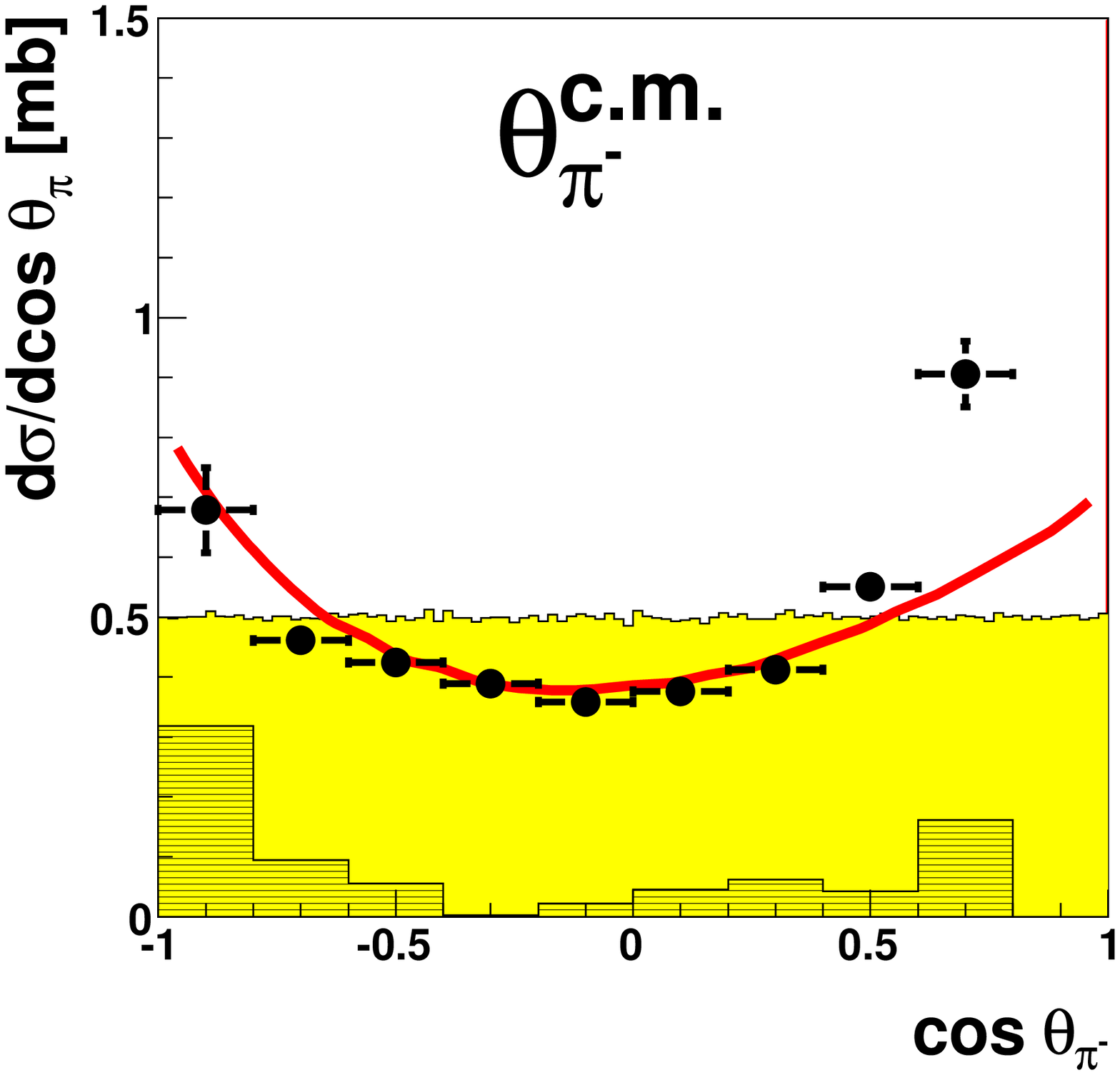}
\includegraphics[width=0.49\columnwidth]{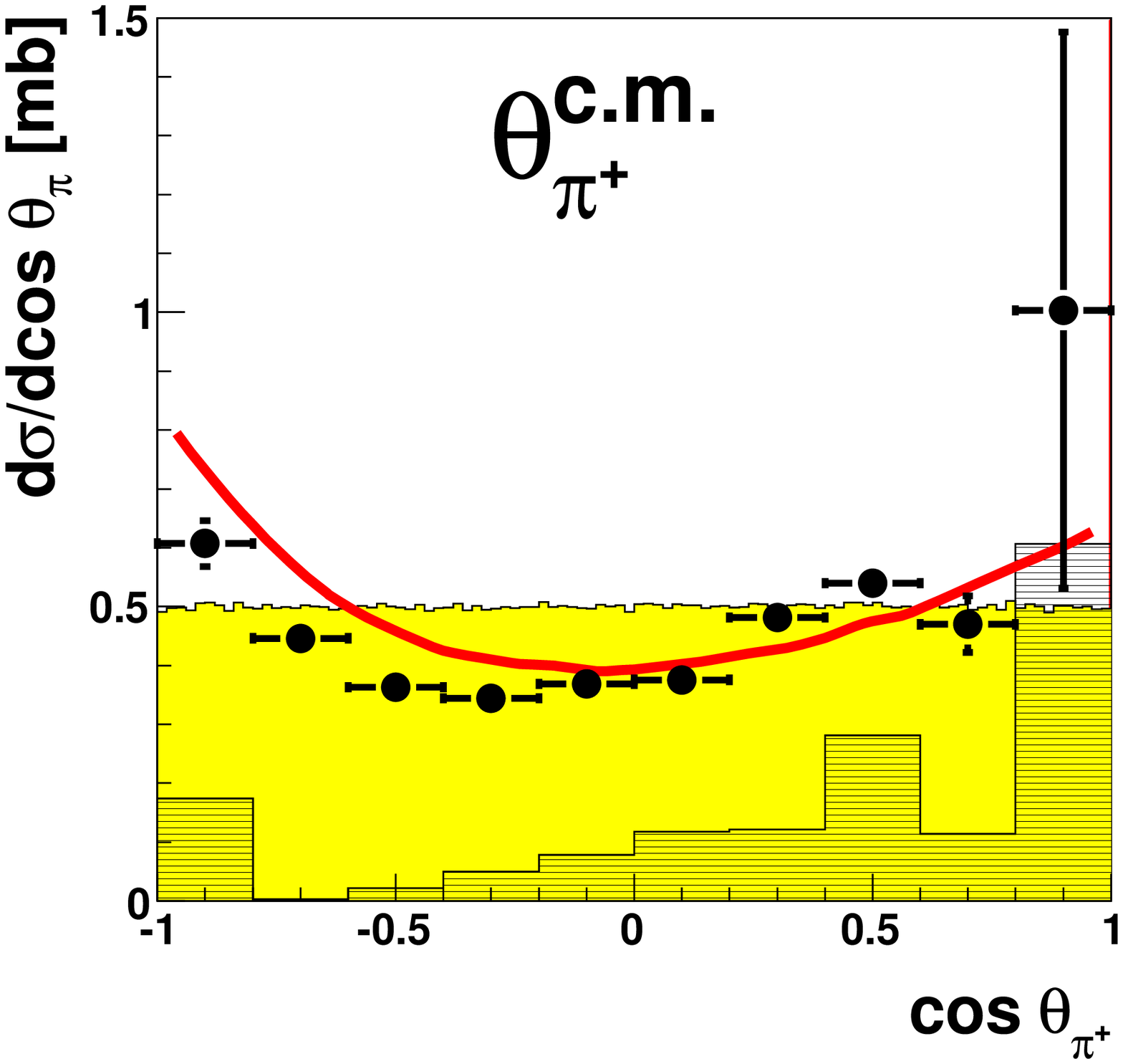}
\caption{\small (Color online)
  Same as Fig~\ref{fig2}, but for the angular distributions of protons
($\Theta_p^{c.m.}$), neutrons ($\Theta_n^{c.m.}$), positive pions
($\Theta_{\pi^+}^{c.m.}$) and negative pions ($\Theta_{\pi^-}^{c.m.}$).
}
\label{fig3}
\end{figure}

For an axially symmetric five-body final state there are eleven independent
differential observables. 
For the non-spectator background reaction $pd \to ppn\pi^+\pi^-$ we show in
Figs.~\ref{fig2} and \ref{fig3} twelve differential observables: the invariant
mass distributions 
$M_{pp}$, $M_{pn}$, 
$M_{p\pi^+}$, $M_{n\pi^-}$, $M_{p\pi^-}$, $M_{n\pi^+}$, $M_{\pi^+\pi^-}$ and
$M_{pn\pi^+\pi^-}$ as well as the angular distributions for protons
($\Theta_p^{c.m.}$), neutrons ($\Theta_n^{c.m.}$), positive pions
($\Theta_{\pi^+}^{c.m.}$) and negative pions ($\Theta_{\pi^-}^{c.m.}$).

The obtained differential distributions deviate partly substantially from
pure phase-space distributions. This is the case in particular for the
distributions of the 
invariant masses $M_{p\pi^+}$ and $M_{n\pi^-}$ exhibiting the excitations
of $\Delta^{++}$ and $\Delta^-$, as well as of the angles
$\Theta_p^{c.m.}$, $\Theta_{n}^{c.m.}$, $\Theta_{\pi^-}^{c.m.}$ and
$\Theta_{\pi^+}^{c.m.}$. However, all differential distributions fit well
to a modeling of the 
process $pd \to N\Delta\Delta \to ppn\pi^+\pi^-$. Since it proceeds dominantly
via the $\Delta^{++}\Delta^-$ configuration due to isospin selection,
the $M_{p\pi^+}$ and $M_{n\pi^-}$ spectra peak at the $\Delta$ mass,
as we observe in Fig.~\ref{fig2}. 
The pion angular distributions are as expected from the $p$-wave decay of the
intermediate $\Delta$ resonances.
Proton and neutron angular distributions are strongly curved as expected from 
a peripheral collision. In comparison to the neutron angular distribution, the
proton angular distribution appears to be less anisotropic, since only
one of the two protons is dominantly active. 

The success of such a background modeling is of no great
surprise, since the $pn\pi^+\pi^-$ channel has the by far largest two-pion
production cross section. Also we know from the $pd \to ^3$He$\pi\pi$
reaction, where the $ppn$ system has fused to $^3$He, that for $T_p >$ 1 GeV
the $t$-channel $\Delta\Delta$ process is by far dominating 
\cite{EP}. As in the latter case we observe also here the $\Delta$ signals in
the invariant mass spectra to be somewhat broadened, which may be traced back
to the Fermi motion of the participating nucleons and may be accounted for
most easily by increasing the $\Delta$ width from 120 to 140 MeV by a fit to
the data. 

Having achieved a quantitative description of the non-quasifree background
process for $p_n >$  0.25 GeV/c, we may extrapolate its
contribution reliably also for $p_n <$ 0.15 GeV/c and subtract it from the
measured neutron momentum distribution (Fig.~\ref{fig1}), in order to obtain
the pure quasi-free part, which is of main interest here.

\section{The Quasifree Reaction $pp \to pp\pi^+\pi^- + n_{spectator}$}

\subsection{Total cross section}

\begin{figure} 
\centering
\includegraphics[width=0.8\columnwidth]{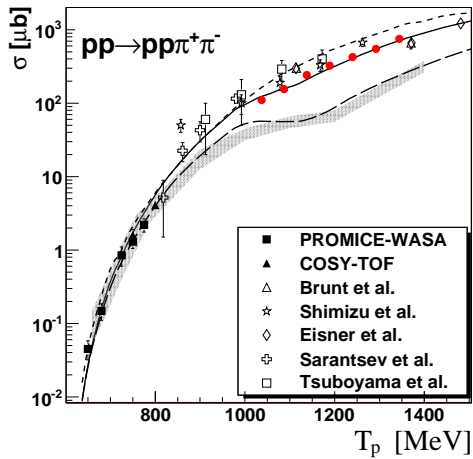}
\caption{\small (Color online) 
  Total cross section in dependence of the incident proton energy $T_p$ for
  the reaction $pp \to pp\pi^+\pi^-$. The solid dots show results from this
  work. Other symbols denote results from previous measurements
  \cite{Brunt,Shimizu,Sarantsev,WB2,JJ,JP,AE}. The dotted line gives the
  original 
  Valencia calculation \cite{Luis}, the dashed one the so-called "modified
  Valencia" calculation \cite{deldel}. The solid line is obtained, if to the
  latter an associatedly produced $D_{21}$ resonance is added according to the
  process $pp \to D_{21}\pi^- \to pp\pi^+\pi^-$ with the strength of this
  process being fitted to the total cross section data. The grey shaded band
  exhibits the results of eq. (1) based on the $pp\pi^0\pi^0$ channel.
}
\label{fig4}
\end{figure}

For the determination of the energy dependence of total and differential cross
sections  we have divided the background subtracted data for the quasi-free
process into bins of 50 MeV width in the incident energy $T_p$. The resulting
total cross sections are shown in Fig.~\ref{fig4} (solid dots) together with
results from earlier measurements (other symbols)
\cite{Brunt,Shimizu,Sarantsev,WB2,JJ,JP,AE}. Our data are in reasonable
agreement with the earlier measurements in the overlap energy region.

For comparison to theoretical expectations we first plot in
Fig. ~\ref{fig4} the results of the original Valencia calculations \cite{Luis}
by the dotted line. At first glance the agreement with the data appears
remarkable. However, as 
mentioned in the introduction, these calculations are far off for the
$pp\pi^0\pi^0$ channel. The so-called "modified Valencia" calculations, which
account reasonably well for the latter channel, are shown in Fig.~\ref{fig4}
by the dashed line. These calculations do very well at low energies, but yield
a much too low cross section at higher energies. The reason is that by isospin
relations the energy dependences of $pp\pi^0\pi^0$ and $pp\pi^+\pi^-$
channels have to be qualitatively  similar, if only $t$-channel Roper and
$\Delta\Delta$ processes contribute. In that case the matrix element
$M_{I_{pp}^fI_{\pi^+\pi^-}I_{pp}^i} = M_{111}$ ($\rho$-channel in the $\pi^+\pi^-$
subsystem) vanishes \footnote{neglecting a very small contribution from the
  Roper decay branch $N^* \to \Delta \pi$}  \cite{iso,e+e-}. So, if the kink
around $T_p \approx$ 1.1 GeV in the $pp\pi^0\pi^0$ data \cite{deldel} got to
be reproduced by such model calculations, then also the $pp\pi^+\pi^-$ channel
has to behave similarly, if only these two processes are at work. 

In the total cross section the $t$-channel Roper and $\Delta\Delta$
excitations interfere only weakly (see, {\it e.g.}, Fig.~3 in
Ref.~\cite{Luis}, where the cross sections of the individual processes are
seen to just add up in good approximation). Hence we may neglect their
interference 
in good approximation and obtain thus from isospin decomposition,
eqs. (1) - (5) in Ref.~\cite{iso} for the total cross sections of
$pp\pi^0\pi^0$ and $pp\pi^+\pi^-$ channels:

\begin{eqnarray} 
&\sigma_{pp\pi^0\pi^0} \approx ~~\sigma^{N^*}& +~~~\sigma^{\Delta\Delta}\\
&\sigma_{pp\pi^+\pi^-} \approx 2 \sigma^{N^*}& +~\frac{9}{8}\sigma^{\Delta\Delta} +
\frac 1 8 |M_{111}|^2\nonumber
\end{eqnarray}

where $\sigma^{N^*}$ and $\sigma^{\Delta\Delta}$ denote the cross sections of
  $t$-channel Roper and $\Delta\Delta$ processes, respectively. Since the
  relative weight of the $\Delta\Delta$ process is less than that of the
  Roper process in $\sigma_{pp\pi^+\pi^-}$, the kink near $T_p \approx$  1.1 GeV is
  smaller than in $\sigma_{pp\pi^0\pi^0}$, but still present, because the
  $\Delta\Delta$ process provides a much bigger cross section than the Roper
  process does. In Fig.~\ref{fig4} the isospin-based prediction according to
  eq. (1) is plotted by the grey shaded band. At low energies it agrees
  perfectly with the "modified Valencia" calculation, which was tuned to the
  data in the $pp\pi^0\pi^0$ channel. At higher energies the band deviates
  slightly from the model calculation. The reason  for it is that the model
  calculation includes interference between Roper and $\Delta\Delta$ processes,
  which is neglected in the isospin result, and also includes contributions
  from $\Delta(1600)$ not included in the isospin based result. 

We note in passing that in the article about isospin decomposition \cite{iso}
the missing strength in the $pp\pi^+\pi^-$ channel appeared still less
dramatic, since at that time full interference between the isospin matrix
elements for Roper and $\Delta\Delta$ excitations was assumed. But as later
model calculations showed, this interference is very small, since both
excitations act on quite different phase-space volumes. For that reason
interferences between the various resonance excitations have been omitted at
all in the model calculations of Ref. \cite{Zou}.

The $\Delta(1600)$ excitation also
contributes to $M_{111}$ albeit much too little in order to heal the deficit
in the cross section. The "modified Valencia" calculations (dashed line in
Fig. ~\ref{fig4}) do include this contribution.

In this context we also have to ask, whether possibly other higher-lying $N^*$
and $\Delta$ resonances provide substantial contributions in the energy region
of interest here. This has been comprehensively investigated in Ref. \cite{Zou}
with the result that all of these (including also $N^*(1520)$) give only
negligible contributions to the two-pion production cross sections.

\begin{figure} [t]
\begin{center}
\includegraphics[width=0.49\columnwidth]{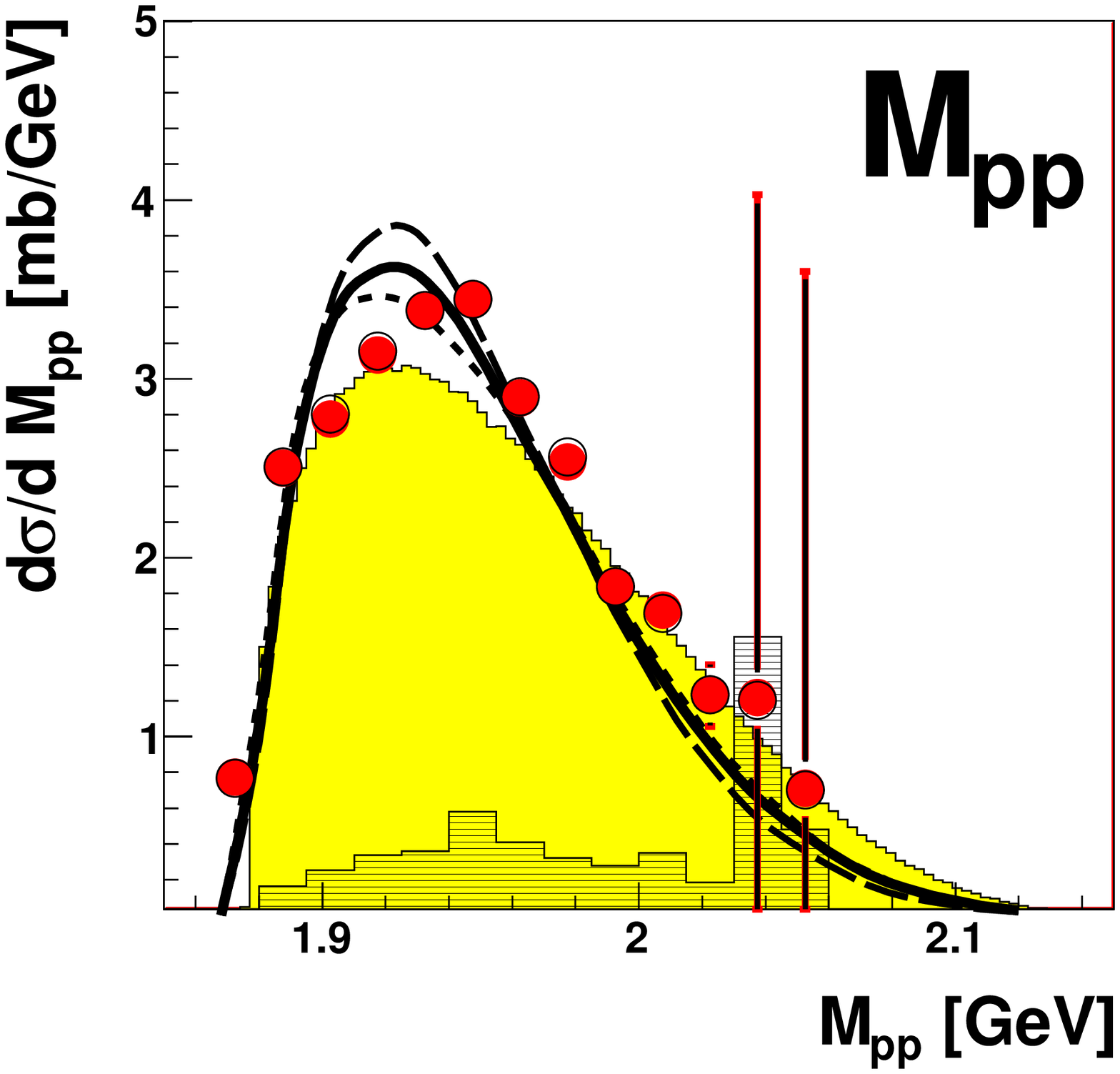}
\includegraphics[width=0.49\columnwidth]{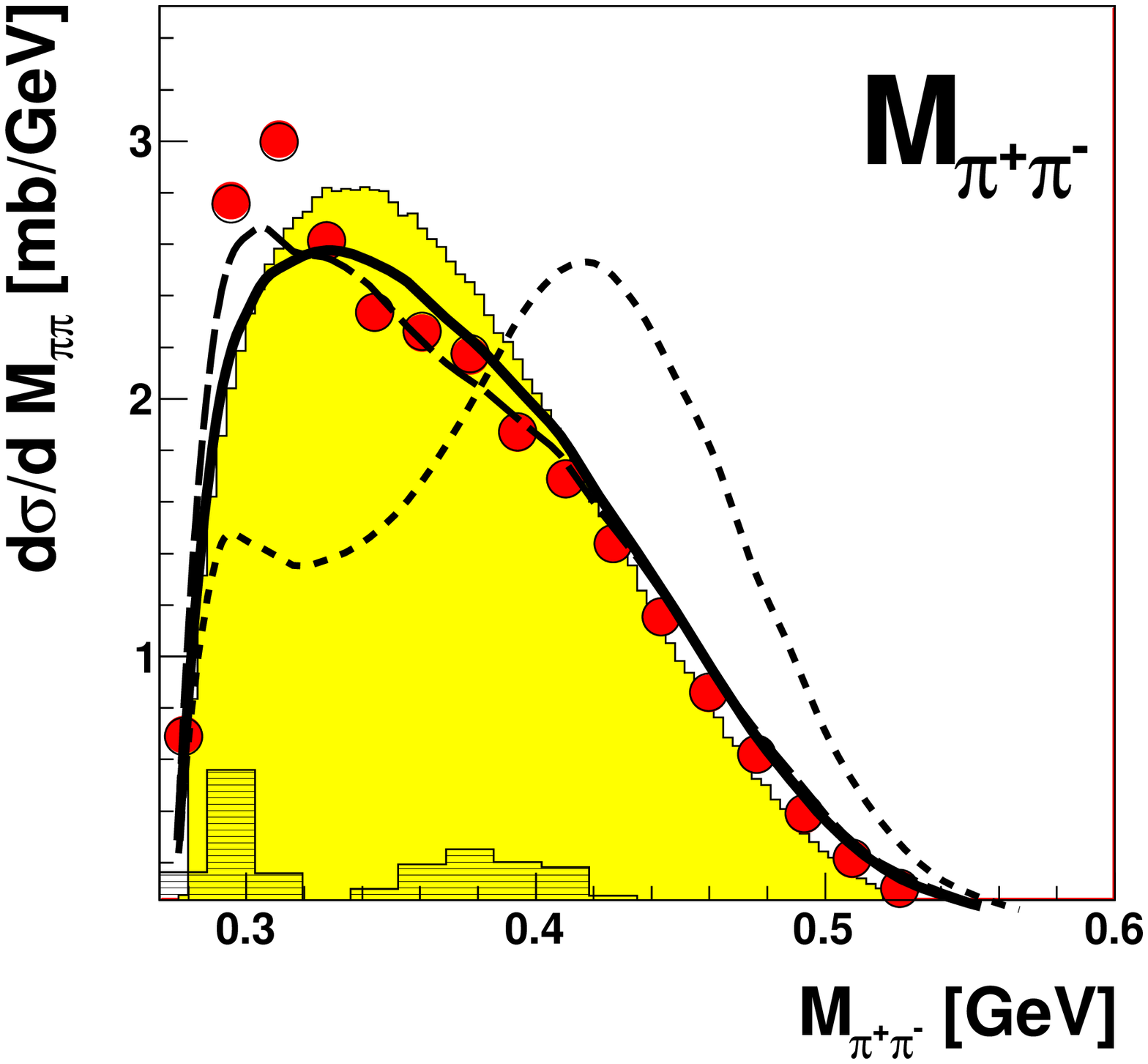}
\includegraphics[width=0.49\columnwidth]{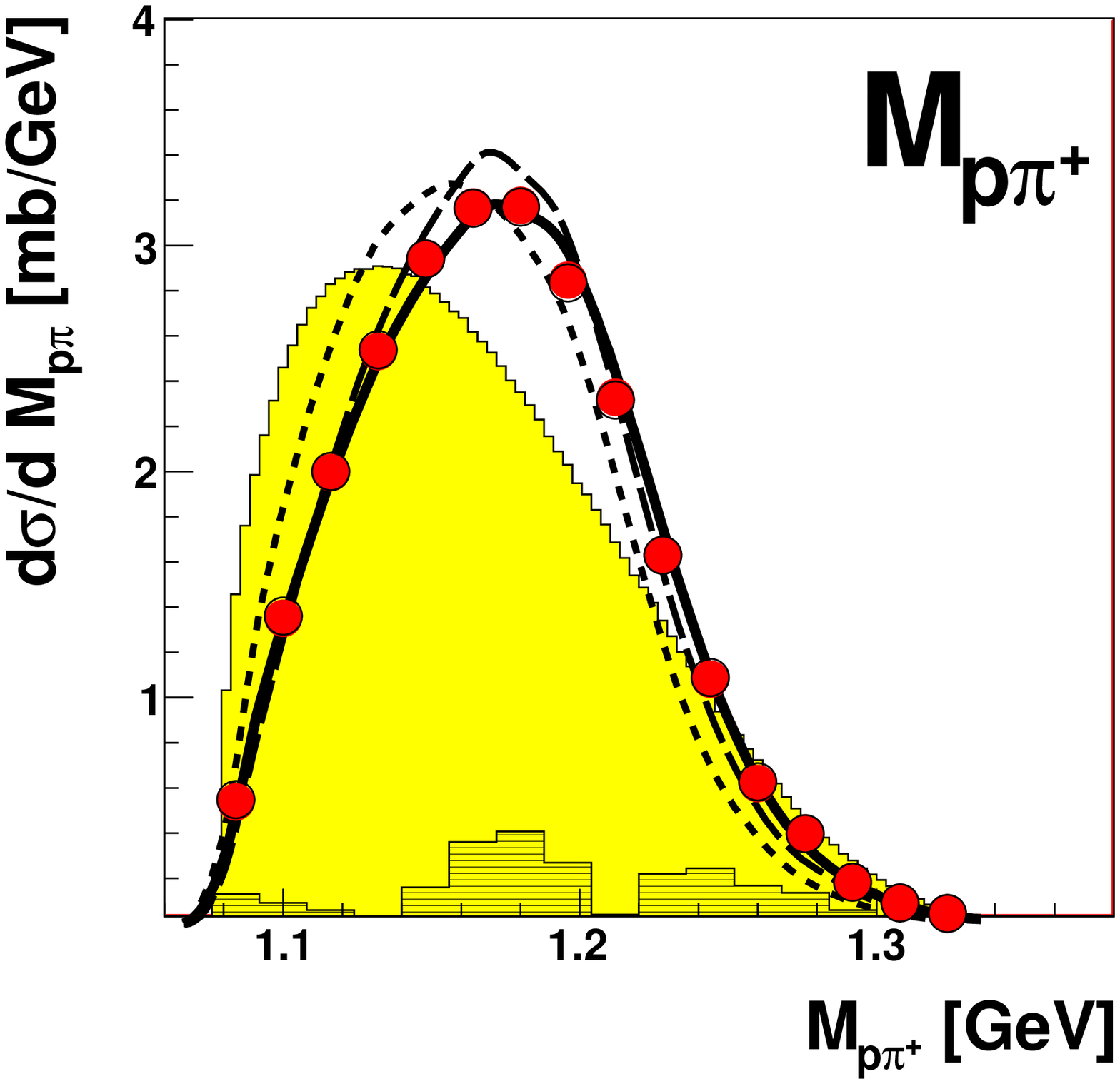}
\includegraphics[width=0.49\columnwidth]{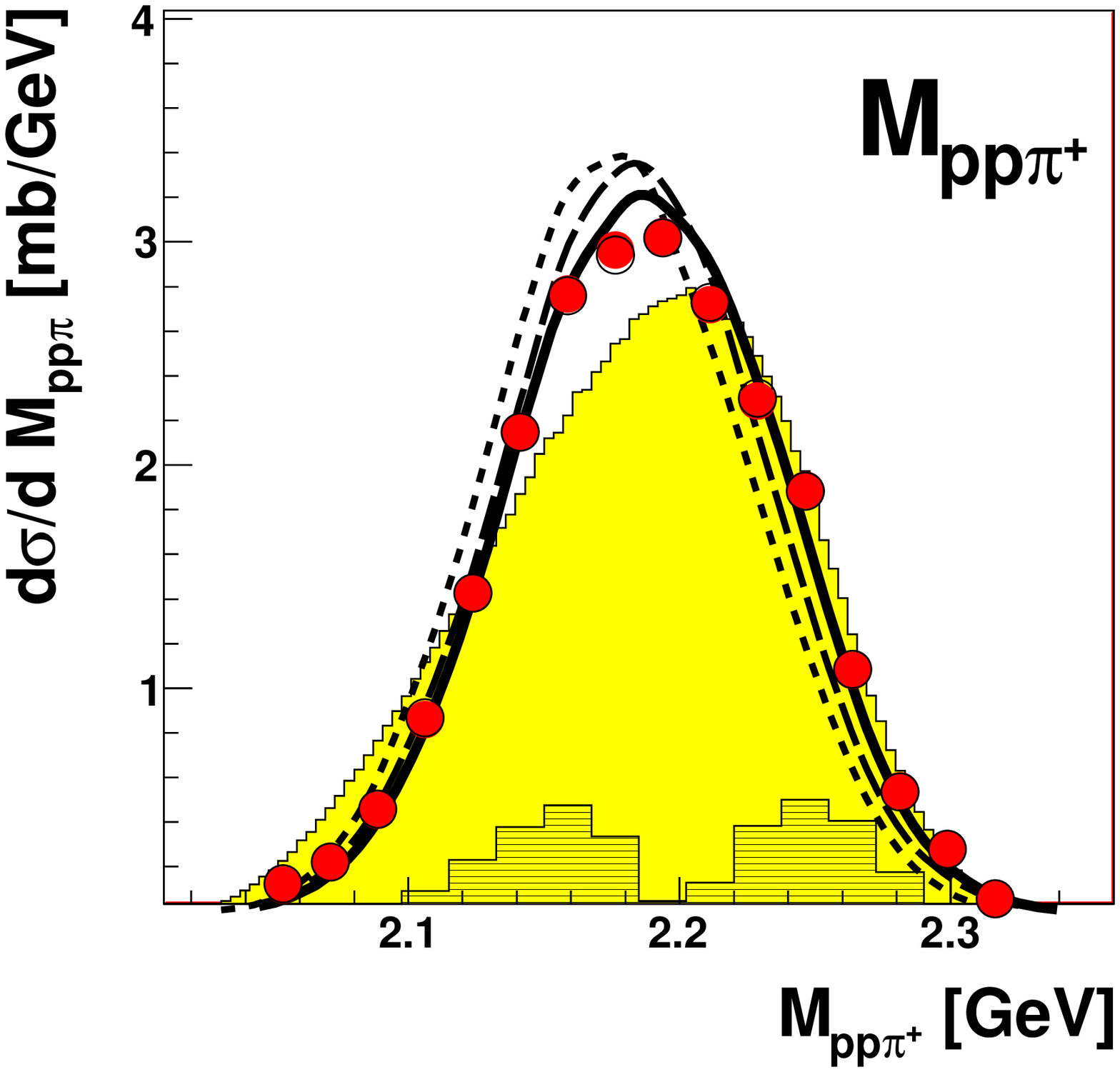}
\includegraphics[width=0.49\columnwidth]{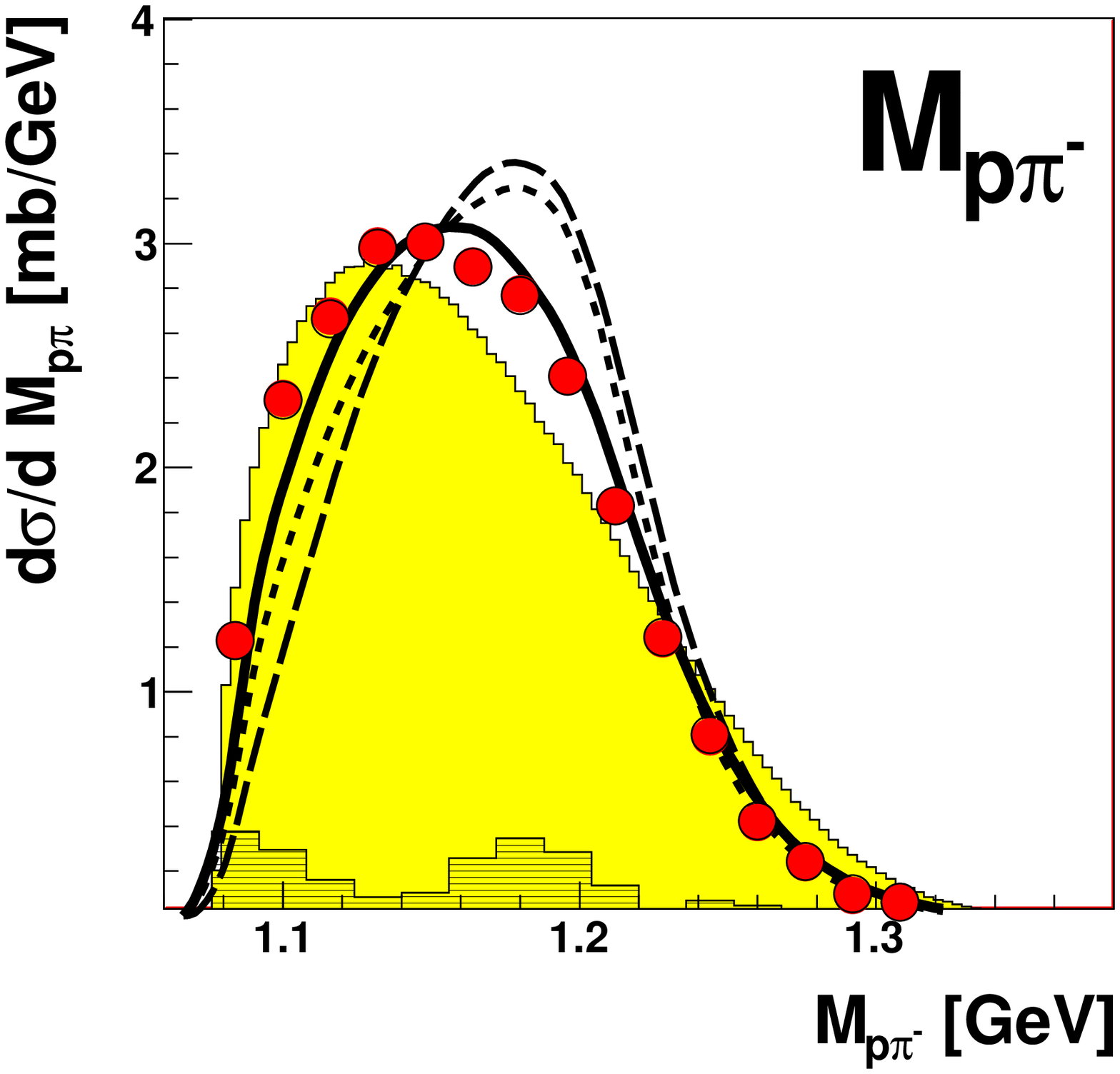}
\includegraphics[width=0.49\columnwidth]{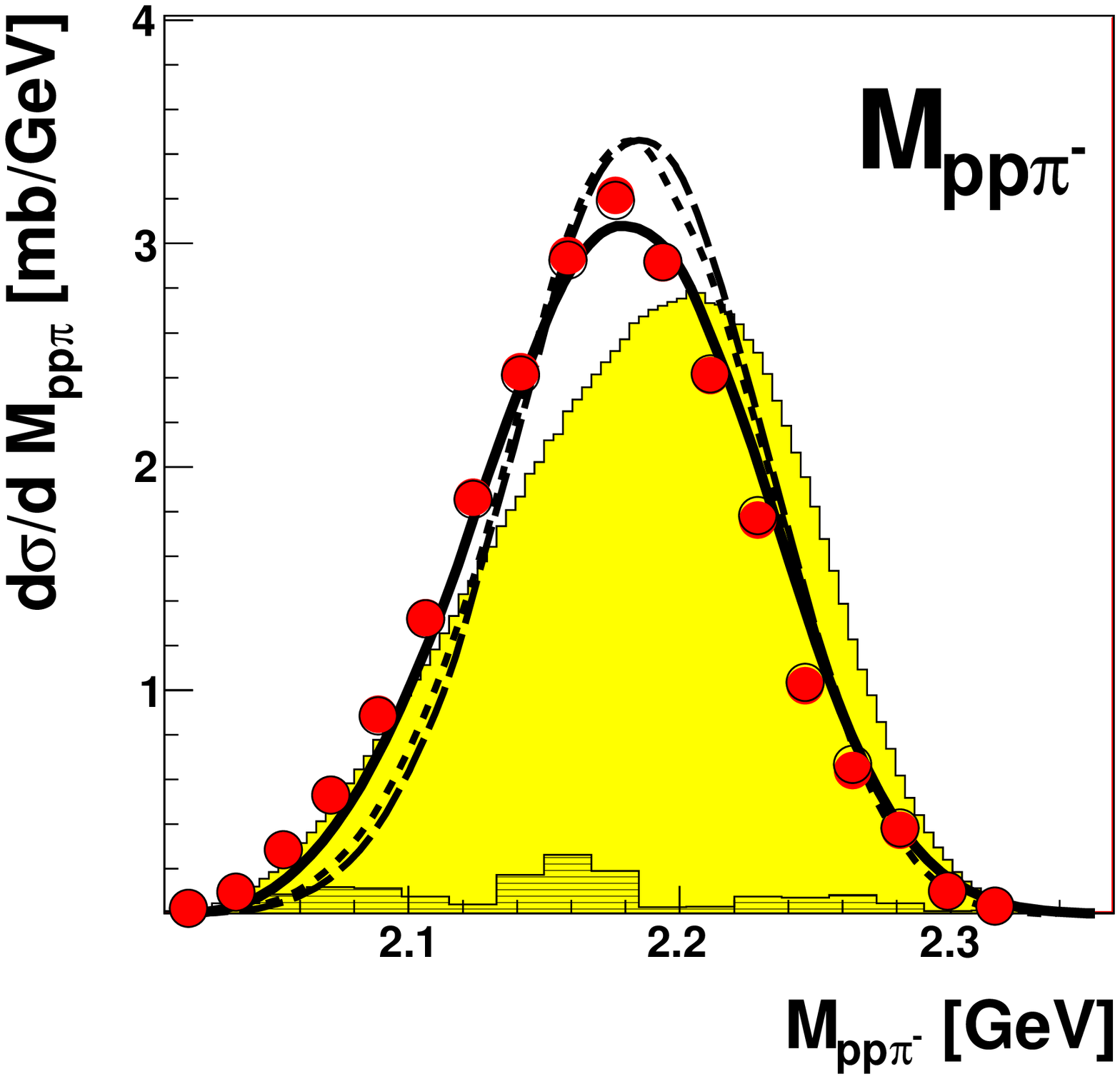}
\caption{(Color online) 
  Differential distributions of the $pp \to pp\pi^+\pi^-$ reaction in the
  region $T_p$ = 1.08 - 1.36 GeV for the invariant-masses $M_{pp}$ (top
  left), $M_{\pi^+\pi^-}$ (top right), $M_{p\pi^+}$ (middle left), $M_{pp\pi^+}$
  (middle right), $M_{p\pi^-}$ (bottom left), $M_{pp\pi^-}$ (bottom right). 
  Filled (open) circles denote the results from this work after (before)
  background subtraction. In most cases these symbols lie on top of each
  other. The hatched histograms indicate 
  systematic uncertainties due to the restricted phase-space coverage of the 
  data. The shaded areas represent pure 
  phase-space distributions, dotted (dashed) lines represent original
  (modified) Valencia calculations \cite{Luis} (\cite{deldel}).
  The solid lines include the process $pp \to D_{21}¸\pi^- \to
  pp\pi^+\pi^-$. All calculations are normalized in area to the data.
}
\label{fig5}
\end{center}
\end{figure}

\begin{figure} 
\begin{center}
\includegraphics[width=0.49\columnwidth]{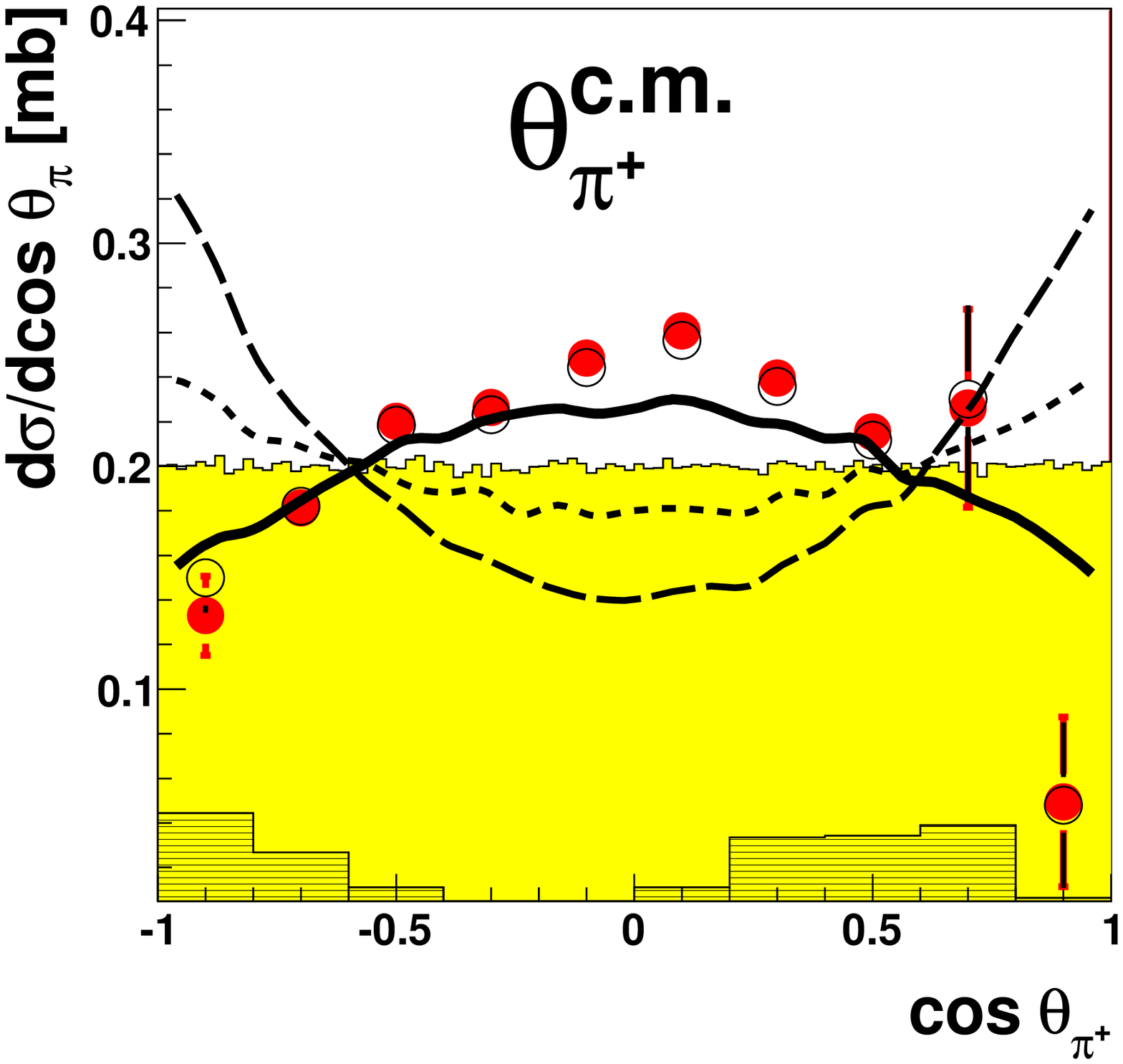}
\includegraphics[width=0.49\columnwidth]{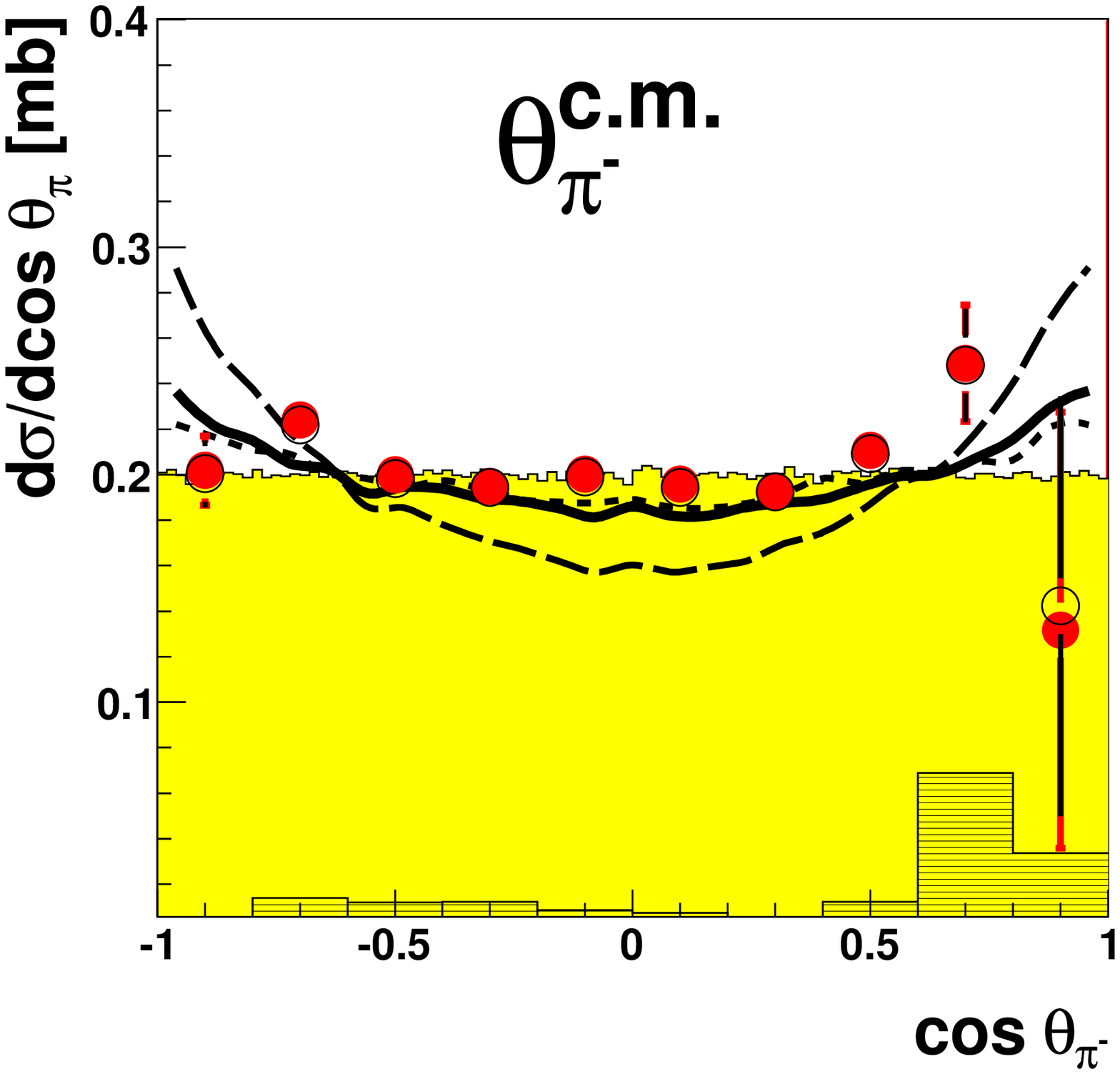}
\includegraphics[width=0.49\columnwidth]{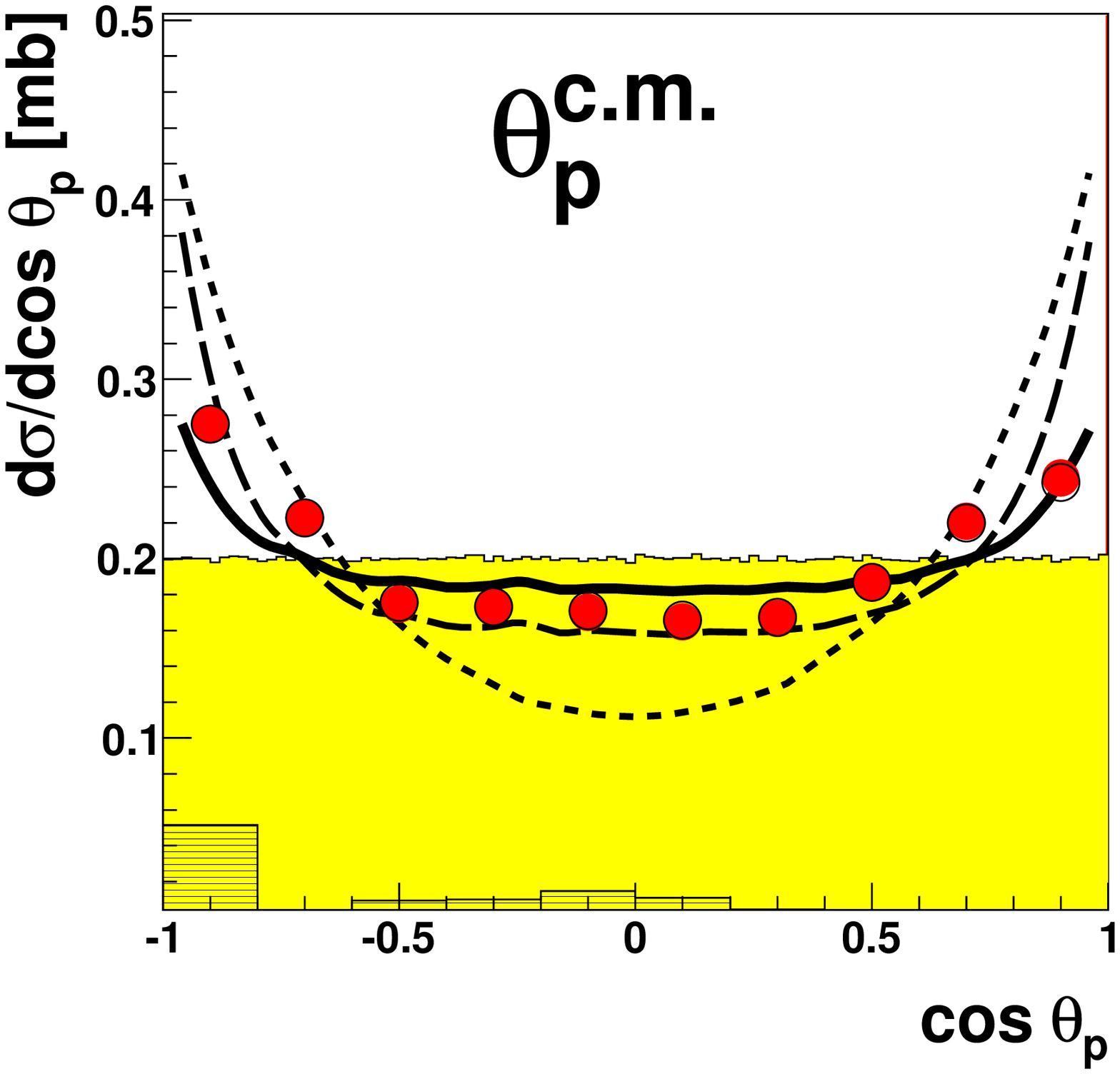}
\caption{(Color online) 
   The same as Fig.~\ref{fig5}, but for the c.m. angles of  positive and
   negative pions $\Theta_{\pi^+}^{c.m.}$ and $\Theta_{\pi^-}^{c.m.}$,
   respectively, as well as protons $\Theta_p^{c.m.}$.
}
\label{fig6}
\end{center}
\end{figure}

\subsection{Differential cross sections}
 
The differential distributions do not exhibit any particularly strong energy
dependence in their shapes, when binned into $T_p$ bins of 50~MeV width-- which
is of no 
surprise, since the energy region covered in this measurement is dominated by
$\Delta\Delta$ and Roper excitations with very smooth energy dependencies due to
their large decay widths. Hence we discuss the differential distributions at
first unbinned, {\it i.e.} averaged over the full covered energy range.

For an axially symmetric four-body final state there are seven independent
differential observables. But for a better understanding of the underlying
physics we  decided to show more, namely nine differential distributions. These
are the ones 
for the center-of-mass (c.m.) angles for protons and pions denoted by 
$\Theta_p^{c.m.}$, $\Theta_{\pi^+}^{c.m.}$ and $\Theta_{\pi^-}^{c.m.}$,
respectively, as well as those for the invariant masses $M_{pp}$, $M_{\pi^+\pi^-}$,
$M_{p\pi^+}$, $M_{pp\pi^+}$, $M_{p\pi^-}$ and $M_{pp\pi^-}$. These
distributions are shown in Figs.~\ref{fig5} and \ref{fig6}.  

There are no data to compare with from previous experiments in the energy
range considered here. Except for $\Theta_{\pi^-}^{c.m.}$ all measured differential distributions differ markedly
in shape from pure phase-space distributions (shaded areas in
Figs.~\ref{fig5} - \ref{fig6}). With the exception of $\Theta_{\pi^+}^{c.m.}$,
$M_{p\pi^-}$ and 
$M_{pp\pi^-}$ spectra, the differential distributions are reasonably well
reproduced by the "modified Valencia model" calculations (dashed curves). 
For the original
Valencia calculation (dotted lines), which contains substantial contributions
from the Roper excitation still in this energy region, large discrepancies get
apparent in 
addition for the $M_{\pi^+\pi^-}$ distribution. For better comparison all
calculations are adjusted in area to the data in Figs.~\ref{fig5} - \ref{fig9}.

Because of identical particles in the entrance channel all c.m. angular
distributions have to be symmetric about $90^\circ$. Within uncertainties this
requirement is met by the data.
The proton angular distribution is forward-backward peaked as
expected for a peripheral reaction process. The $\pi^-$ angular distribution is
flat, in tendency slightly convex curved as also observed in the other
$NN\pi\pi$ channels at these energies. But 
surprisingly, the $\pi^+$ angular distribution exhibits a strikingly
concave shape. Such a strange behavior, which is in sharp contrast to the
theoretical expectation, has been observed so far in none of the two-pion
production channels.

Also the  $M_{p\pi^-}$ spectrum is markedly different from the
$M_{p\pi^+}$ spectrum. The same is true for the $M_{pp\pi^-}$ spectrum with
respect to the  $M_{pp\pi^+}$ distribution. In case of the $t$-channel
$\Delta\Delta$ process, which is thought to be the dominating one at the
energies of interest here, $\Delta^{++}$ and $\Delta^0$ get excited 
simultaneously and with equal strengths. Hence, the $M_{p\pi^+}$
($M_{pp\pi^+}$) spectrum should be equal to the $M_{p\pi^-}$ ($M_{pp\pi^-}$)
one and also the $\pi^+$ angular distribution should be identical to the 
$\pi^-$ angular distribution.

So the failure of the "modified Valencia" calculation to describe properly
the total cross section and the differential distributions underlines
the suspicion that the $t$-channel $\Delta\Delta$ process is not the leading
process here.

Since the total cross section is grossly underpredicted above $T_p \approx$
1.0 GeV, it
appears that an important piece of reaction dynamics is missing, 
which selectively affects the $pp\pi^+\pi^-$ channel. Furthermore, the
discrepancy between data and "modified Valencia" description opens up
scissor-like around $T_p \approx$ 0.9 GeV, which suggests the opening of a new
channel, where a $\Delta N$ system is produced associatedly with another
pion. Such a state with the desired properties could be the isotensor $D_{21}$
state with $I(J^P) = 2(1^+)$ predicted already by Dyson and Xuong \cite{Dyson}
with a mass close to that of its isospin partner $D_{12}$ with $I(J^P) =
1(2^+)$.  Whereas $D_{12}$ can be reached directly by the initial $pp$
channel, $D_{21}$ cannot be reached that way because of its isospin $I$ =
2. However, it can be produced in initial $pp$ collisions associatedly 
with an additional pion. 

\subsection{$D_{12}$ resonance}

In several partial-wave analyses of $pp$ and $\pi d$ scattering as well as of
the $pp \to d\pi^+$ reaction the $D_{12}$ resonance has been identified at a
mass of 2144 - 2148 MeV \cite{Hoshizaki,SAID}, {\it i.e.} with a
binding energy of a few MeV relative to the nominal $\Delta N$ threshold --
and with a width compatible to that of the $\Delta$ resonance. For a recent
discussion 
about the nature of this $D_{12}$ state see, {\it e.g.}, Ref. \cite{hcl} and
references therein. Also recent Faddeev calculations for the $NN\pi$ system find
both $D_{12}$ and $D_{21}$ dibaryon resonances with masses slightly below the
$\Delta N$ threshold and with widths close to that of the $\Delta$ resonance
\cite{GG}. 
The decay of the $D_{12}$ resonance proceeds dominantly into $d\pi$ and $np\pi$
channels, since there the $np$ pair can reside in the $^3S_1$ partial wave,
which readily couples with the $p$-wave pion (from $\Delta$ decay) to $J^P =
2^+$. In contrast, its decay into $pp\pi$ is heavily suppressed, since the
$pp$ pair can couple only to $^1S_0$ in relative $s$-wave and hence needs at
least relative $d$-waves for building a $J^P = 2^+$ state in the $pp\pi$
system. Since it does not show up in the $pp\pi$ system, it also will not
appear in the $pp \to pp\pi^+\pi^-$ reaction.

\subsection{$D_{21}$ resonance}

The hypothetical isotensor state $D_{21}$, on the other hand, strongly favors
the purely isotensor channel $pp\pi^+$ in its decay. In addition, $J^P = 1^+$
can be easily reached by adding a $p$-wave pion (from $\Delta$ decay) to a
$pp$ pair in the $^1S_0$ partial wave. Hence -- as already suggested by Dyson
and Xuong \cite{Dyson} -- the favored production process should be in the
$pp \to pp\pi^+\pi^-$ reaction. 


If we use the formalism outlined in Ref. \cite{ABC}, then the resonance
process $pp \to D_{21}\pi \to \Delta p \pi \to pp \pi\pi$ can be described by
the transition amplitude    
\begin{eqnarray}
M_R(m_{p_1},m_{p_2},m_{p_3},m_{p_4},{\hat k_1},{\hat k_2}) =~~~~~~~~~~~~~~~~\\
\nonumber 
~~~~~~~~~~~~~~~~~~~~M_R^0~~\Theta_R(m_{p_1},m_{p_2},m_{p_3},m_{p_4},{\hat
  k_1},{\hat k_2}), 
\end{eqnarray}
where the function $\Theta$ contains the substate and angular dependent part,
and 
\begin{equation}
M_R^0 =  D_{D_{21}} * D_{\Delta}
\end{equation}
with $D_{D_{21}}$ and $D_{\Delta}$ denoting the corresponding resonance
propagators. Here $p_1$, $p_2$ and $p_3$, $p_4$ denote the ingoing and
outgoing protons, respectively. $k_1$ is the momentum of the
associatedly produced pion and $k_2$ that of the pion resulting from the decay
$D_{21} \to \Delta p \to pp\pi$

If the coordinate system is chosen to be the standard one with the z-axis
pointing in beam direction (implying $m_L$ = 0 and $(\Theta_i, \Phi_i) =
(0,0)$),  
then the function $\Theta_R(m_p,m_n,m_{p_3},m_{p_4},{\hat k_1},{\hat k_2})$
defined in eq. (2) is built up by the corresponding vector coupling coefficients
  and spherical harmonics representing the angular dependence due to the orbital
  angular momenta involved in the reaction: 
\begin{eqnarray}
\Theta_R(m_{p_1},m_{p_2},m_{p_3},m_{p_4},{\hat k_1},{\hat k_2}) =~~~~~~~~~~~~~~~~~~~~~~~~\\ \nonumber
~~~~~~~\sum (\frac {1}{2} \frac {1}{2}
m_{p_1} m_{p_2} | S m_s) ~~ (S L m_s0 | J M) ~~\\ \nonumber
(J M | J_{D_{21}} l m_{D_{21}} m_1) ~~
(J_{D_{21}} m_{D_{21}} | \frac {3}{2} \frac {1}{2}  m_{\Delta} m_{p_3})~~ \\\nonumber
(\frac {3}{2} m_{\Delta} | \frac {1}{2} 1 m_{p_4} m_2) ~~
Y_{L0}(0,0) ~~Y_{l m_1}(\hat {k_1}) ~~Y_{1 m_2}(\hat {k_2}).
\end{eqnarray}

The $D_{21}$ resonance can be formed together with an associatedly produced
pion either in relative $s$ or $p$ wave. In the first instance the initial $pp$
partial wave is $^3P_1$, in the latter one it is $^1S_0$ or $^1D_2$. The
first case is special, since here $(S L 0 0 | J M) = (1 1 0 0 | 1
0)$~=~0. Only in this case eq. (4) yields  a $sin \Theta_\pi$ dependence for
the angular distribution of the pion originating  from the $D_{21}$ decay  ---
exactly what is needed for the description of the data for the $\pi^+$ angular
distribution.

In fact, if we add such a resonance with the processes 
\begin{eqnarray}
pp \to D_{21}^{+++}\pi^- \to \Delta^{++}p\pi^- \to pp\pi^+\pi^- \\ \nonumber
pp \to D_{21}^{+}\pi^+ \to \Delta^{0}p\pi^+ \to pp\pi^+\pi^-  
\end{eqnarray}

with fitted mass $m_{D_{21}}$ = 2140 MeV  and width $\Gamma_{D_{21}}$ = 110 MeV, we
obtain a good 
description of the total cross section by adjusting the strength of the
assumed resonance process to the total cross section data (solid line in
Fig.~\ref{fig4}). Simultaneously, the 
addition of this resonance process provides a quantitative description of all 
differential distributions (solid lines in Figs.~~\ref{fig5} - \ref{fig9}), in
particular also 
of the $\Theta_{\pi^+}^{c.m.}$, $M_{p\pi^-}$ and $M_{pp\pi^-}$ distributions. 
Due to isospin coupling the branch via $\Delta^0$ is very small and yields only
   marginal contributions to the observables.
Since therefore
the $D_{21}$ decay populates practically only $\Delta^{++}$, its reflexion in the
$M_{p\pi^-}$ spectrum shifts the strength to lower masses -- as required by
the data. The same holds for the $M_{pp\pi^-}$ spectrum. We are not aware of
any other mechanism, which could provide an equally successful description of
the observables of the $pp \to pp\pi^+\pi^-$ reaction at the energies of
interest here. 

We note that the only other place in pion production, where a
concave curved pion angular distribution has been observed, is the $pp \to
pp\pi^0$ reaction in the region of single $\Delta$ excitation
\cite{Jozef,Jozef1,ED,Colinreview,ANKE}. Also in this case it turned out that
the reason was the 
excitation of resonances in the $\Delta N$ system \cite{ANKE} causing a proton
spinflip situation. In general, t-channel resonance excitations are connected
with pions emerging in s- or p-waves in non-spinflip configurations and hence
lead to flat-to-convex shaped angular distributions.

Also the description of the $\pi^-$ angular distribution improves by inclusion
of the $D_{21}$ resonance scenario. Whereas the "modified" Valencia
calculations predict still a distribution, which is significantly convex, the
full calculations, which include the $D_{21}$ reaction amplitude with $\pi^-$
particles emerging in relative $s$-wave, predict a much flatter angular
distribution in agreement with the measurements.

\subsubsection{$D_{21}$ subsystem representations}

\begin{figure} 
\begin{center}
\includegraphics[width=0.49\columnwidth]{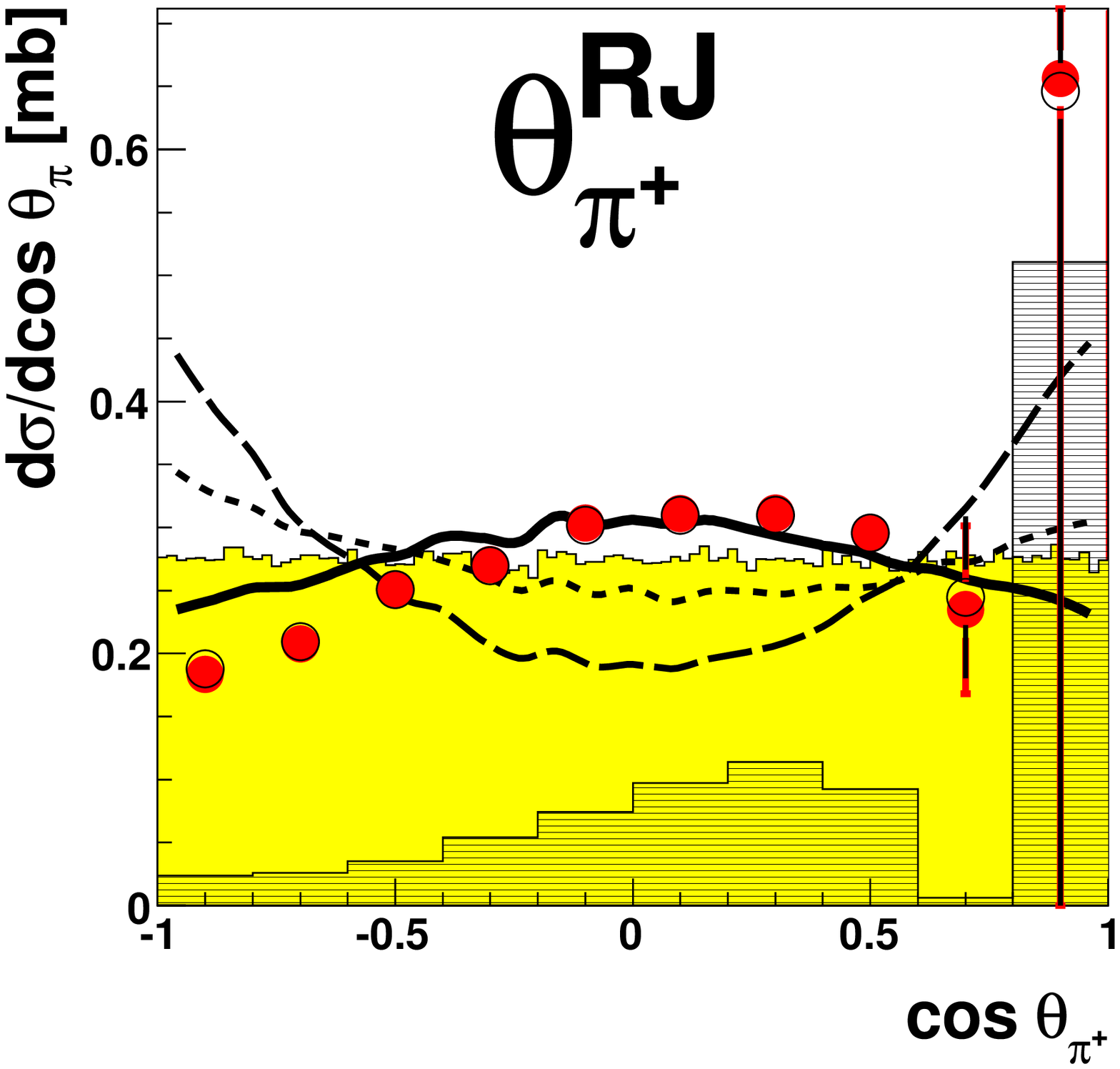}
\includegraphics[width=0.49\columnwidth]{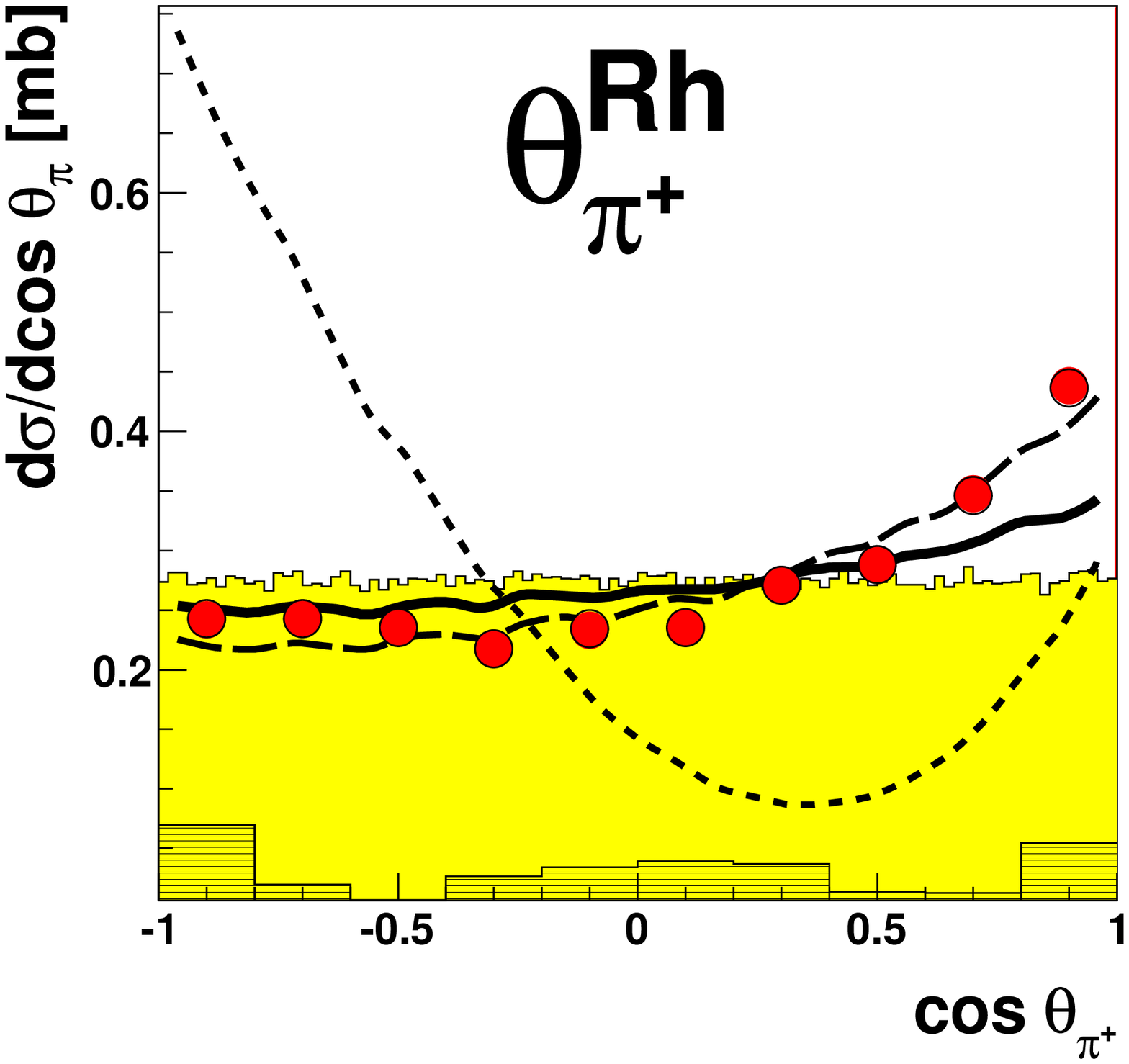}
\caption{(Color online) 
   The same as Fig.~\ref{fig6}, but for the angles of positive pions in the
   $D_{21}$ resonance subsystem either in the Jackson frame
   ($\Theta_{\pi^+}^{RJ}$, left) or in the helicity frame
   ($\Theta_{\pi^+}^{Rh}$, right).
}
\label{fig6a}
\end{center}
\end{figure}

Next, we look on the differential distributions in the subsystem of interest
here, namely the $D_{21}$ resonance system. Since the width of the $\Delta$ excitation
is not small compared to the available phase space energy range, the Dalitz
plot of invariant masses in the resonance subsystem is just overwhelmingly
dominated by the $\Delta^{++}$ excitation as already seen in its square-root
projections displayed in Fig.~\ref{fig5}. Hence we do not need to show the
Dalitz plot here. However, since the $\pi^+$ angular distribution
turned out to be special in the overall center-of-mass system, we look on it
once more in the resonance subsystem by plotting it both in the Jackson and in
the helicity frame \cite{Byckling} in Fig.~\ref{fig6a}.

In the Jackson frame
the reference axis for the polar angle $\Theta_{\pi^+}^{RJ}$ is still the beam
axis, {\it i.e.} the same as in the center-of-mass system. That way entrance
and exit channel systems stay connected in this representation. Since in
addition resonance and center-of-mass systems deviate only by the additional
$\pi^-$, the mass of which is small compared to the residual mass of the other
ejectiles, the angular distributions in these two frames are very similar. 

The situation is very different in the helicity frame, where the reference
axis for the 
polar angle $\Theta_{\pi^+}^{Rh}$ is given by the direction of the $\pi^-$
momentum in the resonance subsystem. Thus this reference frame has no longer a
connection to the initial system and is only based on the emitted particles
representing the opening angle between the two pions in the resonance frame.
In consequence the information about the proton spinflip during the production
process is absent in this representation 
and with it the sinus shape for the $D_{21}$ contribution. Instead, the
$D_{21}$ contribution is flat in this case, since the resonance is in relative
$s$-wave with the associatedly produced $\pi^-$ particle and hence the
directions of the two pions originating from two different sources appear to
be uncorrelated. The situation is more complex for the background of
conventional $t$-channel processes. For the 
$\Delta\Delta$ process the emerging pions originate again from two different
largely uncorrelated sources, since the well-known conventional ABC
effect causing large $\pi\pi$ correlations and giving rise to an
enhancement near $cos \Theta_{\pi^+}^{Rh}$ = 1 (and at low masses in the
$M_{\pi\pi}$ spectrum) is only substantial, if the nucleons in the exit
channel are bound in a nucleus \cite{Risser,abc}. In contrast, the $t$-channel
excitation and decay of the Roper resonance produces a highly correlated pion
pair originating from the same source. Hence the distribution of the $\pi\pi$
opening angle is strongly anisotropic \cite{deldel,ts} in this case. And since
in the original Valencia model the Roper process is assumed to be still large
at the energies of interest here, this calculation predicts a very anisotropic
distribution for the helicity angle $\Theta_{\pi^+}^{Rh}$ (dotted line in
Fig.~\ref{fig6a}, right panel).   

Whereas the original Valencia calculations (dotted lines in Fig.~\ref{fig6a})
are grossly at variance with the data in both reference frames, the
calculations including the $D_{21}$ resonance process (solid lines) give a
good description of the data both in the Jackson and in the helicity frame. For
the modified Valencia calculations the situation is split. Whereas they are
again at variance with the data in the Jackson frame, they fit even perfect to
the data in the helicity frame, in particular at small angles, {\it i.e.} near
$cos\Theta_{\pi^+}^{Rh}$ = 1, where the data show a slight enhancement. This
enhancement in the opening angle is strictly correlated with a corresponding
enhancement in the $M_{\pi^+\pi^-}$ distribution at low masses constituting
the ABC effect. Though this enhancement is small --- see Fig.~\ref{fig5}, top
right ---  it is not fully accounted for by the modified Valencia calculations,
as already apparent in the analysis of the $pp \to pp\pi^0\pi^0$ reaction
(Fig. 2 in Ref. \cite{deldel}. So the small failure of the model calculation
including the $D_{21}$ resonance to describe the ABC enhancement quantitatively
could be traced back to the fact that the background description by the
modified Valencia model is not perfect in the ABC effect region.

\subsubsection{$\Delta$ subsystem representation}

\begin{figure} 
\begin{center}
\includegraphics[width=0.49\columnwidth]{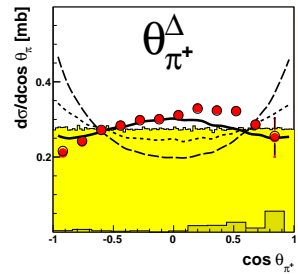}
\caption{(Color online) 
   The same as Fig.~\ref{fig6}, but for the angles of positive pions in the
   $\Delta$ resonance subsystem.
}
\label{fig6b}
\end{center}
\end{figure}

We now want to check, whether the concave shape of the $\pi^+$ angular
distribution really originates from $\Delta$ excitation and decay associated
with a proton spinflip. For this purpose we boost the distribution further
from the $D_{21}$ into the $\Delta$ reference frame, see Fig.~\ref{fig6b}.
The concave shape persists also in this case
though somewhat washed out due to the fact that we do not know, which of the
two emerging protons originates from the $\Delta$. The pure $D_{21}$
process gives a sinus-shaped distribution due to the proton spinflip in
the $\Delta$ excitation and decay process, whereas the convex shaped results
from original (dotted) and modified (dashed) Valencia calculations provide a
convex distribution due to the dominance of the cosine-shaped non-spinflip $
\Delta$ process. The original Valencia calculation is less anisotropic than
the modified one, since in the former the Roper process providing a flat
angular dependence plays a larger role. 

That way, we have traced back the origin of the concave shape
of the $\pi^+$-distribution to the proton spinflip in the $\Delta$ process as
required for the $D_{21}$ production process.

\subsection{Energy dependence of differential distributions}

\begin{figure} [t]
\begin{center}
\includegraphics[width=0.49\columnwidth]{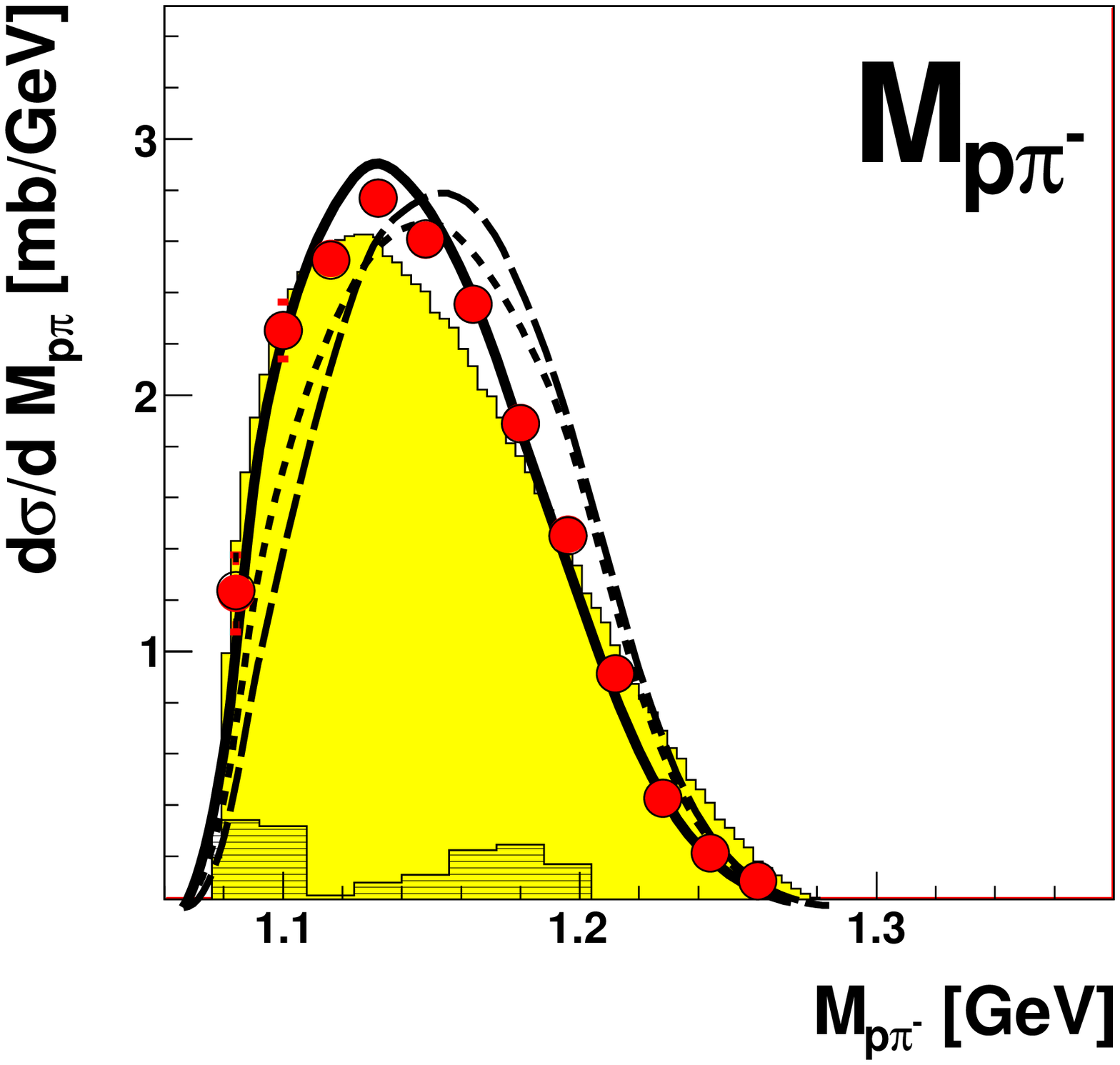}
\includegraphics[width=0.49\columnwidth]{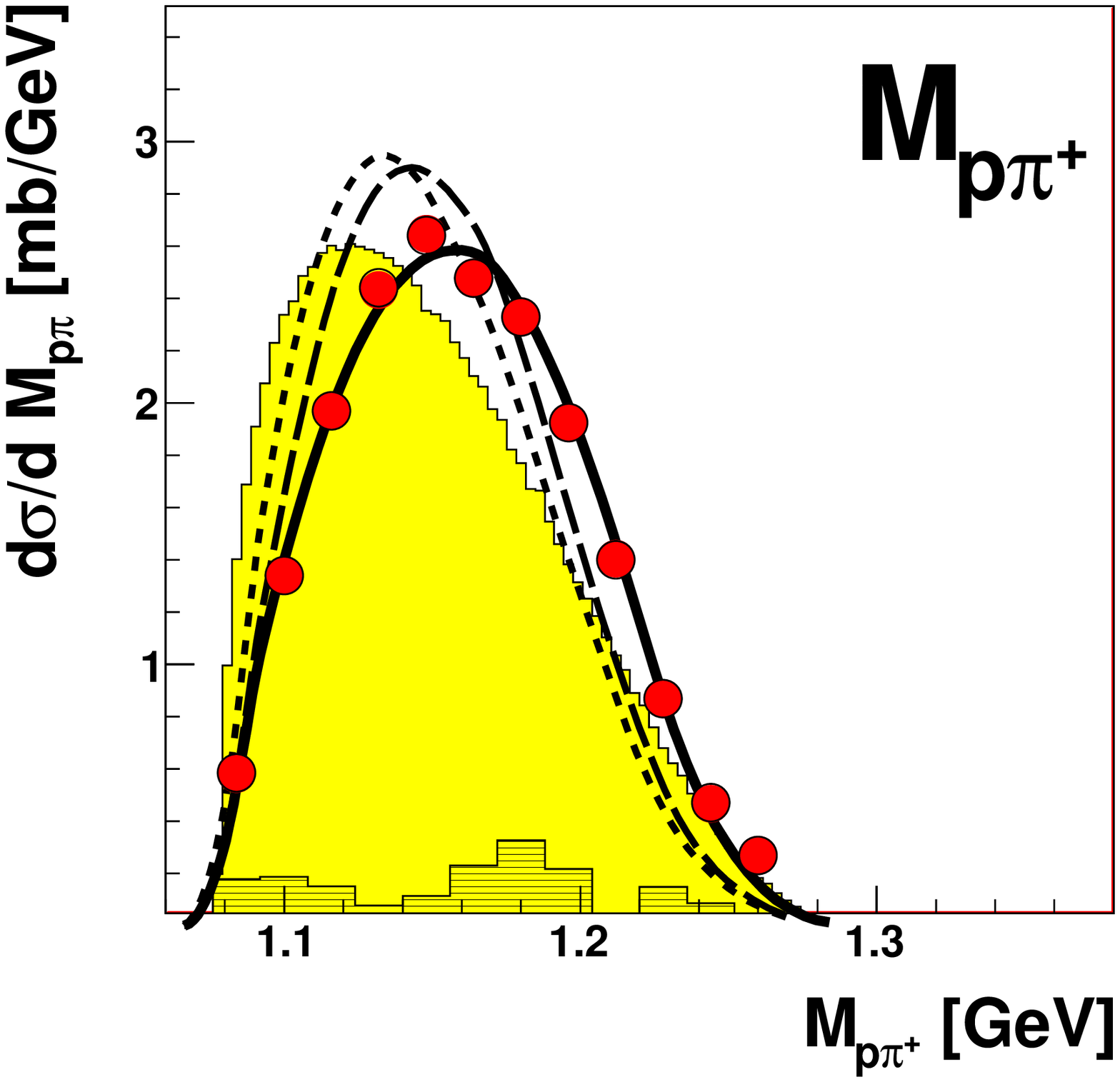}
\includegraphics[width=0.49\columnwidth]{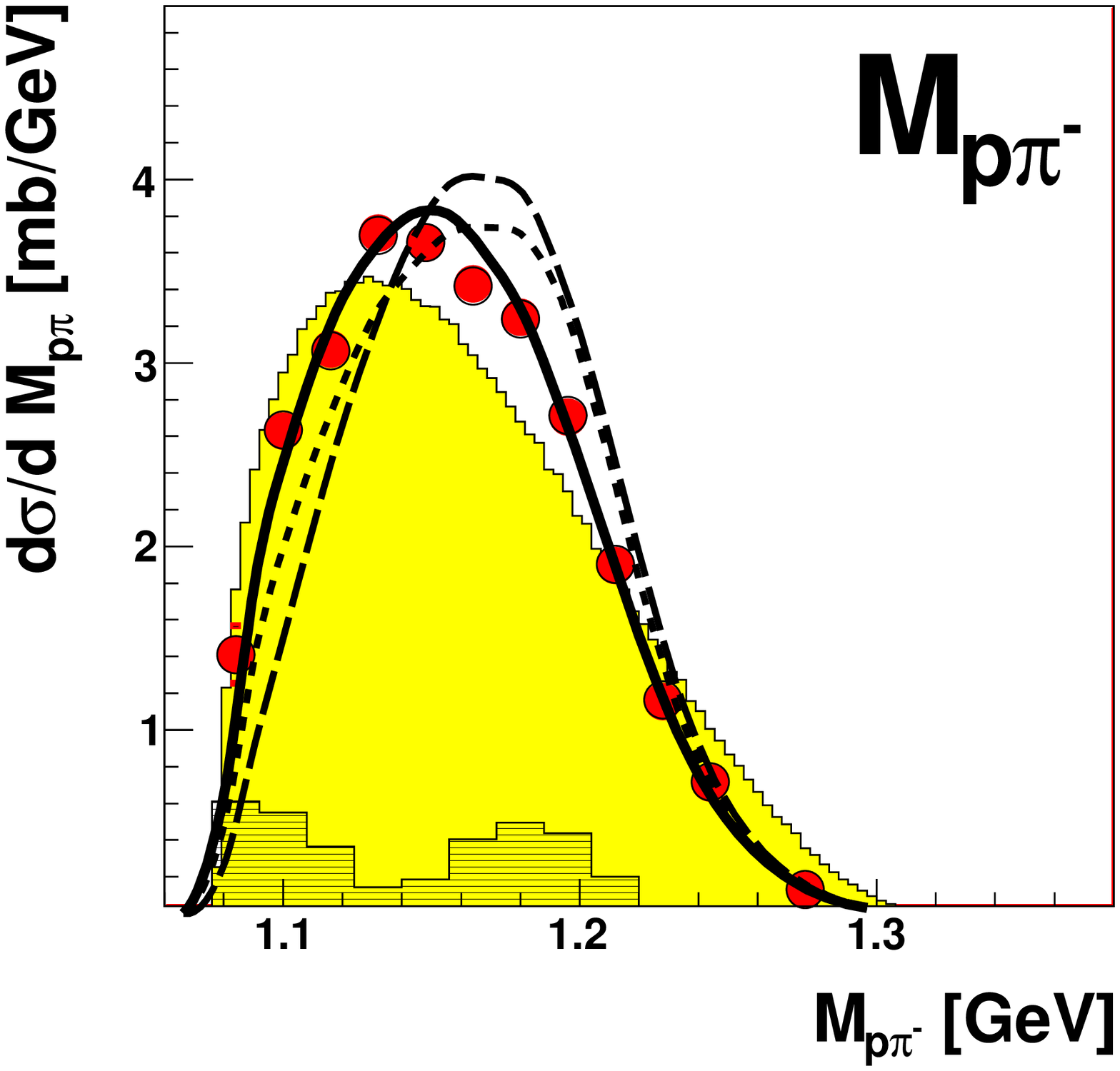}
\includegraphics[width=0.49\columnwidth]{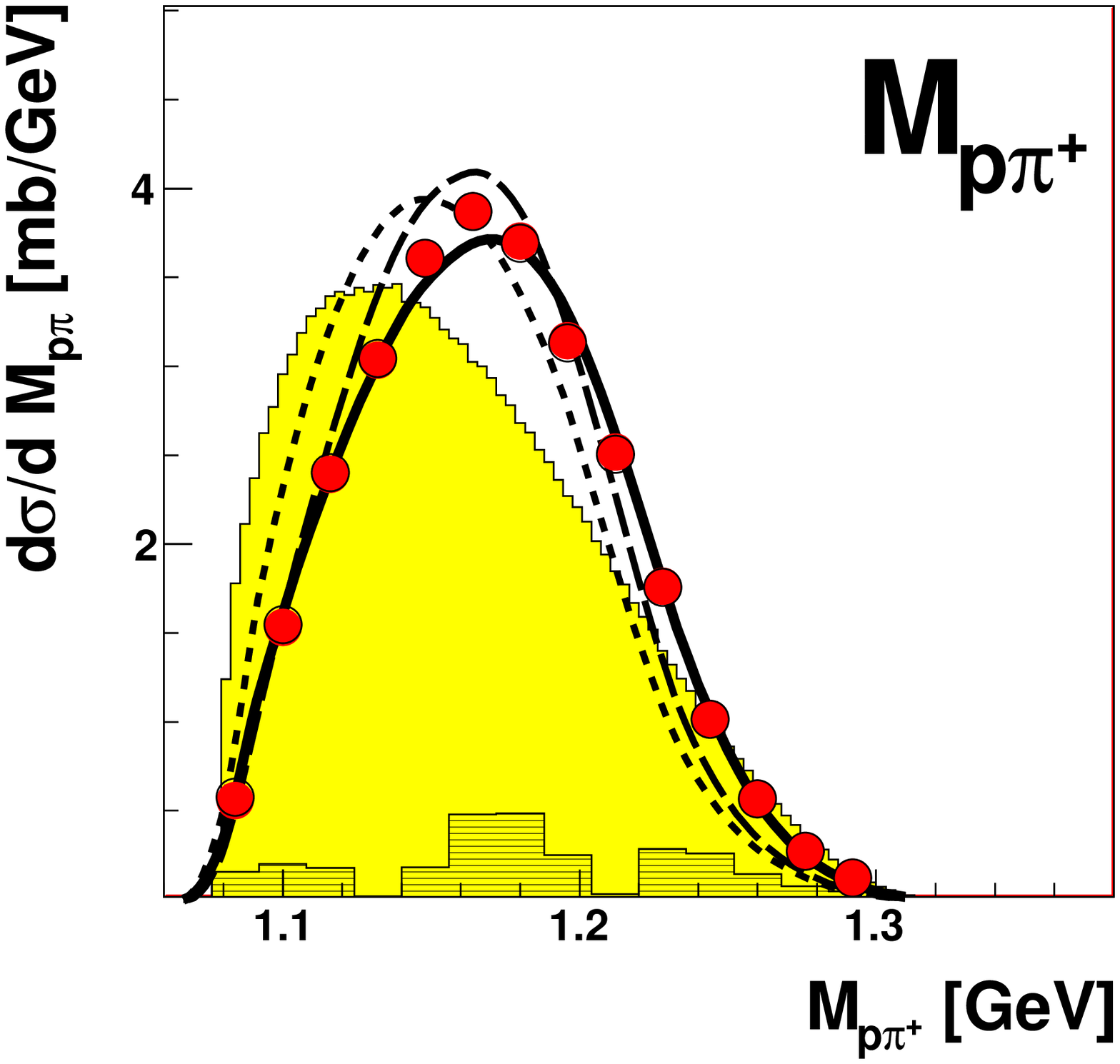}
\includegraphics[width=0.49\columnwidth]{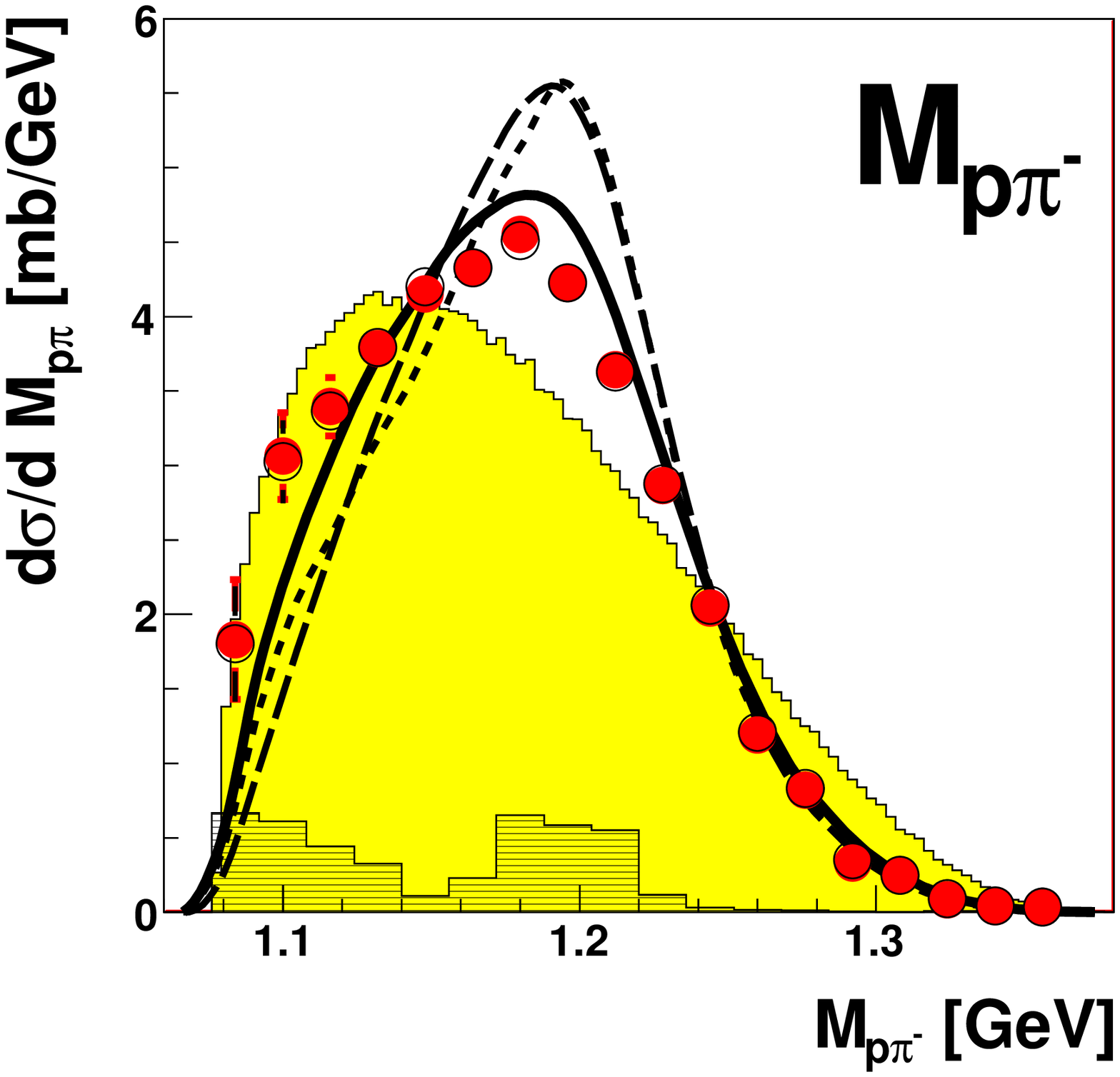}
\includegraphics[width=0.49\columnwidth]{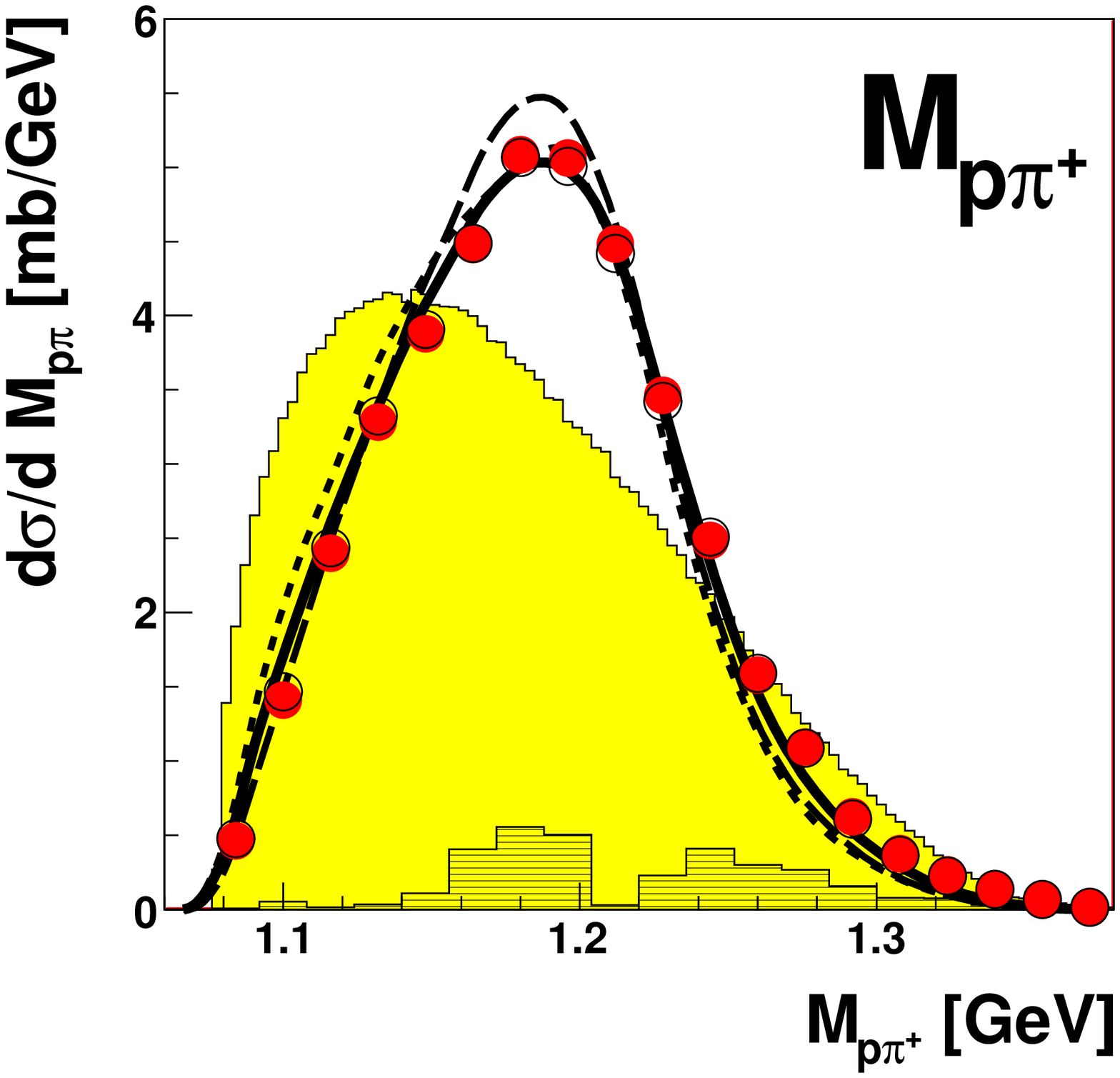}
\caption{(Color online) 
  Same as Fig.~\ref{fig5}, but for the differential distributions of the
  invariant-masses $M_{p\pi^-}$ (left) and $M_{p\pi^+}$ (right) for the
  energy bins at $T_p$ = 1.10 (top), 1.18 (middle) and 1.31 GeV (bottom).
}
\label{fig7}
\end{center}
\end{figure}

\begin{figure} [t]
\begin{center}
\includegraphics[width=0.49\columnwidth]{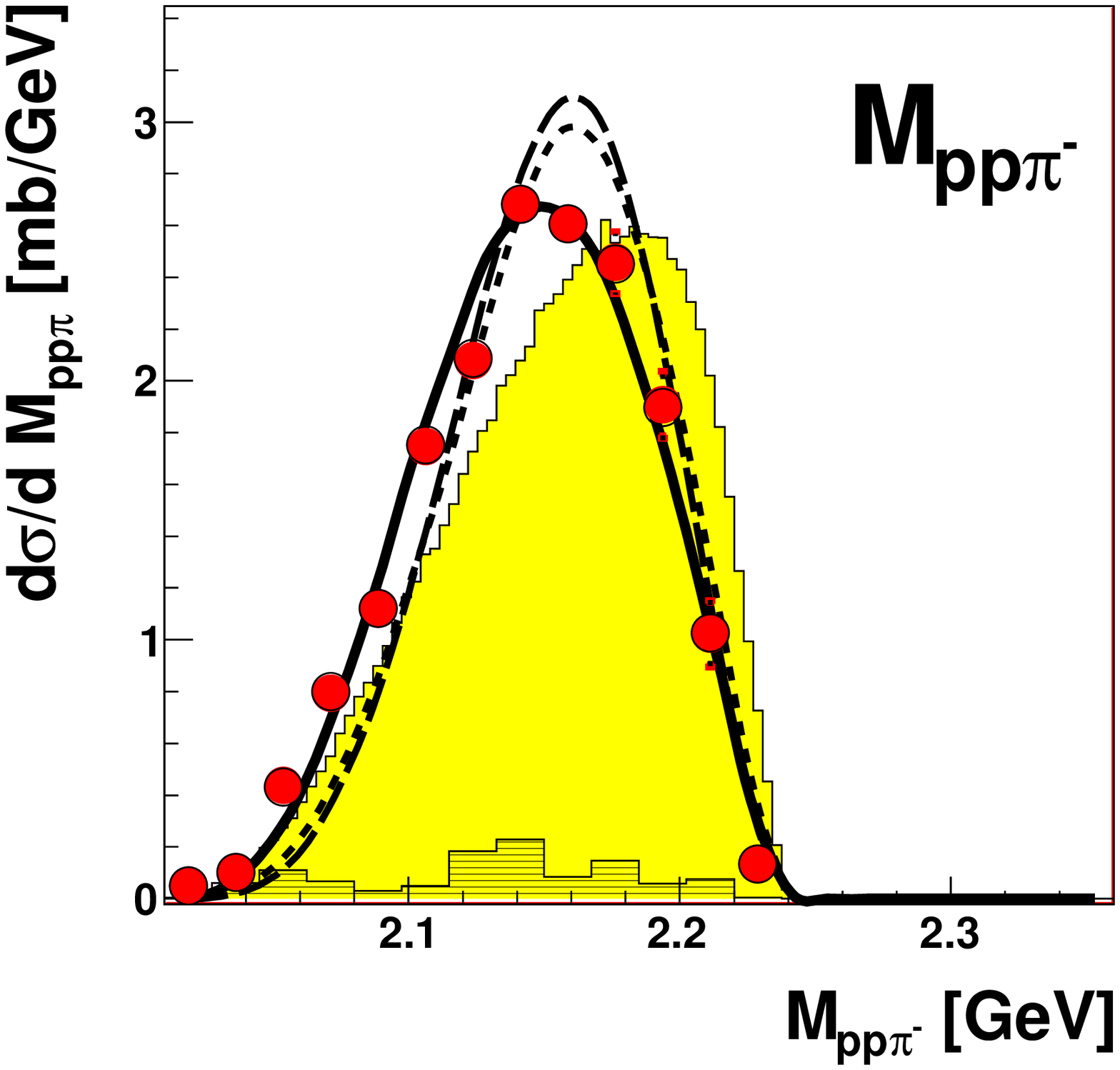}
\includegraphics[width=0.49\columnwidth]{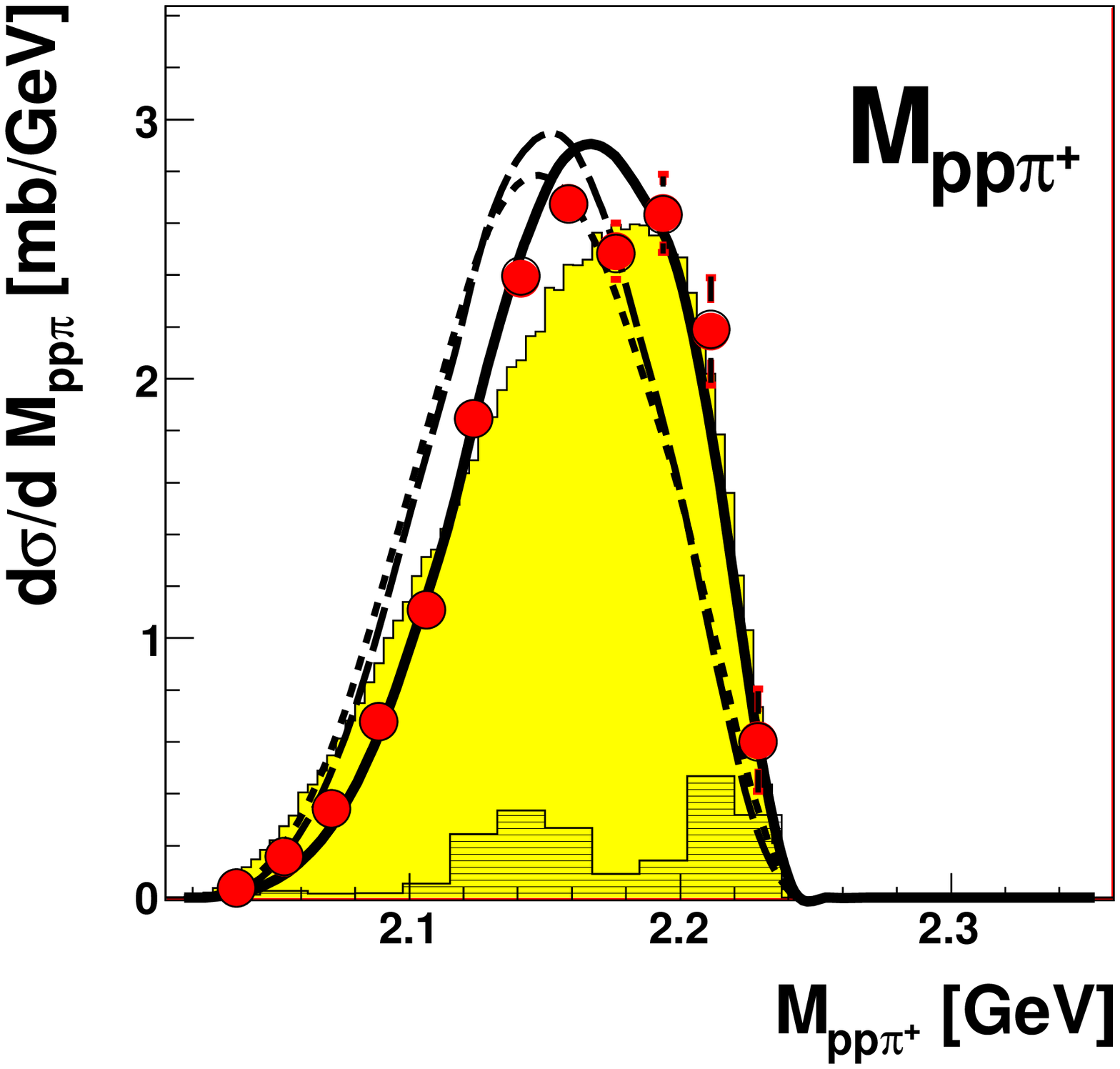}
\includegraphics[width=0.49\columnwidth]{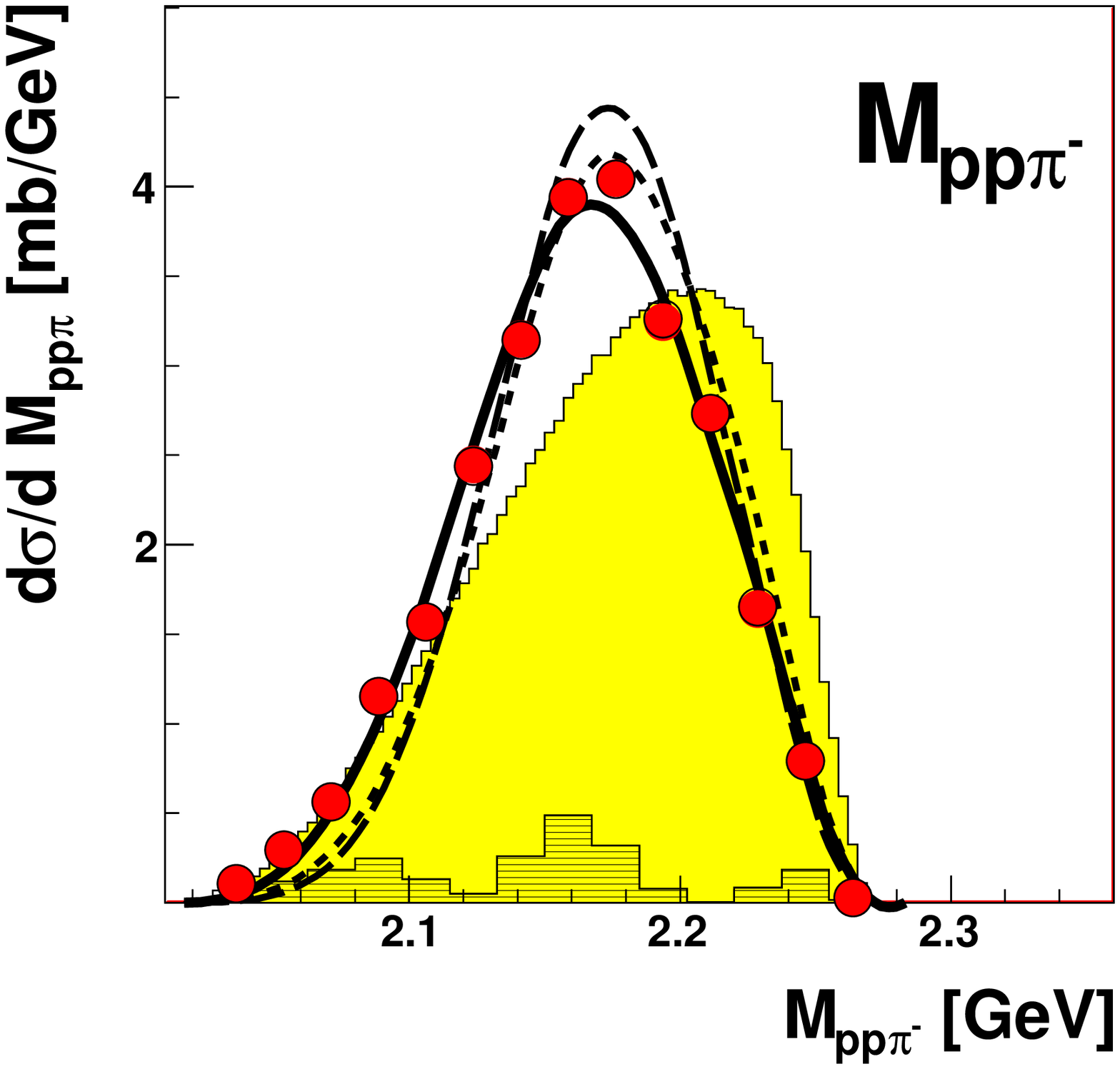}
\includegraphics[width=0.49\columnwidth]{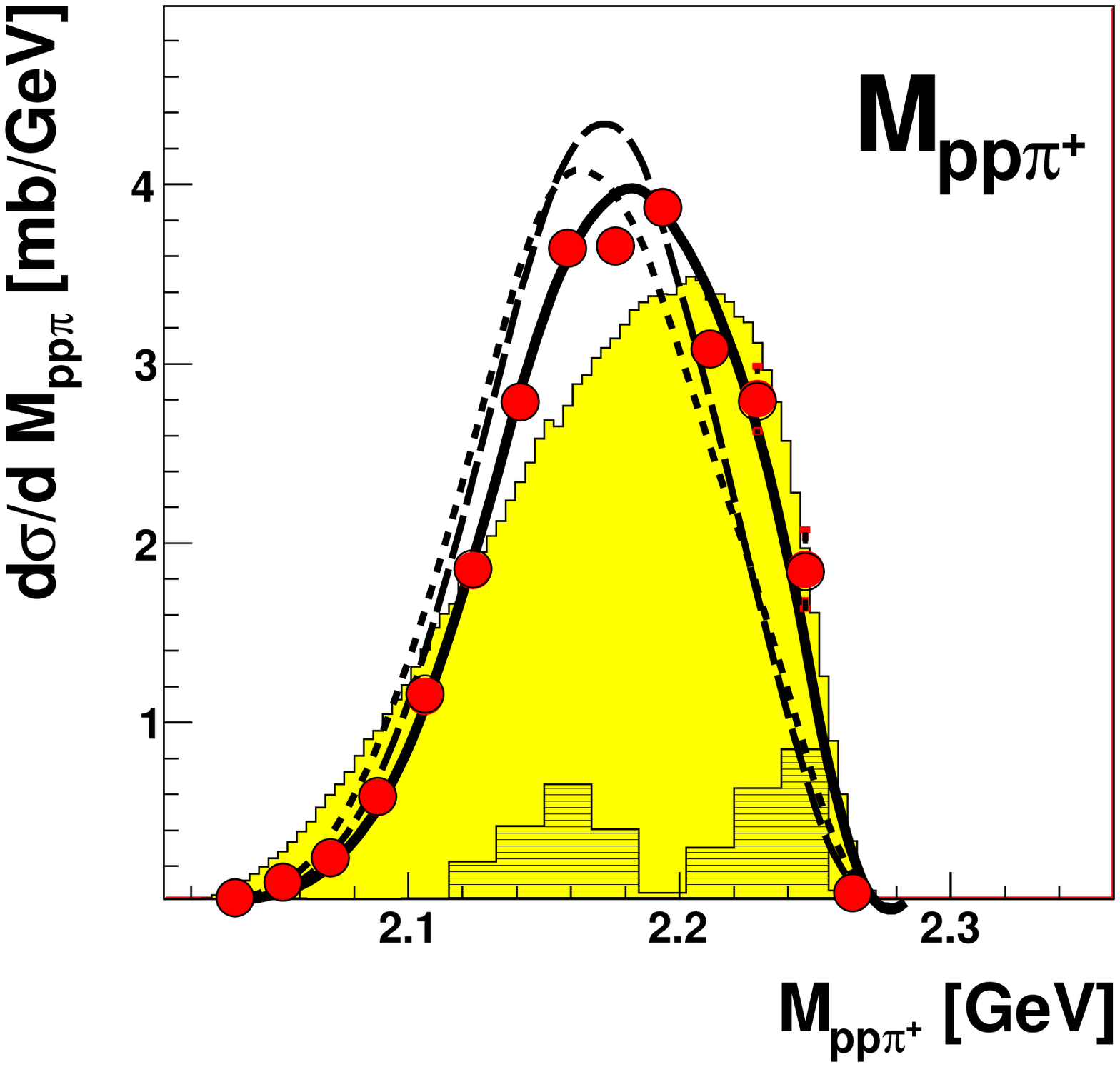}
\includegraphics[width=0.49\columnwidth]{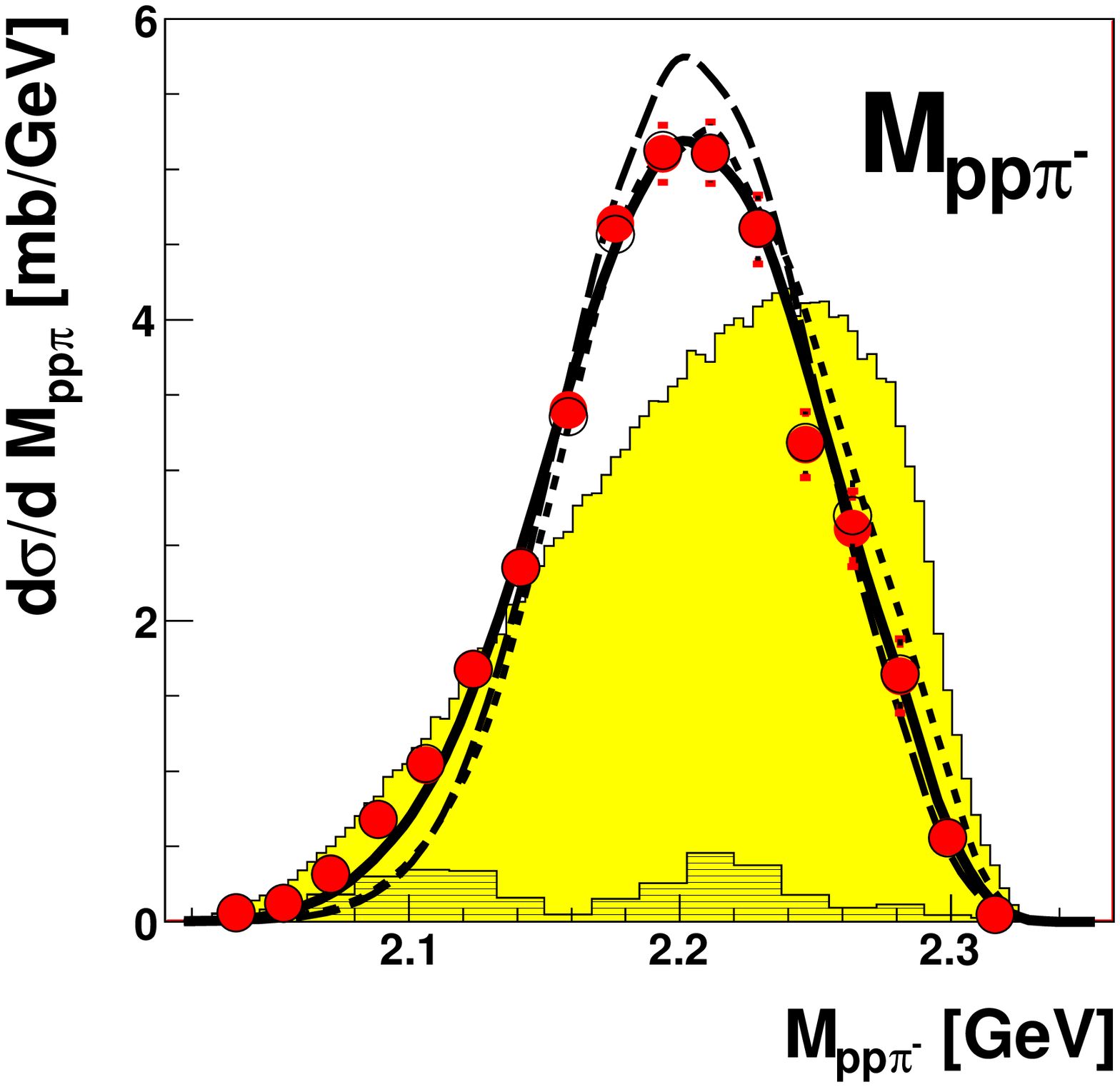}
\includegraphics[width=0.49\columnwidth]{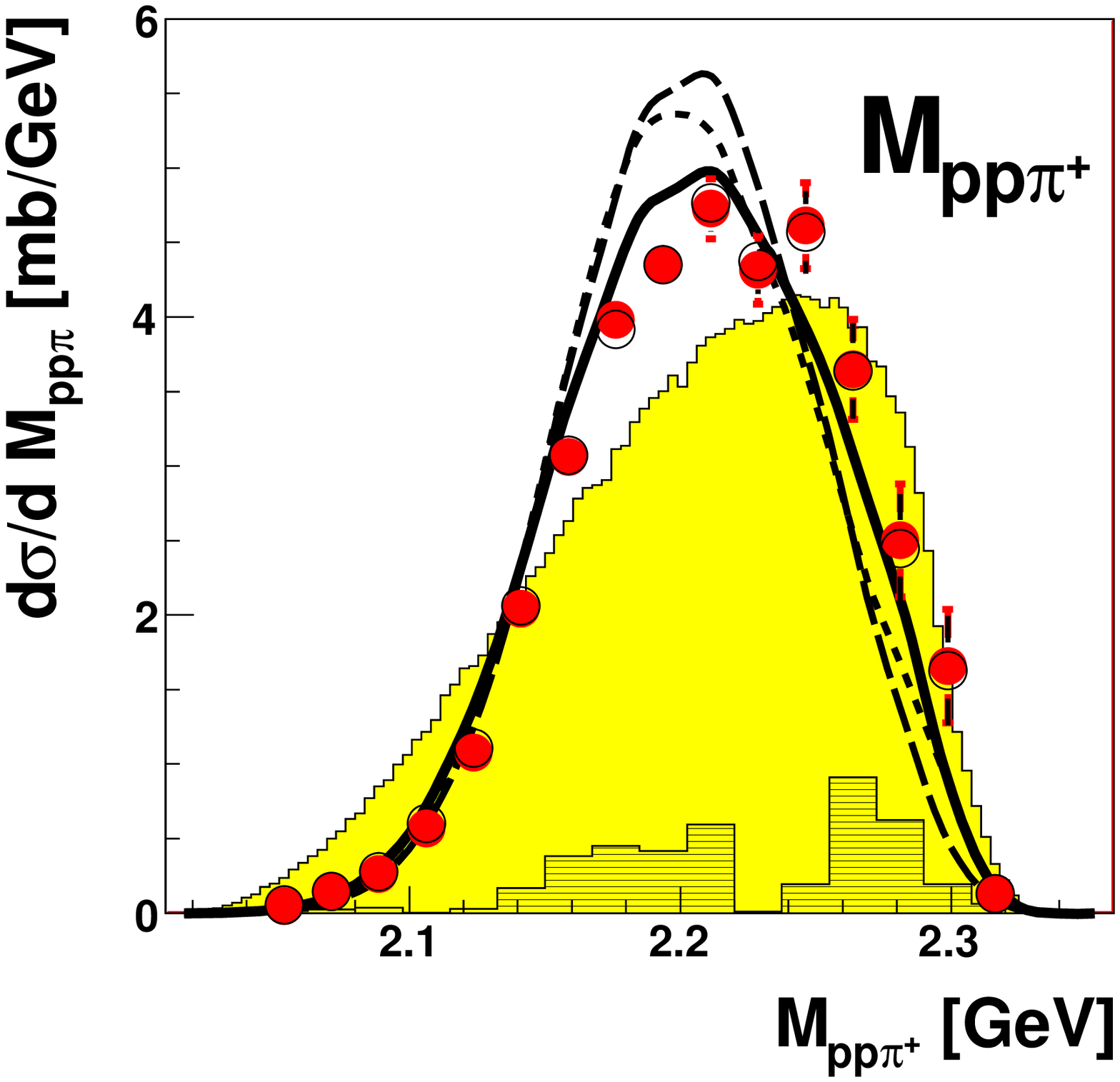}
\caption{(Color online) 
  Same as Fig.~\ref{fig5}, but for the differential distributions of the
  invariant-masses $M_{pp\pi^-}$ (left) and $M_{pp\pi^+}$ (right) for the
  energy bins at $T_p$ = 1.10 (top), 1.18 (middle) and 1.31 GeV (bottom).
}
\label{fig8}
\end{center}
\end{figure}

\begin{figure} [t]
\begin{center}
\includegraphics[width=0.49\columnwidth]{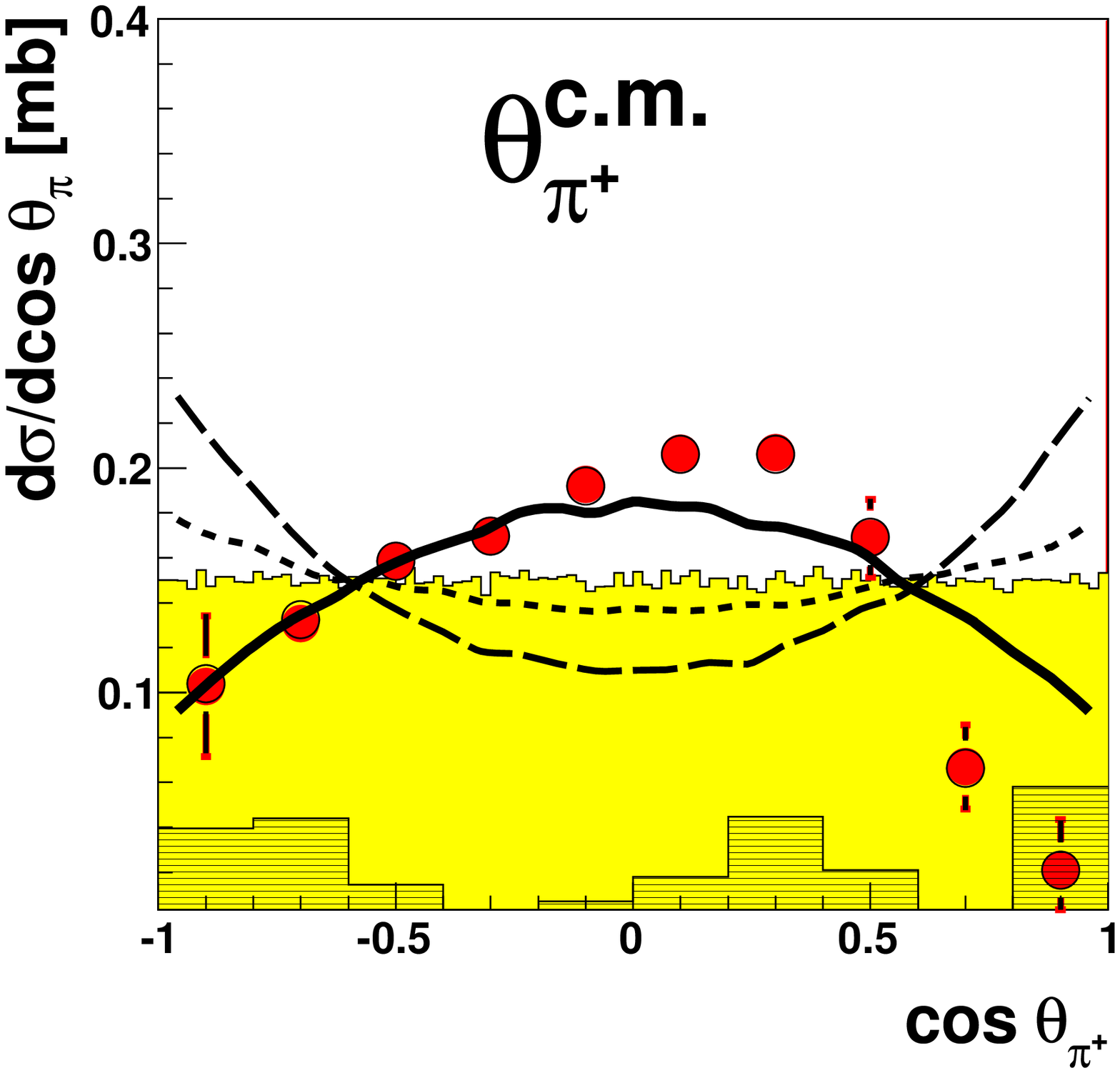}
\includegraphics[width=0.49\columnwidth]{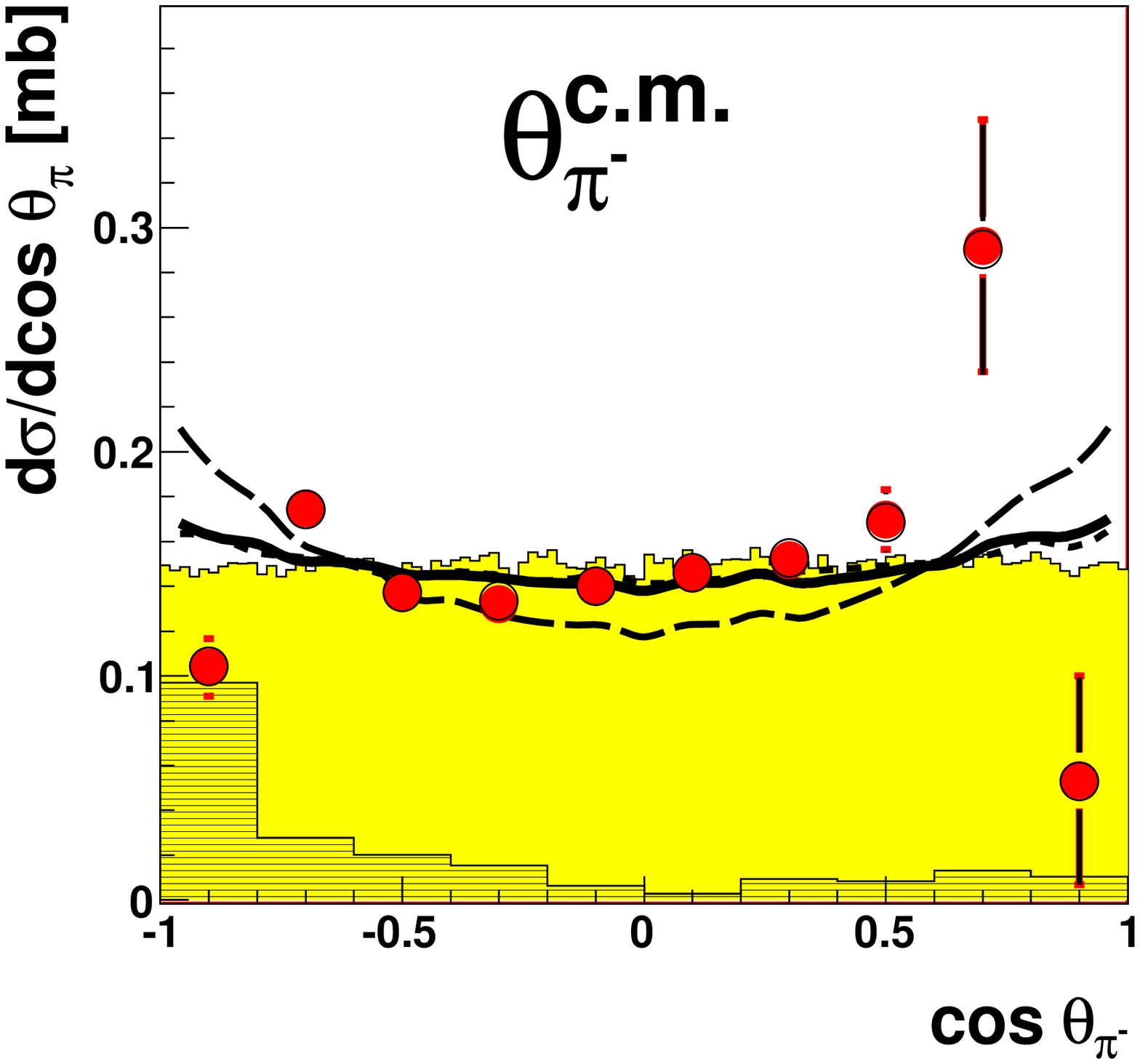}
\includegraphics[width=0.49\columnwidth]{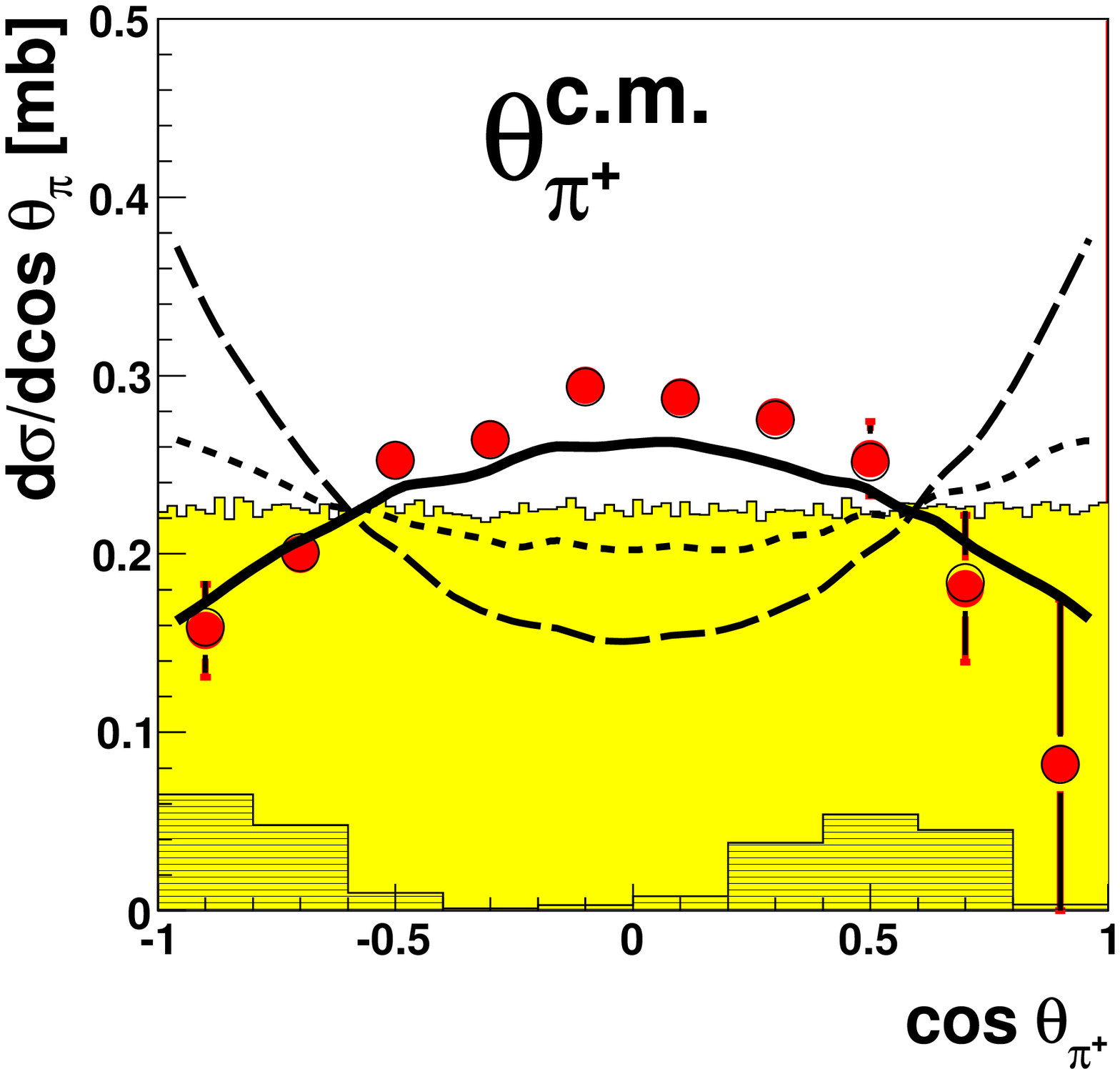}
\includegraphics[width=0.49\columnwidth]{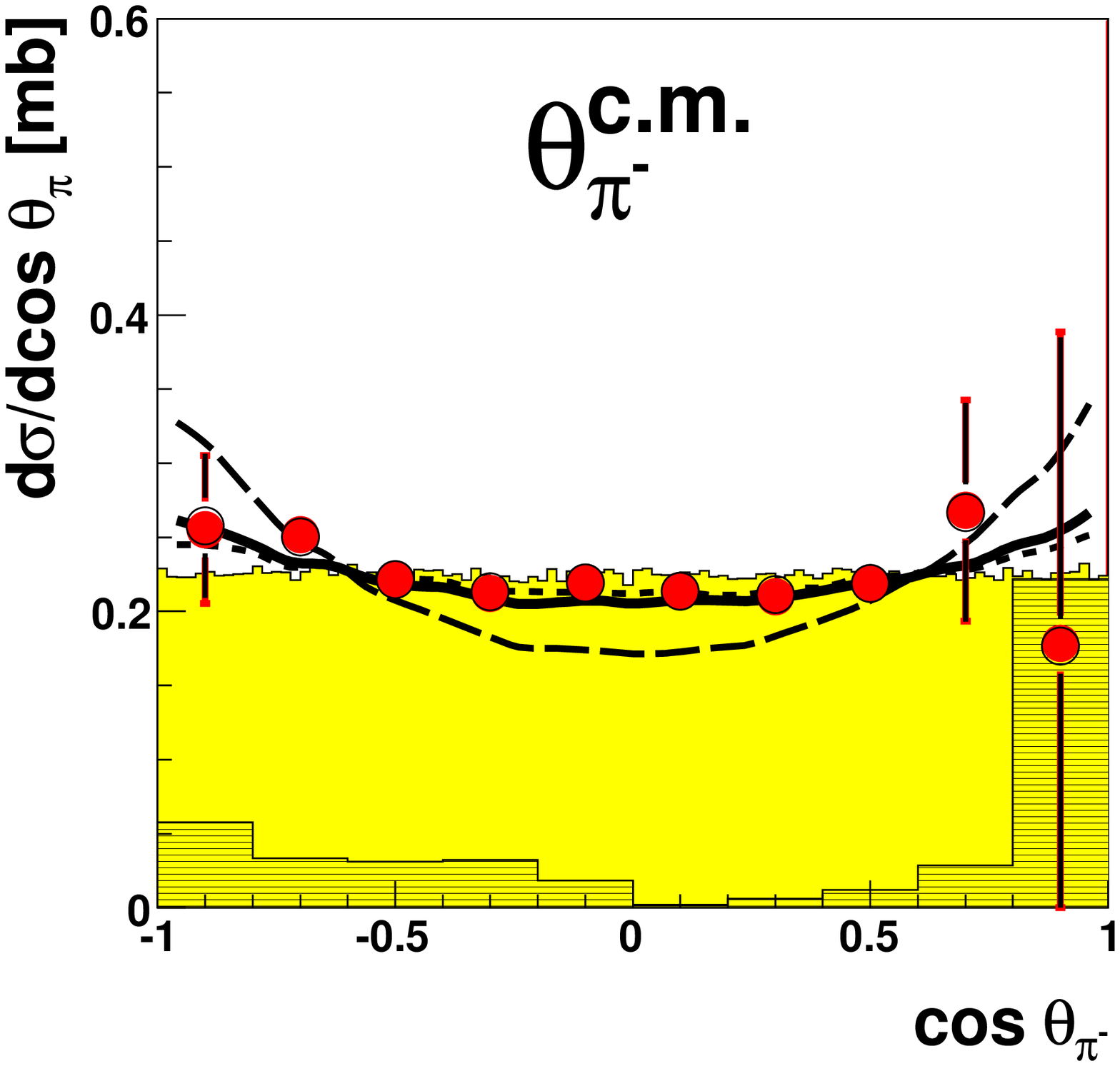}
\includegraphics[width=0.49\columnwidth]{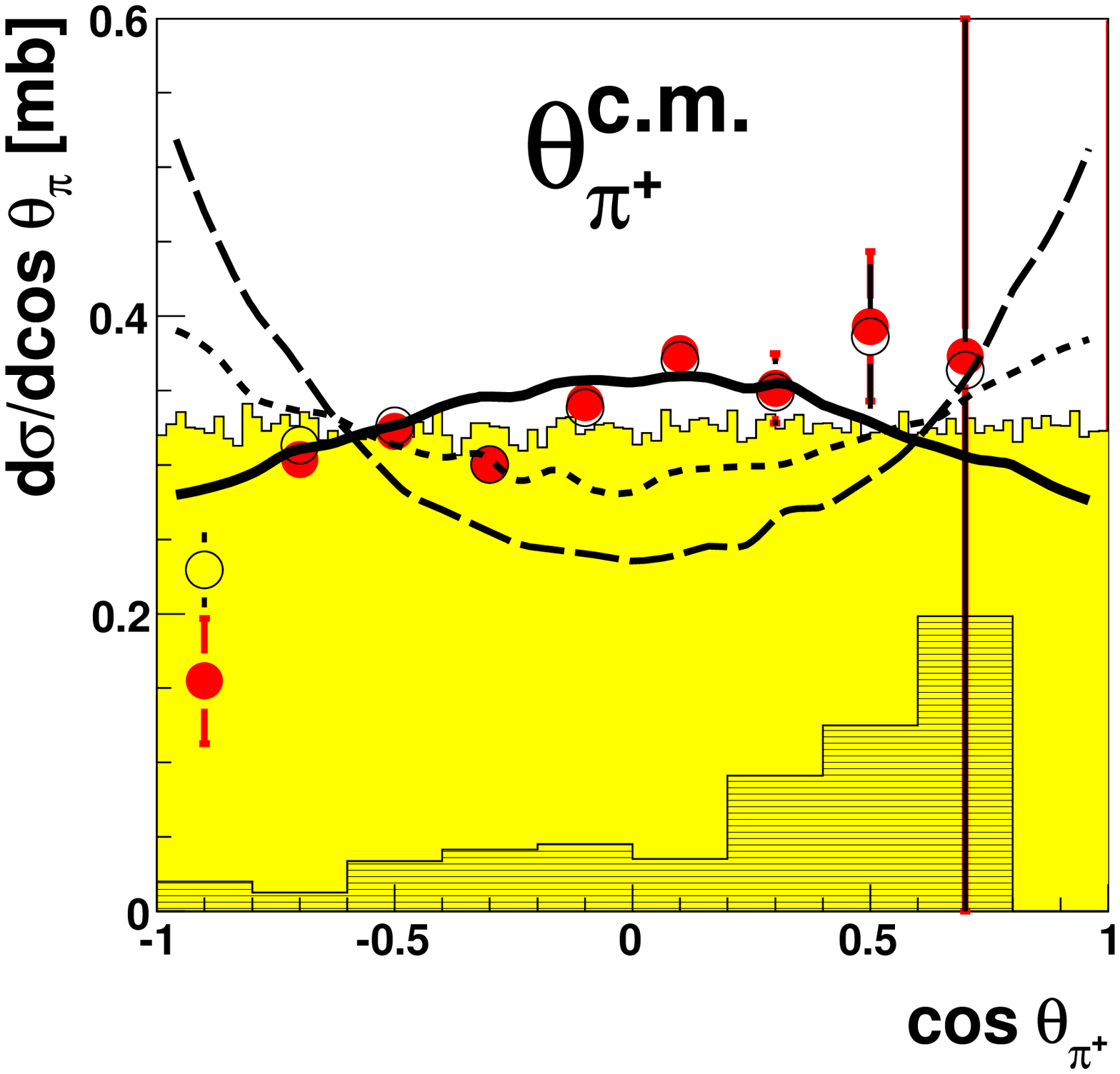}
\includegraphics[width=0.49\columnwidth]{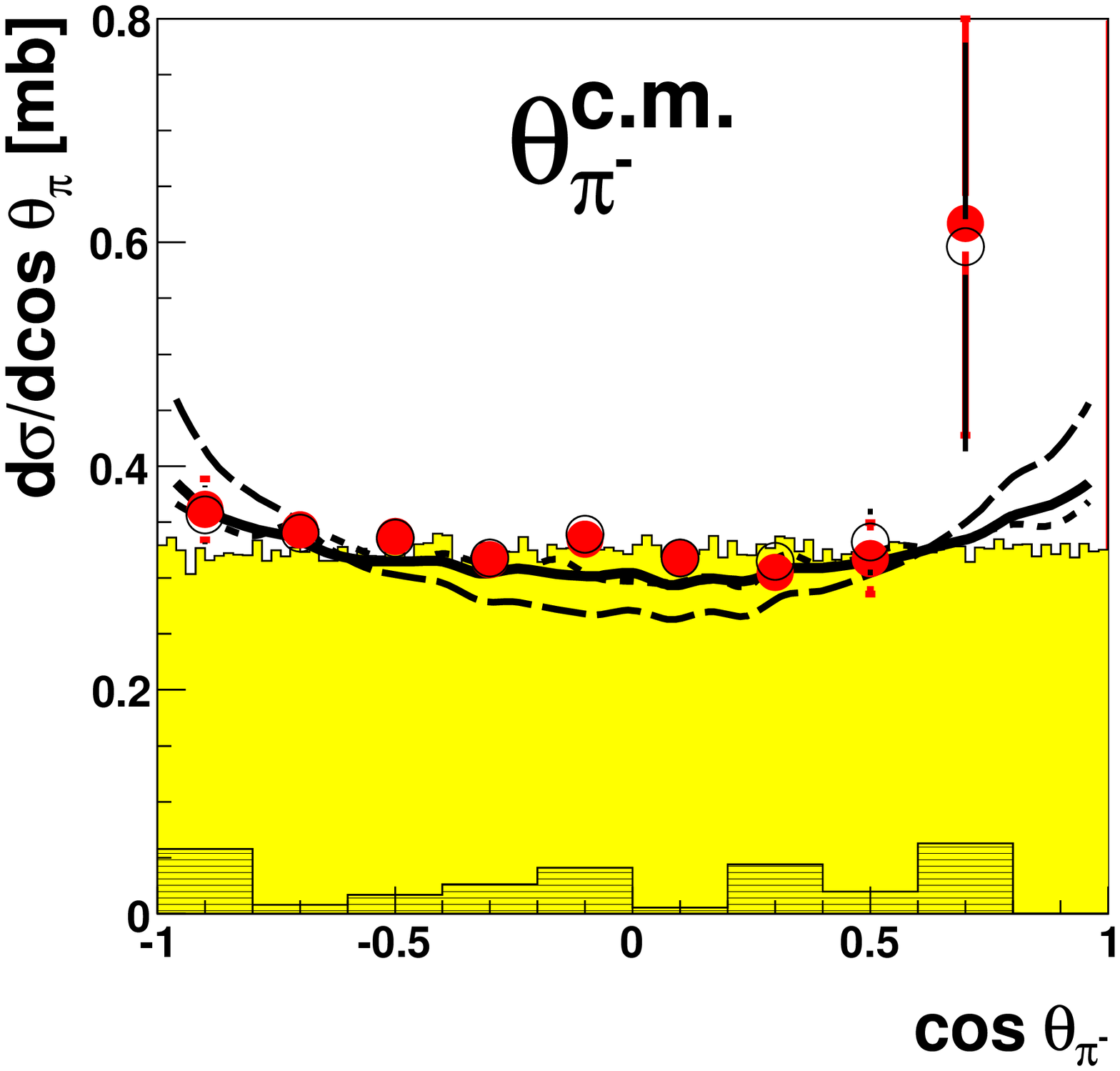}
\caption{(Color online) 
  Same as Fig.~\ref{fig5}, but for the differential distributions of the
  pion angles $\Theta_{\pi^+}^{c.m.}$ (left) and $\Theta_{\pi^-}^{c.m.}$
  (right) for the energy bins at $T_p$ = 1.10 (top), 1.18 (middle) and 1.31
  GeV (bottom). 
}
\label{fig9}
\end{center}
\end{figure}

\begin{figure} 
\begin{center}
\includegraphics[width=0.49\columnwidth]{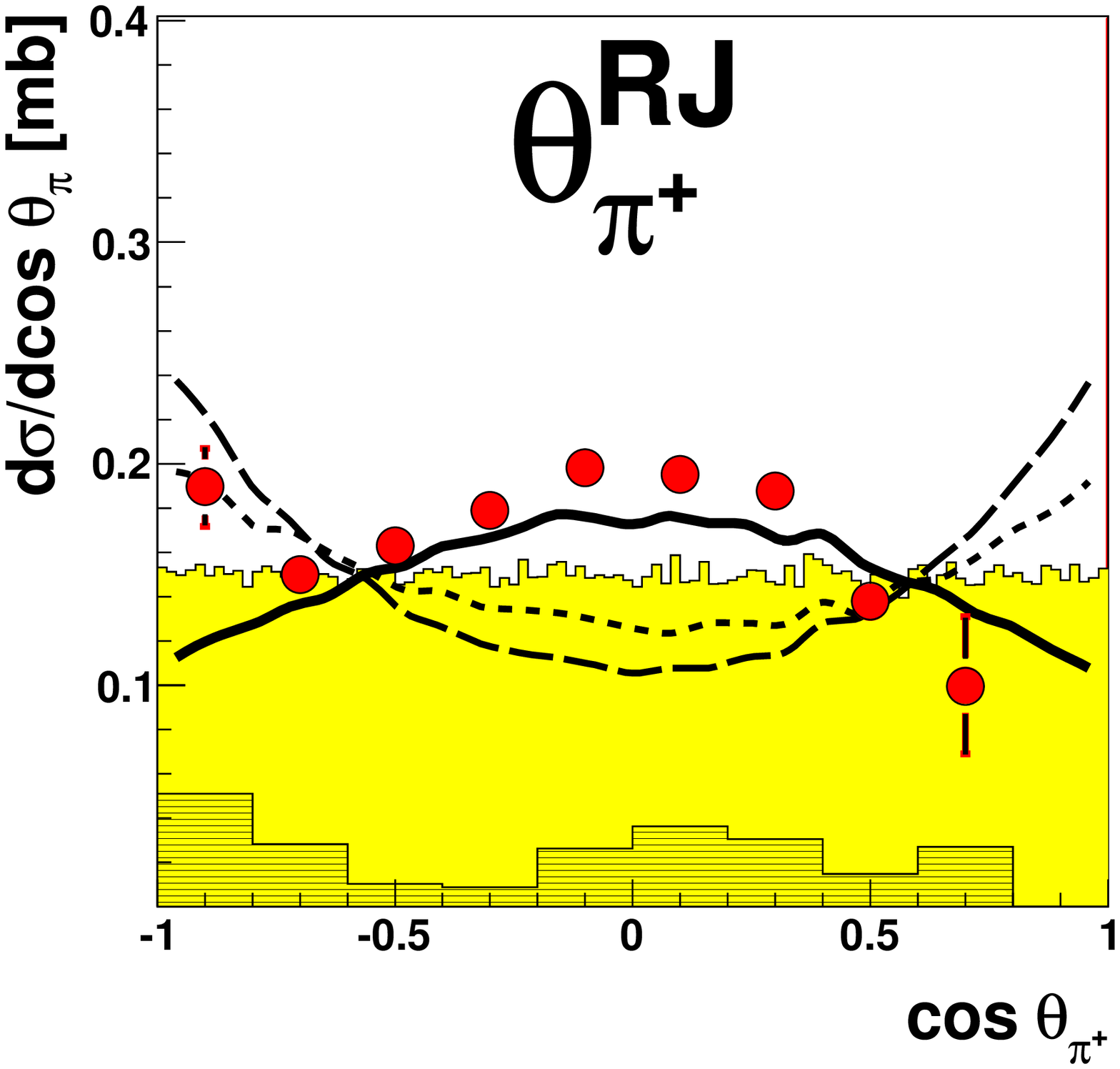}
\includegraphics[width=0.49\columnwidth]{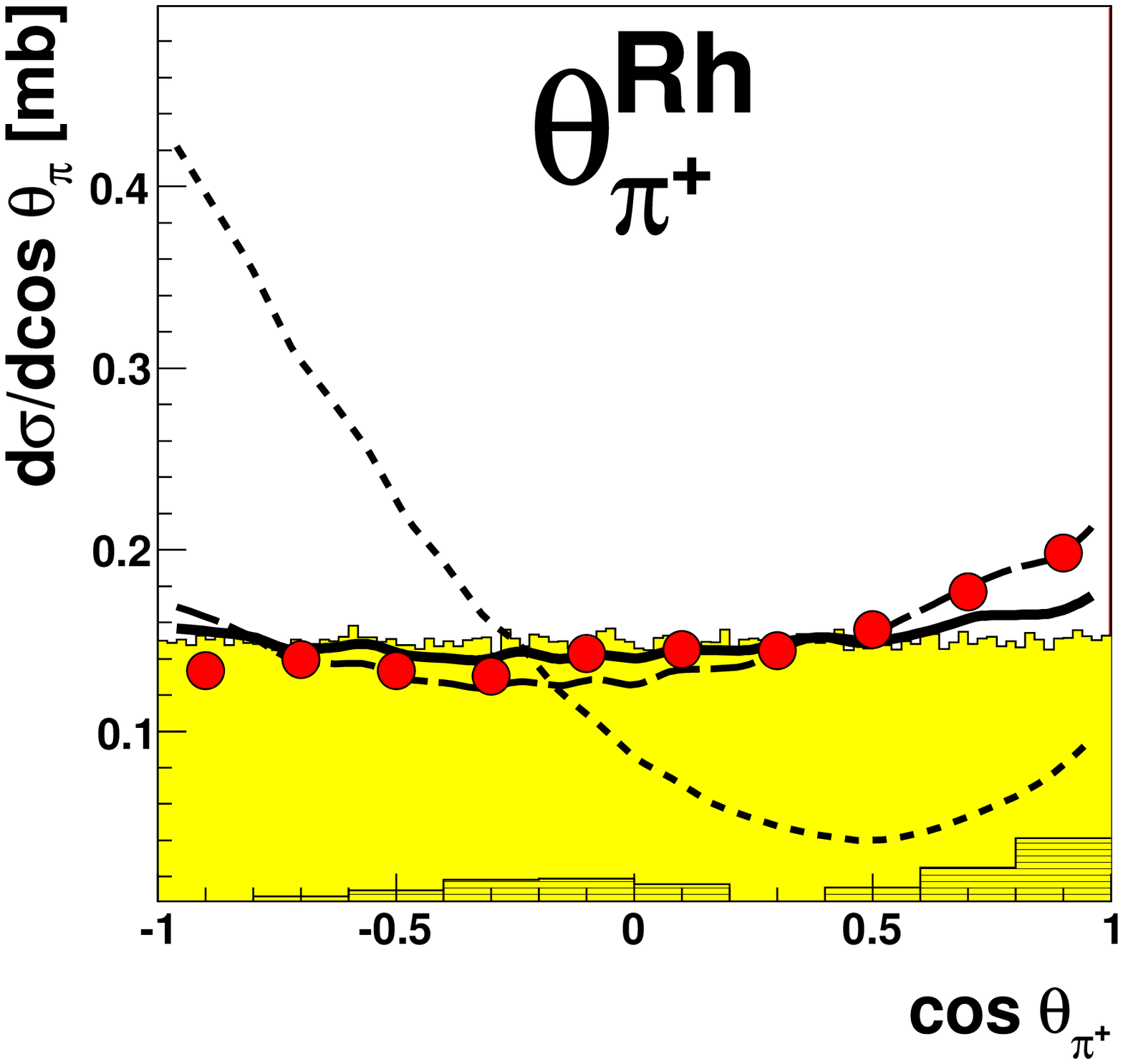}
\includegraphics[width=0.49\columnwidth]{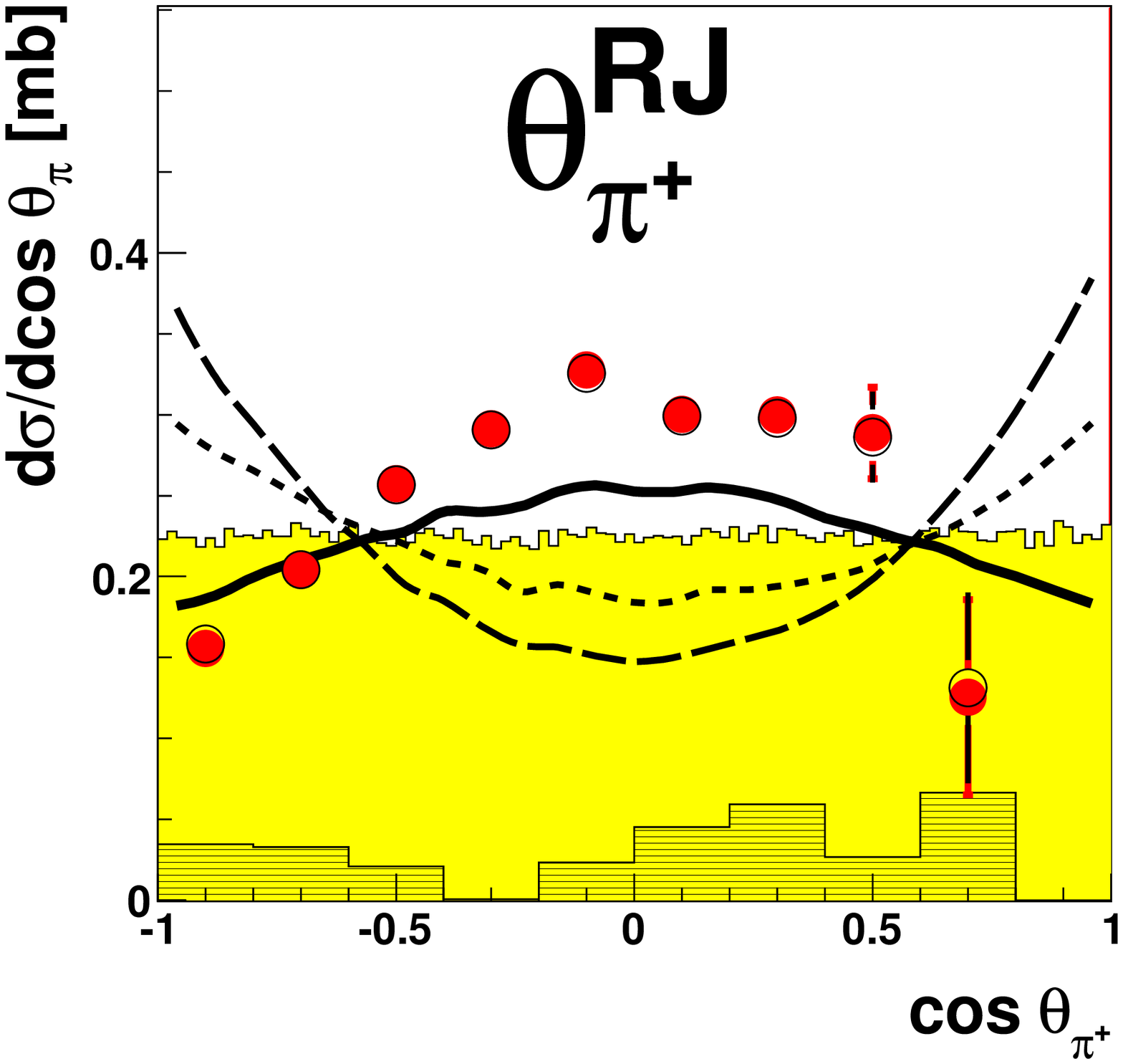}
\includegraphics[width=0.49\columnwidth]{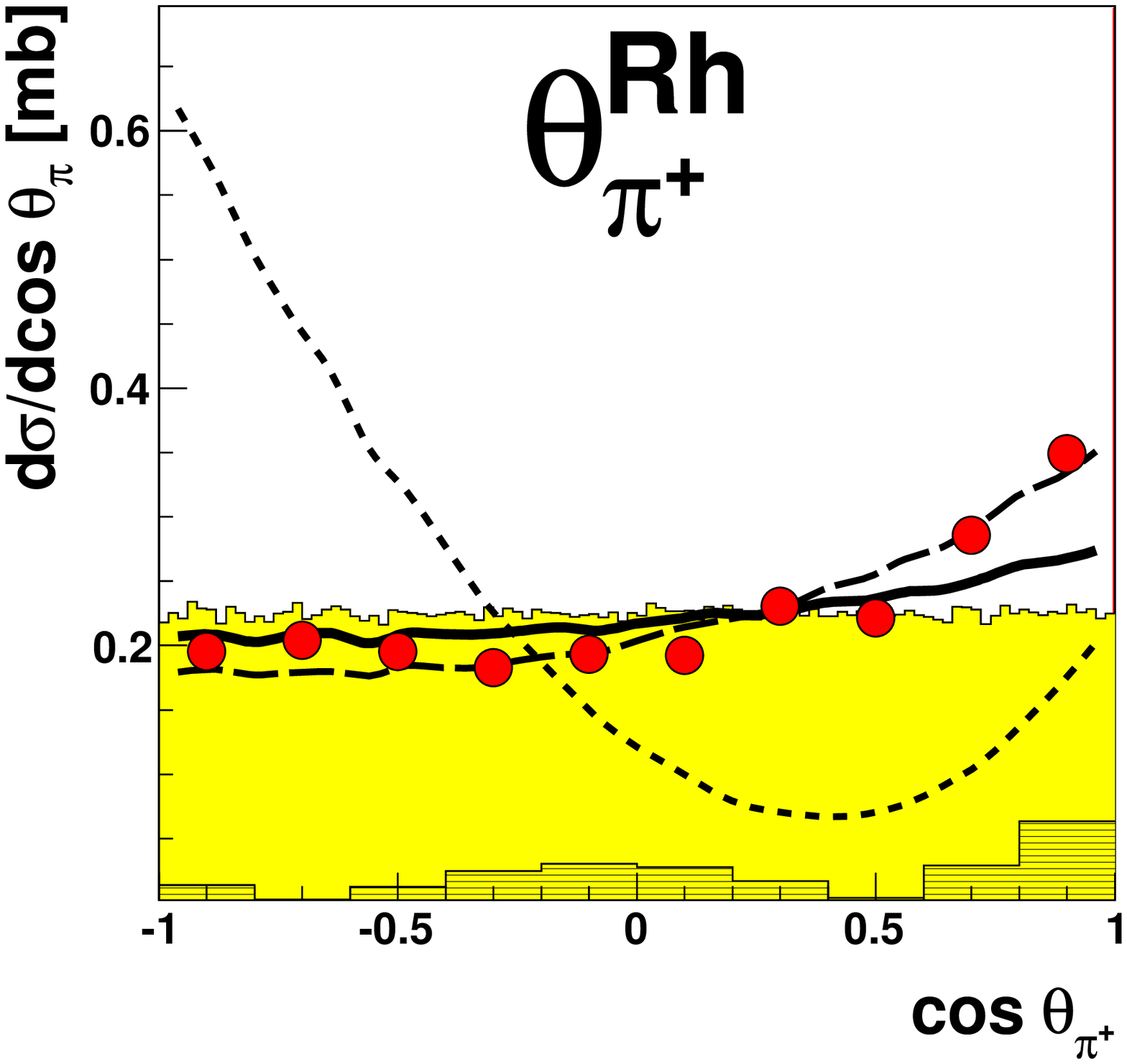}
\includegraphics[width=0.49\columnwidth]{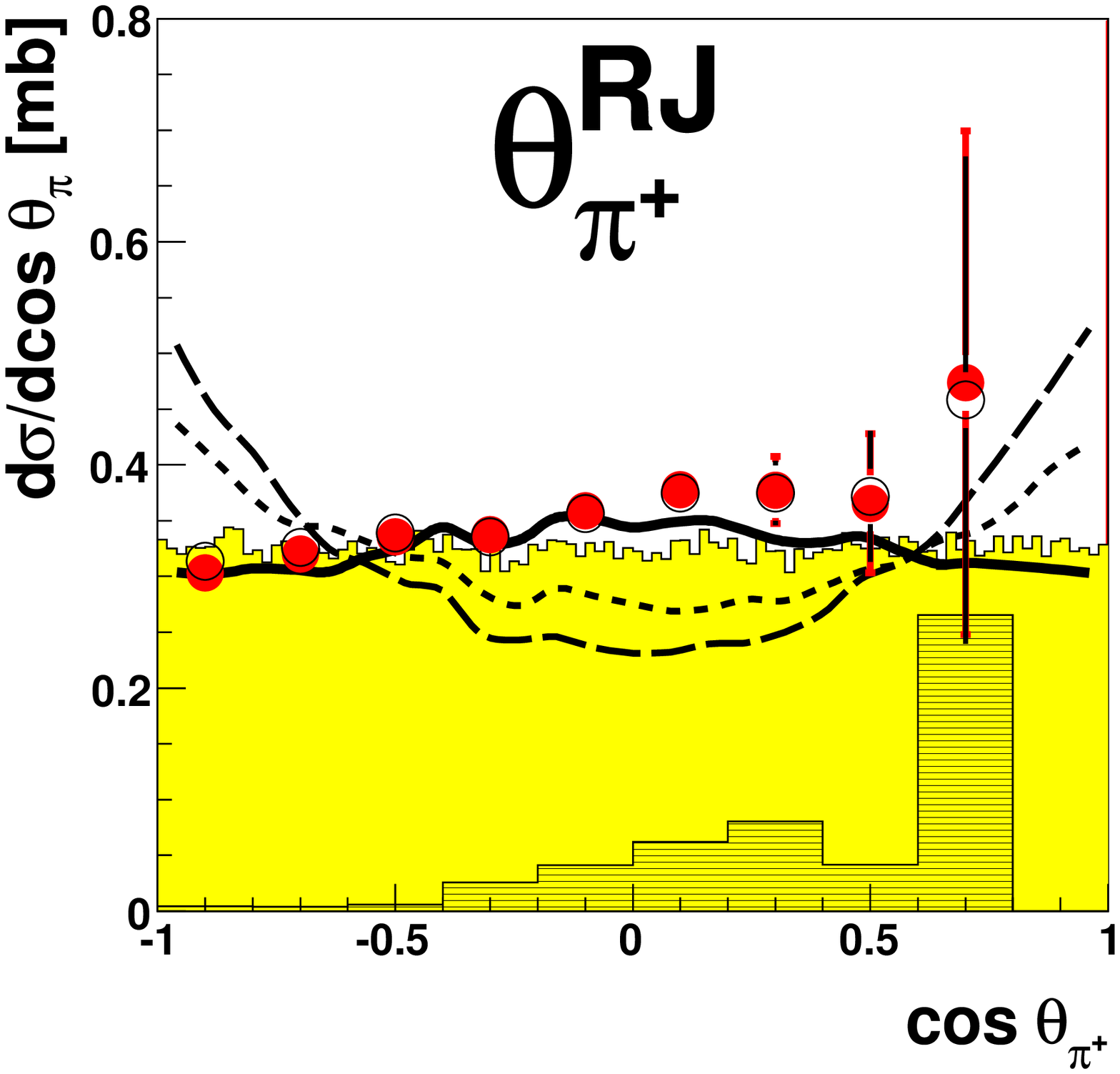}
\includegraphics[width=0.49\columnwidth]{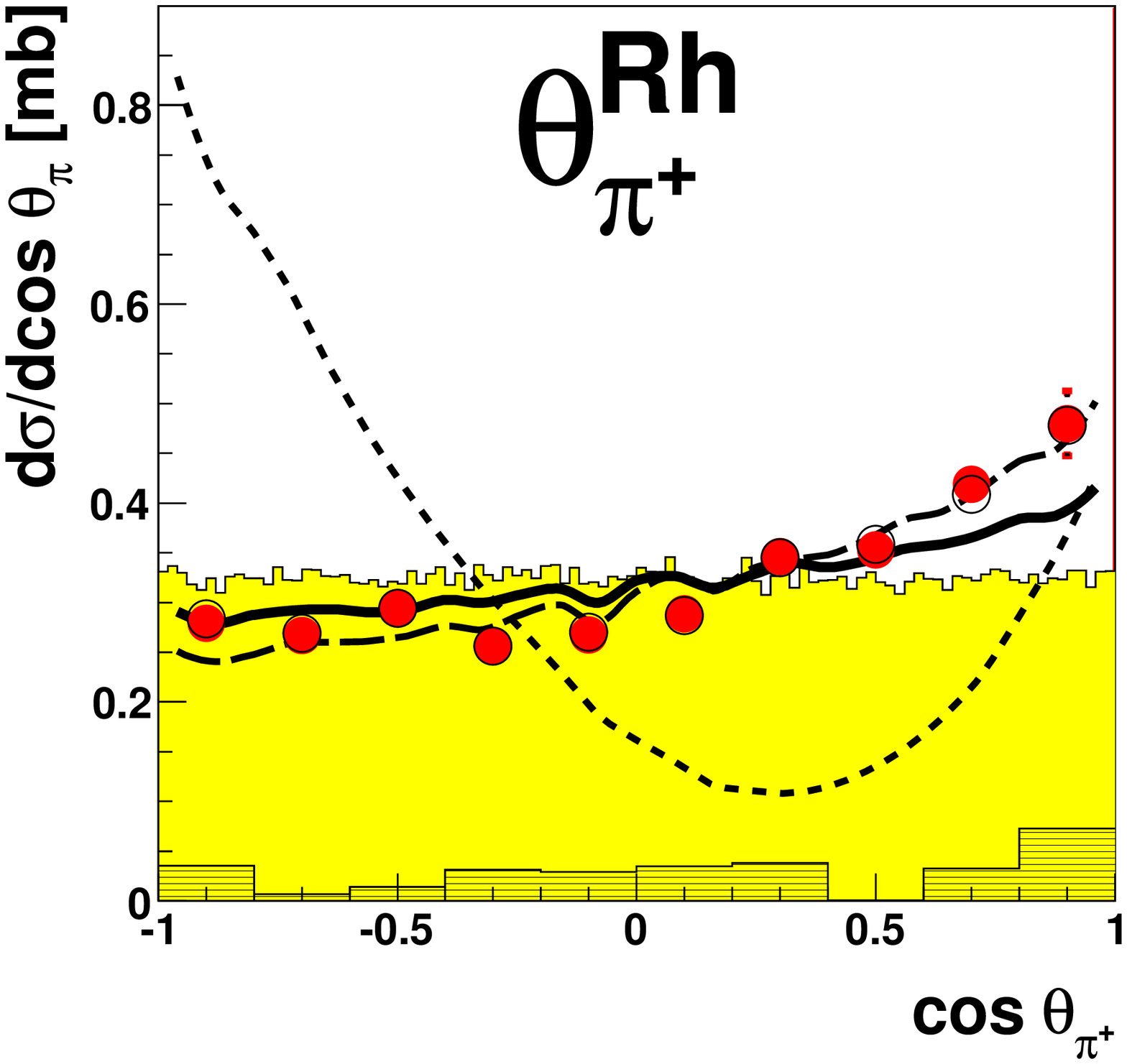}
\caption{(Color online) 
   Same as Fig.~\ref{fig5}, but for the the angles of positive pions in the
   $D_{21}$ resonance subsystem either in the Jackson frame
   ($\Theta_{\pi^+}^{RJ}$, left) or in the helicity frame
   ($\Theta_{\pi^+}^{Rh}$, right) for the energy bins at $T_p$ = 1.10 (top),
   1.18 (middle) and 1.31 GeV (bottom).
}
\label{fig10}
\end{center}
\end{figure}

At the low-energy side of the beam-energy interval covered by our data the
$D_{21}$ resonance contributes nearly 60$\%$ to the total cross section 
shrinking slightly to less than 50$\%$ at the high-energy end. Hence we expect
to observe 
no substantial changes in the differential distributions of
$\Theta_{\pi^+}^{c.m.}$, $M_{p\pi^-}$ and $M_{pp\pi^-}$ , just a smooth
transition from a more to a somewhat less $D_{21}$ dominated scenario. 
In Figs.~\ref{fig7} - \ref{fig9} we plot the three crucial distributions
together with their counterparts for the bins at lowest, central and highest
energy. For the $\pi^+$ angular distribution we show also their $D_{21}$
resonance subsystem representations for the three energy bins in
Fig.~\ref{fig10}. 

Indeed, we observe no significant changes, just a smooth transition of
strength to higher masses in the  $M_{p\pi}$ and $M_{pp\pi}$ spectra. 
Simultaneously we observe for the
$\Theta_{\pi^+}^{c.m.}$ distribution the transition from a pronounced concave
shape at the low-energy bin to a slightly flatter distribution at the
high-energy bin.
The observed smooth energy dependence of differential
distributions is in accord with the $D_{21}$ hypothesis (solid lines in
Figs.~\ref{fig7} - \ref{fig10}).  Unfortunately, there are no such data
available for the energy region $T_p$ = 0.9 - 1.0 GeV, where due to the
opening of the $D_{21}$ channel the changes in these spectra are expected to
be much bigger.

If this scenario is correct, then the $D_{21}$ contribution of
nearly 50$\%$ should also persist to higher energies and impress its specific
features on the differential observables. Though there are no
high-statistics data, there exist at least two bubble-chamber measurements at
$T_p$ = 1.37 \cite{Brunt} and 2.0 GeV \cite{Pickup}, which show a few 
differential distributions. Despite limited statistics their $M_{p\pi^+}$
and $M_{p\pi^-}$ spectra clearly exhibit the same trend as we observe,
namely a strongly excited $\Delta^{++}$ resonance in combination with a much
reduced $\Delta^0$ excitation.

\subsection{resume}

As we have demonstrated, the addition of an isotensor dibaryon resonance is
able to settle the shortcomings of the "modified Valencia" calculations for
the $pp \to pp\pi^+\pi^-$ reaction. However, before we can take this as an
evidence  we for the existence of an isotensor $\Delta N$ resonance, we 
have to investigate, whether this dibaryon hypothesis leads to inconsistencies
in the description of other two-pion production channels. The reason is that
such a state may decay also into $NN\pi$ channels other than the $pp\pi^+$
channel, albeit with  much reduced branchings due to their much inferior isospin
couplings. Consequently, such a resonance may also contribute to other
two-pion production channels. In particular we have to consider, whether it
can affect the $pp \to pp\pi^0\pi^0$ reaction with its comparatively small cross
section at the energies of interest here. However, the $D_{21}$ production via
the $^3P_1$ partial wave leaves the two emitted pions in relative $p$-wave to
each other. Therefore they must be in an isovector state by Bose
symmetry. Such a $\rho$-channel situation is not possible for identical pions
in line with the isospin relations for the various two-pion production
channels. Hence there are no contributions from
$D_{21}$ in $pp\pi^0\pi^0$ and $nn\pi^+\pi^+$ channels, {\it i.e.} there is no
consistency problem.

From a fit to the data we obtain a mass $m_{D_{21}}$ = 2140(10) MeV and
a width 
$\Gamma_{D_{21}}$ = 110(10) MeV. The mass is in good agreement with the
prediction of Dyson and Xuong \cite{Dyson}. From their Faddeev calculations
Gal and Garcilazo \cite{GG} obtain slightly larger values for mass and width.
Within uncertainties the extracted mass and width of the $D_{21}$ state
coincide with those for the 
$D_{12}$ state. This means that the masses of this dibaryon doublet do not
exhibit any particular isospin dependence -- just as assumed in the work of
Dyson and Xuong after having noted the near mass degeneracy of the deuteron
groundstate $D_{01}$ with the virtual $^1S_0$ state $D_{10}$. Obviously, the
spin-isospin splitting for dibaryons is different from that of baryons.

Possibly this resonance was sensed already before in the pionic double charge
exchange reaction on nuclei. There the so-called non-analog transitions
exhibit an unexpected resonance-like behavior in the region of the $\Delta$
resonance \cite{hcl,hclDCX, JohnsonMorris}. For its explanation the DINT
mechanism \cite{Johnson,Johnson1,JohnsonKisslinger,Wirzba} was introduced,
which in essence can be imagined as representing a $\Delta N$ system with $I(J^P)
= 2(1^+)$ in the intermediate state \cite{JohnsonKisslinger}.





\section{Summary and Conclusions}

Total and differential cross sections of the $pp \to pp\pi^+\pi^-$ reaction
have been measured exclusively and kinematically complete in the energy range
$T_p = 1.08 - 1.36$ GeV ($\sqrt s$ = 2.35 - 2.46 GeV) by use of the quasi-free
process $pd \to pp\pi^+\pi^- + n_{spectator}$. The results for the total
cross section are in good agreement with previous bubble-chamber data. For the
differential cross sections there are no data available from previous
measurements in the considered energy range.

The original Valencia calculations describing Roper and $\Delta\Delta$
excitations by $t$-channel  meson exchange account well for the total
cross section, but fail badly for the differential distributions of the $pp
\to pp\pi^+\pi^-$ reaction. These calculations also have been shown to fail in
other two-pion production channels, both for total and differential cross
sections. 
 
The differential cross sections for the $pp \to pp\pi^+\pi^-$ reaction are
somewhat better accounted for by the "modified Valencia" calculations, but
still fail strikingly for the $M_{p\pi^-}$, $M_{pp\pi^-}$ and $\Theta_{\pi^-}^{c.m.}$
distributions. These calculations, which were tuned to the $pp \to 
pp\pi^0\pi^0$ and $pp \to nn\pi^+\pi^+$ reactions, have been shown to provide
a good description of the other two-pion channels both in total and in
differential cross sections. However, these so far very
successful calculations predict also a much too small total cross section for
the $pp\pi^+\pi^-$ channel at energies above $T_p \approx$ 0.9 GeV. 
 
This failure can be cured, if there is an opening of a new reaction
channel near $T_p \approx$ 0.9 GeV, {\it i.e.}, near the $\Delta N\pi$
threshold, which nearly exclusively feeds the
$pp\pi^+\pi^-$ channel. Such a process is given by the associated production of
the isotensor $\Delta N$ state $D_{21}$ with specific signatures in invariant
mass spectra and in the $\pi^+$ angular distribution. We have
demonstrated that such a process provides a quantitative description of the data
for the $pp \to pp\pi^+\pi^-$ reaction --- both for the total cross section
and for all differential distributions. 

This $D_{21}$ state has been predicted already in 1964 by Dyson and Xuong
\cite{Dyson} and more recently by Gal and Garcilazo \cite{GG}, who also
calculated its decay width. It is remarkable that five out of the six dibaryon
states predicted in 1964 by considering $SU(6)$ symmetry breaking have now
been verified with masses very close to the predicted ones. For the sixth
state, $D_{30}$, only upper limits have been found so far \cite{D30search},
but this subject deserves certainly further, more detailed investigations.


\section{Acknowledgments}

We acknowledge valuable discussions with A. Gal, Ch. Hanhart, V. Kukulin and
G. J. Wagner on this issue. We are particularly indebted to L. Alvarez-Ruso
for using his code. We are also grateful to the anonymous referee of the
Letter version of this topic, who suggested us to look also into subsystem
distributions. 
This work has been supported by DFG (CL214/3-1 and 3-2) and STFC
(ST/L00478X/1) as well as by the Polish National Science Centre through the
grants 2016/23/B/ST2/00784 and 2013/11/N/ST2/04152.

\end{document}